\documentclass{aa}
\usepackage{psfig}
\usepackage{times}
\usepackage{color}
\newfont{\vssn}{cmss10 scaled 1050}
\newfont{\vsss}{cmss10 scaled 450}

\def\ha{H$\alpha$}
\def\lsb{LSB$^{\star}$}
\def\hst{{\sl HST}}
\def\nw{{\rm NW}}
\def\se{{\rm SE}}

\newcommand{\sbb}{mag/$\sq\arcsec$}
\def\hh{\ion{H}{ii}}
\def\H1{\ion{H}{i}}
\def\ha{H$\alpha$}

\def\p25{{\it P$_{25}$}}
\def\e25{{\it E$_{25}$}}
\def\ew{{\sl EW}}
\def\kato{\rule[-1.25ex]{0cm}{1.25ex}}
\def\pano{\rule[0.0ex]{0cm}{2.5ex}}
\def\msun{$M_{\odot}$}
\def\zsun{$Z_{\odot}$}

\newcounter{qub}
\begin{document}
\title{The blue compact dwarf galaxy I Zw 18: a comparative study of its 
low-surface-brightness component}

\author{P. Papaderos \inst{1}
\and Y. I. Izotov \inst{2}
\and T. X. Thuan \inst{3}
\and K. G. Noeske \inst{1}
\and K. J. Fricke \inst{1}
\and N. G. Guseva \inst{2}
\and R. F. Green \inst{4} }

\offprints{P. Papaderos; papade@uni-sw.gwdg.de}
\institute{      Universit\"ats--Sternwarte, Geismarlandstra\ss e 11,
                 D--37083 G\"ottingen, Germany
\and
                 Main Astronomical Observatory,
                 National Academy of Sciences of Ukraine,
                 27 Zabolotnoho str., 03680 Kyiv, Ukraine
\and
                 Astronomy Department, University of Virginia, Charlottesville, 
                 VA 22903, USA
\and
                 National Optical Astronomy Observatory, Tucson, AZ 85726, USA}
\date{Received \hskip 2cm; Accepted}

\abstract{Using \hst\thanks{Based on observations with the NASA/ESA \emph{Hubble Space Telescope}, 
obtained at the Space Telescope Science Institute, which is operated by AURA, Inc., under
NASA contract No. NAS 5--26555.} and ground-based optical and 
NIR imaging data\thanks{Obtained at the German--Spanish Astronomical 
Center, Calar Alto, operated by the Max--Planck--Institute for Astronomy, 
Heidelberg, jointly with the Spanish National Commission for Astronomy.}$^,$
\thanks{Obtained at the Kitt Peak National Observatory, 
operated by the Association of Universities for Research in Astronomy, 
Inc., under cooperative agreement with the National Science Foundation.}, we
investigate whether the blue compact dwarf (BCD) galaxy I Zw 18 possesses 
an extended low-surface-brightness (LSB) old stellar population underlying 
its star-forming regions, as is the case in the majority of BCDs.
This question is central to the long-standing debate on the evolutionary
state of I Zw 18. 
We show that the exponential intensity decrease observed in the filamentary LSB 
envelope of the BCD out to $\ga$18\arcsec\ ($\ga$1.3 kpc assuming a distance of 15 Mpc)
is not due to an evolved stellar disc underlying its star-forming regions, but
rather, due to extended ionized gas emission.
Ionized gas accounts for more than 80\% of the line-of-sight emission at a galactocentric 
distance of $\sim$ 0.65 kpc ($\sim$ 3 effective radii), and for 
$\ga$ 30\% to 50\% of the $R$ light of the main body of I Zw 18. 
Broad-band images reveal, after subtraction of nebular line emission,
a relatively smooth stellar host extending slightly beyond 
the star-forming regions. 
This unresolved stellar component, though very compact, is not exceptional for intrinsically 
faint dwarfs with respect to its structural properties. 
However, being blue over a radius range of $\sim$ 5 exponential scale lengths 
and showing little colour contrast to the star-forming regions,
it differs strikingly from the red LSB host of standard BCDs.
This fact, together with the comparably blue colours of the 
faint C component, $\sim$ 1.6 kpc away from the main body of I Zw 18,
suggests that the formation of I Zw 18 as a whole has occurred
within the last 0.5 Gyr, making it a young BCD candidate.
Furthermore, we show that the ionized envelope of I Zw 18 is not 
exceptional among star-forming dwarf galaxies, neither by its 
exponential intensity fall-off nor by its scale length.
However, contrary to evolved BCDs, the stellar LSB component of I Zw 18
is much more compact than the ionized gas envelope. 
In the absence of an appreciable underlying stellar population, 
extended ionized gas emission dominates in the outer parts of I Zw 18, 
mimicking an exponential stellar disc on optical surface brightness profiles.
}
\maketitle


\markboth {Papaderos et al.}{I Zw 18: a comparative study of its low-surface-brightness component}

\section{Introduction \label{S1} }
Since its discovery by Sargent \& Searle (\cite{SS70}), I Zw 18 has been 
looked at as the prototypical blue compact dwarf (BCD) galaxy.
Its low oxygen abundance (Searle \& Sargent \cite{SS72}), established in
numerous subsequent studies (Lequeux et al. \cite{Leq79}; 
French \cite{Fre80}; Kinman \& Davidson \cite{KD81}; Pagel et
al. \cite{Pagel92}; Skillman \& Kennicutt \cite{SK93}; Martin \cite{Martin96};
V{\'\i}lchez \& Iglesias-Par\'amo \cite{ViIp98}; 
Izotov \& Thuan \cite{IT98a,IT98b};
Izotov et al. \cite{Yu99}) to be 12~+~log(O/H)~$\approx$~7.2 makes it the least
chemically evolved star-forming galaxy in the local Universe. 
Whether this is a signature of youth (cf. e.g. Izotov \& Thuan \cite{IT98a}) 
remains, however, a subject of debate. 

I Zw 18 was described by Zwicky (\cite{Fritz66}) as a pair of compact galaxies, 
subsequently shown to be two compact star-forming (SF) regions within the same 
galaxy with an angular separation of 5\farcs8, the brighter northwestern (\nw) and fainter 
southeastern (\se) components.  
Ground-based and \hst\ observations have revealed that both regions are 
embedded in a low-surface brightness (LSB) filamentary envelope extending 
out to a radius $\sim$20\arcsec\ (Davidson et al. \cite{Dav89}; Dufour \& Hester 
\cite{Duf90}; Dufour et al. \cite{Duf96a}; Martin \cite{Martin96};
\"Ostlin et al. \cite{Gor96}), and within an extensive \H1\ halo with a 
projected size of 60\arcsec $\times$ 45\arcsec\ (van Zee et al. \cite{vZ98}; 
see also Viallefond et al. \cite{Via87}). 

The spectrophotometric properties of the fainter detached
component I Zw 18\,C, located $\sim$ 22\arcsec\ northwest of 
the \nw\ region are still poorly known. 
Dufour et al. (\cite{Duf96a}), Petrosian et al. (\cite{Petr97}), 
Izotov \& Thuan (\cite{IT98a}), van Zee et al. (\cite{vZ98}) and 
Izotov et al. (\cite{Yu01a}) have shown it to have the same 
recession velocity as the main body, thus establishing its 
physical association to I Zw 18. The SF activity of I Zw 18\,C is weak, 
its \ha\ equivalent width not exceeding $\sim$60 \AA\ along the major axis 
(Izotov et al. \cite{Yu01a}).
In spite of deep Keck\,II spectroscopy, Izotov et al. (\cite{Yu01a}) 
failed to detect oxygen lines, so its oxygen abundance is not known.

Colour-magnitude diagram (CMD) studies, based on \hst\ WFPC2 images,
suggest for the main body an age between several 10 Myr
(Hunter \& Thronson \cite{HT95}; Dufour et al. \cite{Duf96b}) and
$\sim$~1 Gyr (Aloisi et al. \cite{Alo99}). 
\"Ostlin (\cite{Gor00}) argues from \hst\ NICMOS
$J$ and $H$ images that a fit to the $J$ vs. $J-H$ CMD is best 
achieved with a stellar population with age as high as 5~Gyr.
As for I Zw 18\,C, Dufour et al. (\cite{Duf96b}) and 
Aloisi et al. (\cite{Alo99}) derive an age of a few hundred Myr.
%
\begin{figure} 
\begin{picture}(16,8.7)
\put(0,0){{\psfig{figure=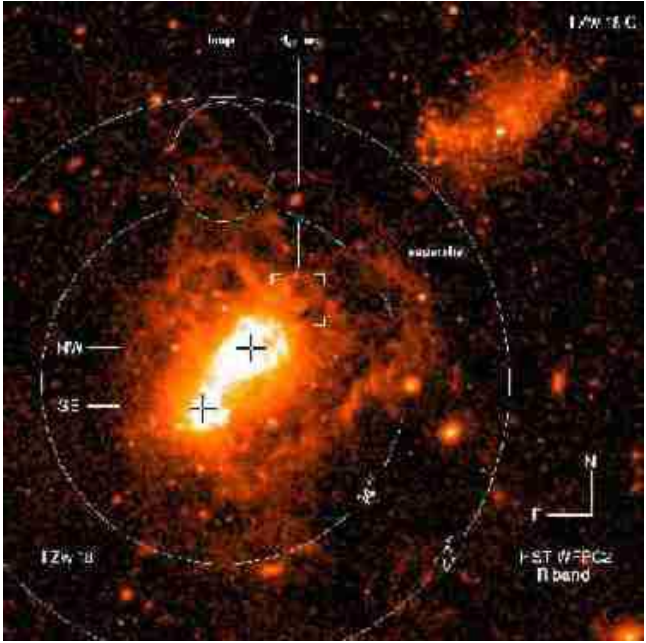,width=8.5cm,angle=0,clip=}}}
\end{picture}
\caption[]{\hst/WFPC2 archival image of I Zw 18 in the $R$ band. 
North is at the top and east to the left. The two star forming 
regions in I~Zw~18 are labelled \nw\ and \se. 
Regions marked {\sl loop} and {\sl H$\alpha$ arc} have been investigated 
spectroscopically in Izotov et al.~(\cite{Yu01a}).
The northwestern supershell in I Zw 18 and the  
detached irregular component I Zw 18\,C are indicated. 
The larger circles with radii 14\arcsec\ and 22\arcsec\ are 
centered between the \nw\ and \se\ star forming regions.}
  \label{f0}
  \end{figure}

Kunth \& \"Ostlin (\cite{KO00}) conclude from optical and NIR 
surface photometry studies that I Zw 18 possesses an evolved 
and spatially extended stellar population underlying its SF regions. 
Their conclusion is based on a nearly constant relatively red
$B-R$ $\sim$ 0.6 mag and outwards increasing $B-J$ colour 
of the LSB envelope.
Because the latter has a surface-brightness profile that can be well fitted 
by an exponential law, Kunth \& \"Ostlin (\cite{KO00}) ascribed the 
LSB emission to a stellar disc with an age of $\sim$~5 Gyr.
 
On the theoretical front, Legrand (\cite{Leg00}) and Legrand et
al. (\cite{Leg00a}) in an attempt to explain the paucity 
of nearby SF dwarf galaxies with oxygen abundance
12~+~log(O/H)~$<$~7.2 proposed that these systems 
form over the Hubble time through a continuous low-level 
SF process. In this scenario, I Zw 18 would have an extended stellar disc 
with $m_V$ $\sim$ 20 mag and a mean surface brightness $\sim$~28~$V$ \sbb.

If the presence of a significant old stellar background can be demonstrated
then I Zw 18 would be a standard BCD, and the hypothesis of it being a
young galaxy must be abandoned. I Zw 18 would then be like the vast majority 
of BCDs which are evolved dwarf galaxies where star formation is occurring 
within an extended, circular or elliptical
stellar host galaxy (Loose \& Thuan \cite{LT86}, hereafter LT86; 
Kunth et al. \cite{KMV88}; Papaderos et al. 1996a, hereafter \cite{P96a}).
Such systems, classified iE/nE by LT86, account for $\sim$ 90\% of the 
local BCD population.
Their red LSB underlying component dominates the surface brightness and 
colour distribution in their outer parts and 
contributes on average $\sim$ 1/2 of their $B$ luminosity 
within the 25 $B$ \sbb\ isophote (Papaderos et al. \cite{P96b}).

The view that I Zw 18 has properties similar to old BCDs, with the 
exception of its low oxygen abundance, is not well established, however.
Stars populating the stellar disc postulated by Legrand et al. (\cite{Leg00a}) 
and Kunth \& \"Ostlin (\cite{KO00}) have not been seen in I Zw 18 
(Izotov \& Thuan \cite{IT01}). 
By contrast, in CMD studies of Local Group dwarfs 
(see e.g. Grebel (\cite{Grebel00}) for a review) 
and of a few nearby BCDs (cf. Sect. \ref{S3a}),
a census of several thousands resolved stars unambiguously 
proves the existence of an evolved and spatially extended 
stellar background.
In fact, the observational evidence for a dominant 1 -- 5 Gyr 
old stellar population in I Zw 18 rests on a sample of only
about one dozen red point sources seen both in the optical 
and NIR ranges by \hst\ (Aloisi et al. \cite{Alo99}; 
\"Ostlin \cite{Gor00}). 
Ages quoted from this tiny statistical probe assume no or a uniform 
extinction, while non-uniform dust absorption in I Zw 18 has been 
discovered recently (Cannon et al. \cite{Cannon01}).

The assumption by Kunth \& \"Ostlin (\cite{KO00}) that ionized 
gas does not dominate the LSB envelope of I Zw 18, thus its colours 
need not be corrected for gaseous emission before deriving ages,
has been disputed by Papaderos et al. (\cite{P01}) and Izotov et al. (\cite{Yu01a}).
The latter authors remarked that the mean $B-R$ and $B-J$ colours 
derived by Kunth \& \"Ostlin (\cite{KO00}) and Papaderos et al. (\cite{P01}) 
in the outer regions of the main body are entirely consistent with pure 
ionized gas emission.
This is also consistent with the results by Hunt et al. (\cite{Hunt02})
who placed from deep NIR data an upper limit of $\la$ 15\%
to the $J$ band light fraction of stars older than 500 Myr in I Zw 18. 

The large range of ages derived previously, together with the 
unexplored role of extended gaseous emission as well as dust 
absorption led us to reexamine the photometric structure of I Zw 18.
Our goal was to determine whether there exists a stellar LSB component 
underlying the filamentary envelope of I Zw 18 and extending well 
beyond the \nw\ and \se\ regions. If present, how does its colour and 
structural properties compare to those of standard BCDs, and what 
implications can be drawn for the evolutionary state of I Zw 18? 

The paper is organized as follows. In Sect. \ref{S2} we 
discuss the set of ground-based and \hst\ data included in this study
and briefly describe the techniques used in the surface photometry
analysis. 
We consider essential to study the photometric structure 
of I Zw 18 not in isolation but in the context of the main class of
evolved BCDs. The photometric structure of these systems is 
discussed in Sect. \ref{S3} on the example of the nearby 
iE BCDs Mkn 178 and VII Zw 403. 
Section \ref{S4} focuses on I Zw 18, the distance of which is 
assumed throughout to be 15 Mpc (Izotov et al. \cite{Yu01a}). 
In Sect. \ref{S4a} we derive surface brightness profiles (SBPs) of its 
main body on the usual assumption that its LSB emission is predominantly 
of stellar origin. 
The properties of the LSB component after subtraction of nebular 
line emission (Sect. \ref{S4b}) are studied in Sect. \ref{S4c}. 
The photometric structure of I Zw 18\,C 
is discussed in Sect. \ref{S4d}. In Sect. \ref{dis1} we compare the 
structural properties of I Zw 18 with those of standard BCDs. 
In Sect. \ref{dis2} we investigate whether spatially extended 
ionized gas emission can mimic the SBP of a red 
exponential stellar disc.
The evolutionary state of I Zw 18 in the light of the present results is 
discussed in Sect. \ref{dis3}. Our conclusions are summarized in Sect. 
\ref{Conclusions}.
%
\section{Observations and data reduction \label{S2}}
\subsection{Data acquisition \label{S2a}}
%
Broad-band Johnson $B,V$ and Cousins $R,I$ images of I Zw 18 were taken during
three observing runs. A first set of exposures in $B$ (20 min) and $R$ (10 min) 
was acquired on March~7~--~10, 1997 with the CAFOS focal reducer attached to the 2.2m 
telescope of the German-Spanish Astronomical Center, 
Calar Alto, Spain. 
CAFOS was equipped with a 2048 $\times$ 2048 SITe CCD operating at a gain 
of 2.3 e$^{-}$ ADU$^{-1}$,
with a read-out noise of $<$ 3 counts (rms). With a focal ratio of f/4.4, 
the instrumental scale was 0\farcs53 pixel$^{-1}$ and the usable field 
of view (FOV) $\sim$~15\arcmin.
Another series of images, each slightly offset from the others,
with a total integration time of 90 min in $B$ and 50 min in $R$ 
was taken with the Calar Alto 1.23m telescope in the period January 
24 to February 22, 2000. A 2048 $\times$ 2048 SITe detector mounted at 
its Cassegrain focus gave an instrumental scale of 0\farcs5 pixel$^{-1}$ and a 
usable FOV of $\sim$ 11\arcmin.
$V$ and $I$ imaging data of I Zw 18 were acquired with the 
Kitt Peak National Observatory (KPNO) 2.1m telescope on April 18, 1999. 
The telescope was equipped with a 
1024 $\times$ 1024 Tektronix CCD detector operating at a gain of 3\,e$^-$\,ADU$^{-1}$, 
giving an instrumental scale of 0\farcs 305 pixel$^{-1}$ and a FOV of 
5\arcmin. 
The total exposure time of 40 and 60 min in $V$ and $I$, respectively, 
was split up into four slightly shifted subexposures.

$J$ images were obtained with the 3.5m telescope at Calar Alto 
during three consecutive nights (May, 12~--~14, 2000).
The Omega camera, mounted at the prime focus (f/3.5) of 
the telescope, consisted of a 1024 $\times$ 1024 pixel Rockwell HAWAII detector
yielding a scale of 0\farcs396 pixel$^{-1}$.
The large FOV (6\farcm 76) as compared to the size of I Zw 18 allowed for 
dithering ON source. The total on object integration time was 66\, min.
The data were calibrated by observing UKIRT NIR standard stars from 
Hunt et al. (\cite{Hunt98}). The photometric accuracy is estimated 
to be better than 0.1\,mag. A detailed presentation of the data 
acquisition and reduction will be given in Noeske et al. (\cite{Kai01b}). 

Data on the iE BCDs Mkn 178 and VII Zw 403 (Sect. \ref{S3a}) are based
on Johnson $B$ and Cousins $R$ images taken at Calar Alto during three 
observing runs. VII Zw 403 was observed with the 2.2m Calar Alto telescope 
in February 1994 for 60 min in $B$ and 20 min in $R$. 
Additional $B$ (20 min), $R$ (10 min) and \ha\ (25 min) images for this galaxy
were taken with the 2.2m/CAFOS during the 1997 run (see above). 
During the same run Mkn 178 was observed in $B$ and $R$ for
17 min and 12 min, respectively.
A third set of broad-band images of Mkn 178 with a total 
exposure of 53 min in $B$ and 63 min in $R$ was taken
with the Calar Alto 1.23m telescope in the period
January--February, 2000.

The FWHM of point sources in all observing runs was in the range between 
1\farcs05 and 1\farcs 6. Bias and flat--field frames were obtained during 
each night of the observations. Calibration was accomplished by observing 
standard fields from Landolt (\cite{Landolt92}) and 
Christian et al. (\cite{Christian85}). 
Our calibration uncertainties are estimated to be below 0.05\,mag in all 
bands. Standard reduction steps, including bias and flat--field correction, 
cosmic ray rejection and image co-alignment were carried out 
using the ESO MIDAS\footnote{Munich Image Data Analysis 
System, provided by the European Southern Observatory (ESO).} 
and IRAF\footnote{IRAF is the Image Reduction and Analysis Facility distributed by the 
National Optical Astronomy Observatory, which is operated by the 
Association of Universities for Research in Astronomy (AURA) under 
cooperative agreement with the National Science Foundation (NSF).} software
packages. 
%
%
\begin{table}
\caption{Summary of the observational data}
\label{data}
\begin{tabular}{lccc}
\hline
\hline
Object     & Year & Telescope        & Observations (min)\\
\hline
I Zw 18    & 1997 & Calar Alto/2.2m & $B$(20), $R$(10)   \\        
                        & 1999 & KPNO/2.1m  & $V$(40), $I$(60)   \\
                        & 2000 & Calar Alto/3.5m & $J$(66)          \\          
                        & 1994 & \hst/WFPC2 & $B$(77), $V$(77), $R$(90) \\
                        &      &           & \ha(77), [\ion{O}{iii}](77)    \\          
\hline
Mkn 178    & 1997 & Calar Alto/2.2m & $B$(17), $R$(12) \\
                              & 2000 & Calar Alto/1.23m& $B$(53), $R$(63) \\
\hline
VII Zw 403 & 1994 & Calar Alto/2.2m  & $B$(60), $R$(20) \\
           & 1997 & Calar Alto/2.2m  & $B$(20), $R$(10), \ha(25) \\
                              & 1995   & \hst/WFPC2 & $V$(70), $I$(70), \ha(40) \\
\hline
\end{tabular}
\end{table}

We have also included in the present study archival \hst/WFPC2 
$B$ (F450W), $V$ (F555W) and $R$ (F702W) images of I Zw 18 (PI: Dufour, 
GO-5434, November 1994). These have the advantage of 
including both the main body and the C component in their FOV (cf. Dufour et al. \cite{Duf96b}). 
In order to estimate the luminosity contribution of ionized gas emission (Sect. \ref{S4b})
we use narrow-band [\ion{O}{iii}] $\lambda$5007 (F502N) and H$\alpha$ (F658N) 
archival \hst/WFPC2 images. In the photometric study of VII Zw 403 we use
archival \hst/WFPC2 $V$, $I$ and \ha\ images (PI: Westphal, GO-6276, July 1995). 
These data have been discussed in detail by Lynds et al. (\cite{Lynds98}), 
Schulte-Ladbeck et al. (\cite{Reg98},\cite{Reg99a}) and Izotov \& Thuan 
(\cite{IT01}). The \hst\ photometry was transformed to the standard 
Johnson-Cousins $UBVRI$ photometric system following the prescriptions of 
Holtzman et al. (\cite{Holtz95}). 
The observational data discussed in Sects. \ref{S3} and \ref{S4} are 
summarized in Table \ref{data}.
%
\subsection{Surface photometry \label{S2b}}
In the surface photometry analysis of the iE BCDs Mkn 178 and VII Zw 403 
(Sect. \ref{S3a}) we use improved versions of the techniques 
described in \cite{P96a}.
\begin{figure*} 
\begin{picture}(16,9.4)
\put(2.8,0){{\psfig{figure=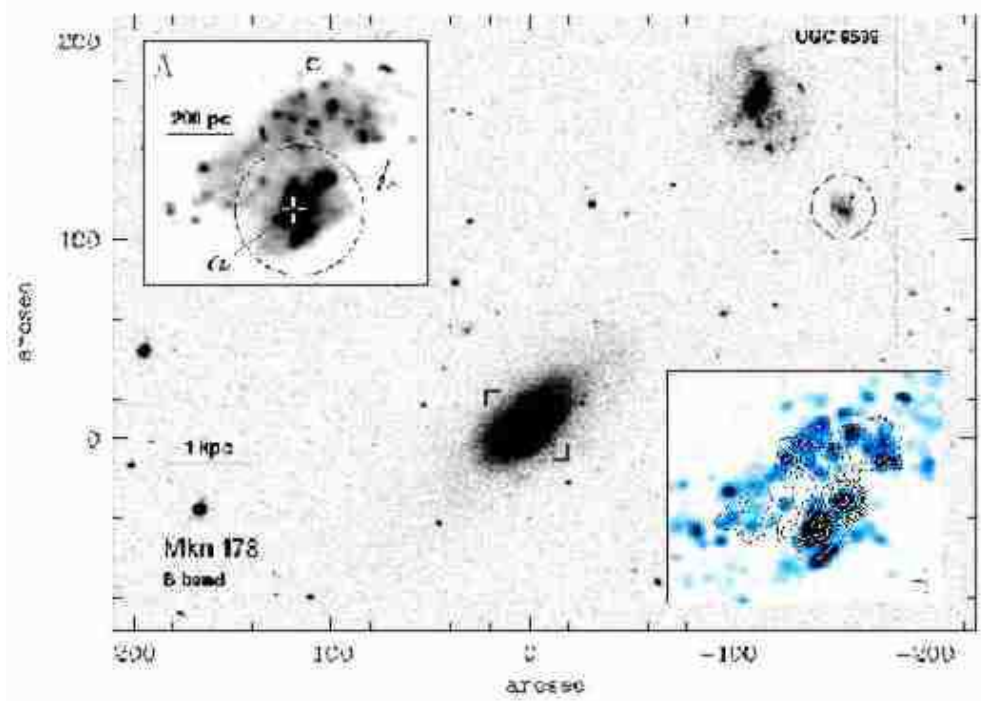,height=9.4cm,angle=0,clip=}}}
\end{picture}
\caption[]{$B$ image of the iE BCD Mkn 178 ($D$ = 4.2 Mpc, Schulte-Ladbeck
et al. \cite{Reg00}). The background galaxy UGC 6538 ($z$~=~0.0106) is indicated.
The redshift of the encircled galaxy is not known. Insets A and B display 
magnified versions of the central 41\farcs5~$\times$~35\arcsec\ region 
of Mkn 178. The encircled region in inset A (diameter = 19\arcsec)
corresponds approximately to the field photographed with \hst/NICMOS 
by Schulte-Ladbeck et al. (\cite{Reg00}). The star-forming region {\it a} 
together with the redder adjacent knot {\it b} and the extended northern region 
{\it c} (notation adopted from Gonz\'alez-Riestra et al. \cite{gr88}) are indicated. 
Inset B shows NIR contours in the $H$ band overlayed on a 
contrast-enhanced version of A. The $H$ emission displays two 
maxima, the main one coinciding with knot {\it b} and the secondary one 
located 8\farcs4 to the northeast (Noeske et al. \cite{Kai01b}).}
  \label{f1}
  \end{figure*}
For intermediate and low surface brightness levels, the equivalent 
radius $R^*=\sqrt{{A(\mu)}/\pi}$ corresponding to the surface brightness 
$\mu$ was computed by a combination of methods {\sf i} (the area $A(\mu)$ 
is determined by fitting an ellipse to the isophote) or {\sf ii} 
(a line integral is calculated along the respective isophote 
in polar and cartesian coordinates). In the high-surface-brightness (HSB) 
regime, where several star forming regions can be distinguished on 
ground-based images, $A(\mu)$ was computed using method {\sf iii}
(summation of all pixels inside a polygonal aperture with a surface 
brightness $\leq\mu$).

By definition, these methods trace the growth 
of the angular size of a galaxy with increasing $\mu$, 
so that the equivalent radius $R^*$ is a monotonic 
function of surface brightness. Because no subjective assumptions 
are made on the geometrical center or the mean position angle 
and ellipticity at any intensity interval, they allow to derive
the surface brightness distribution in a ``standard'' and reproducible 
manner for BCDs that are morphologically very different.

While these techniques work well for most BCDs,
they should be used with caution in the rare case
of an extremely irregular and patchy LSB morphology,
where either ellipses give a poor geometrical 
approach to isophotes or the LSB envelope 
consists of several detached entities.

One such example is the main body of I Zw 18 which shows
an irregular morphology in the LSB outer regions.
As verified from ground-based and \hst\ data, the SBP slope derived for
$\mu\ga$~24 $B$ \sbb\ using method {\sf i}, changes depending on whether 
some compact and diffuse feature in the periphery of the BCD is included 
or not (e.g. features along the northwestern supershell, cf. Fig. \ref{f0}). 
Method {\sf ii} which is best suited when $A(\mu)$ is given by a single 
well-defined isophote, fails when applied to the patchy morphology of 
I Zw 18. As for method {\sf iii}, its applicability is restricted to 
an intensity level typically higher than $\sim$ 3$ \times$ sky noise 
even for adaptively smoothed images.

We, therefore, supplemented methods {\sf i} through {\sf iii}
with another flux-conserving technique which allows for 
a reasonable compromise between the following requirements: 
a) $R^*$ is a nearly monotonic function of $\mu$,
a criterion which guarantees compatibility with methods {\sf i} -- {\sf iii},
b) no implicit assumption is made on the morphology of a BCD 
at any surface brightness level and c) the colour obtained for
a given $R^*$ via SBP subtraction is assigned to one and the 
same region of a BCD in all images. This ensures 
that the colour profile in the central part is not affected when 
the brightest SF region does not coincide
with the bluest one (see discussion in \cite{P96a}).

This method (referred to in the following as {\sf iv}) allows to account
optimally for the morphology of a BCD at all intensity 
levels. The input to method {\sf iv} is a set of coaligned 
multiwavelength frames with the same instrumental scale and 
point spread function (PSF), previously cleaned for bright 
foreground and background sources. 
A reference frame ${\cal F}$ is first computed from the signal-to-noise 
weighted average of the input images. This step incorporates an 
adaptive resolution pattern that allows to conserve small-scale 
intensity enhancements in the SF region while approaching a chosen 
resolution of typically $\la$~3~$\times$~PSF in the LSB regime. 
${\cal F}$ serves as a template for computing a
series of masks ($n_{\rm mask}$) mapping equidistant logarithmic intensity
intervals between $I_{\rm min}$ and $I_{\rm max}$.  A pixel in the mask $m_i$
($i\leq n_{\rm mask}$) is set to unity when its intensity $F$ is such that
$I-\Delta I\leq F \leq I$ and to zero otherwise. Each input frame is then in
turn weighted by the mask $m_i$ and used to compute the mean
surface brightness within the mask area set to unity.

The radius $R^*$($\mu$) corresponding to the mask $m_i$ is given by 
\begin{equation}
R^*{\rm (}\mu{\rm )} = \sqrt{\left(\frac{A_{I}+A_{I-\Delta I}}{2\pi}\right)} ,
\end{equation}
where A$_{I}$ and $A_{I-\Delta I}$ denote respectively 
the areas of frame ${\cal F}$ in $\sq\arcsec$ with intensity levels above 
$I$ and $I-\Delta I$. 

A virtue of method {\sf iv} as compared to {\sf iii}, 
is that the surface brightness $\mu$ is not a user-defined 
input variable, but is computed from {\sl all} pixels within 
a galaxy's region of arbitrary shape. 
This allows to overcome the problem of the artificial SBP 
flattening at low intensities inherent to method 
{\sf iii} (see discussion in \cite{P96a} and Cair\'os et al. \cite{LM01a}). 
On the other hand, method {\sf iv} requires an accurate subtraction
of bright foreground and background sources, especially in the LSB regime.
This is because, in the presence of a faint ($\mu_B\ga$~25 \sbb) LSB 
background, a moderately bright source may affect the photon 
statistics and colours computed within the mask $m_i$.
Tests on the iE BCDs Mkn 178 and VII Zw 403 (Sect. \ref{S3a}) 
as well as on other SF dwarfs with a variety of morphologies 
(cf. Noeske et al. \cite{Kai01b}) have shown that method 
{\sf iv} is reliable down to a faint intensity level 
and gives indistinguishable results from methods {\sf i} -- {\sf iii} for 
the {\sl plateau} (see Sect. \ref{S3a}) and the LSB components. 
Methods {\sf iv} and {\sf iii} were used to derive the 
surface brightness distribution of the main body of I Zw 18 
(Sect. \ref{S4}) in its LSB and HSB regimes.
%
\section{The~photometric~structure~of~standard~BCDs\label{S3}}
%
This section contains a brief overview of the photometric properties of iE BCDs, 
the most common type of BCDs (Loose \& Thuan \cite{LT86}), 
for later comparison with I Zw 18 (Sect. \ref{S5}).
Surface photometry studies of individual BCDs can be found in e.g. \cite{P96a}, 
Vennik et al. (\cite{vennik96},\cite{vennik00}),  Doublier et al. 
(\cite{Doubl97},\cite{Doubl99}), Marlowe et al. (\cite{Mar97},\cite{Mar99}),
Telles et al. (\cite{tmt97}), Chitre \& Joshi (\cite{Chitre99}), 
Makarova (\cite{Makarova99}), Smoker et al. (\cite{Smoker99}) and
Cair\'os et al. (\cite{LM01a}).
%
\begin{figure*} 
\begin{picture}(16,7.95)
\put(0.1,0){{\psfig{figure=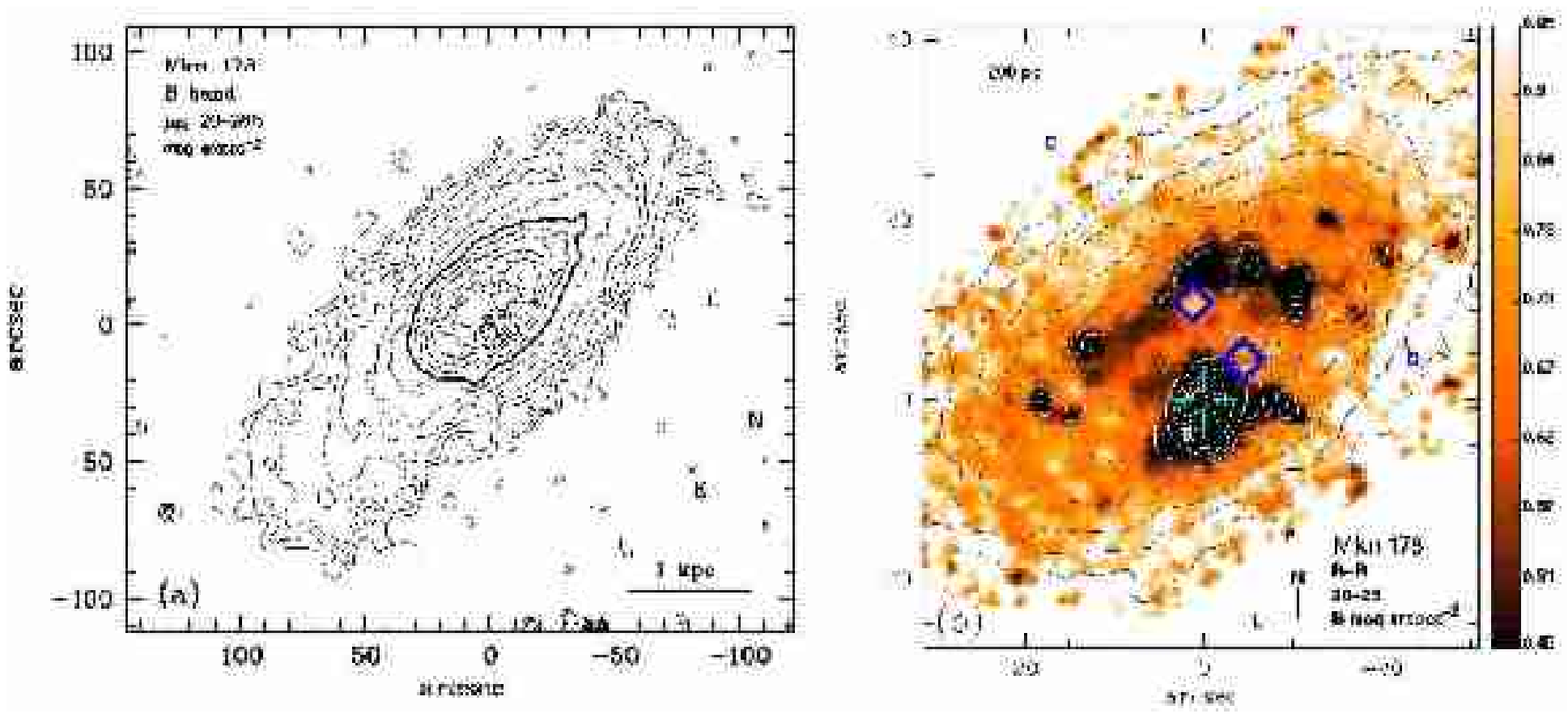,height=7.95cm,angle=0,clip=}}}
%
\end{picture}
   \caption[]{
{\bf (a)} $B$ contour map of Mkn 178. Contours correspond
to surface brightness levels from 20 to 28.5 \sbb\ in steps of 0.5 mag. The 25
$B$ \sbb\ isophote is shown as a thick contour. Note the slight 
southeastern elongation of the LSB component for $\mu\ga$~27 $B$ \sbb.
{\bf (b)} $B-R$ map of the central region of Mkn 178 in the
range 0.45 -- 0.95 mag, overlayed with $B$ contours from 20 to 25 \sbb. 
 The bluest region ($B-R\approx 0.1$ mag) 
is associated with the brightest region {\it a} while the optically 
fainter region {\it b} coincides with the brightest of the $H$ sources 
(depicted by rhombs) detected by Noeske et al. (\cite{Kai01b}, cf. inset B in Fig.\ref{f1}).
Beyond the outermost contour (25 $B$ \sbb), the colour is dominated 
by the red ($B-R\sim 1.1$ mag) stellar LSB component underlying 
the star-forming regions.} 
  \label{f2}
  \end{figure*}
\begin{figure*}
\begin{picture}(16,8.0)
\put(-0.3,0){{\psfig{figure=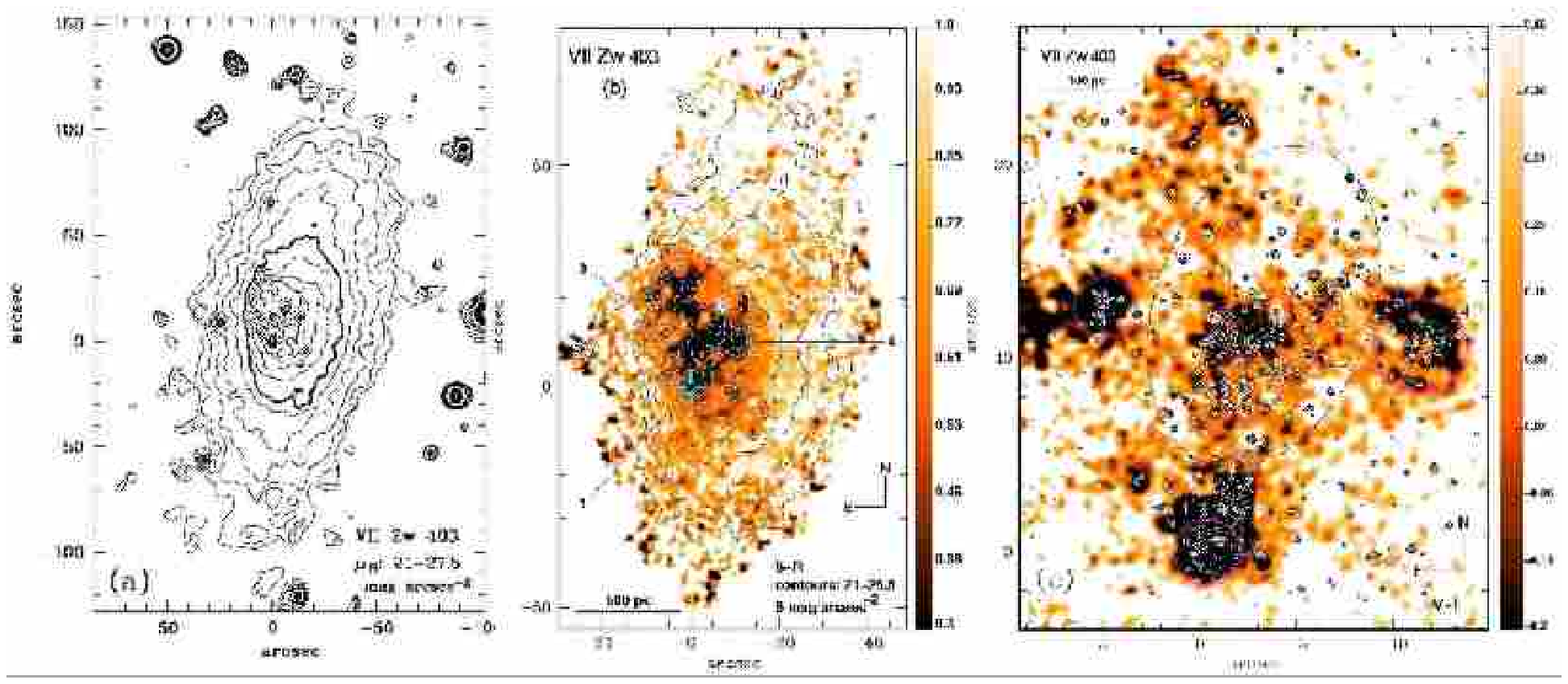,height=8.0cm,angle=0.,clip=}}}
\end{picture}
   \caption[]{
{\bf (a)} Contour map of VII Zw 403 with surface brightness levels from 21 to 
27.5 $B$ \sbb\ in steps of 0.5 mag. The 25 $B$ \sbb\ isophote is shown by the thick contour. 
{\bf (b)} $B-R$ map of VII Zw 403 in the range 0.3 -- 1.0 mag, overlayed with $B$ contours. 
The position of the bright stellar assemblies 1 (cross at axes' origin), 3 and 4 
(notation adopted from Fig. 2 of Lynds et al. \cite{Lynds98}) 
is indicated. Note the blue ($\sim$~0.2 mag) featureless region protruding 
between regions 3 and 4, up to $\sim$~30\arcsec\ to the northeast of region 1.
{\bf (c)} $V-I$ map of the central portion of VII Zw 403 computed from 
archival \hst/WFPC2 data. Contours, derived from a contrast-enhanced $V$ 
image, delineate the spatial distribution of compact sources in the 
high-surface-brightness regime of the BCD.
Comparison with Fig. \ref{f4} shows that the blue ($V-I\leq$~--0.2) 
regions in the periphery of bright young stellar assemblies are spatially 
correlated with the bright \ha\ features, suggesting that ionized gas 
emission produces significant colour changes on scales of a few tens of pc.} 
  \label{f3}
  \end{figure*}

We focus here on two nearby iE systems, VII Zw 403 and Mkn 178.
They were selected for the following reasons:
(a) As indicated by surface photometry (Sect. \ref{S3a}) and CMD studies,
both are {\it bona fide} old BCDs. Furthermore, their distances have been 
accurately determined from the tip of the red giant branch stars 
to be 4.4 Mpc for VII Zw 403 (Lynds et al. \cite{Lynds98}; 
Schulte-Ladbeck et al. \cite{Reg99b}) and
4.2 $\pm$ 0.5 Mpc for Mkn 178 (Schulte-Ladbeck et al. \cite{Reg00}).
(b) Neither galaxy shows signatures of a strong past or ongoing 
interaction (Noeske et al. \cite{Kai01a},\cite{Kai01b}), i.e. they 
are suitable for studying non-tidally triggered SF in BCDs.
(c) The central surface brightnesses and exponential scale lengths 
of their LSB underlying components are intermediate between 
the main class of BCDs and dwarf irregulars (dIs). 
Therefore the discussion below is not concerned with extreme 
ultra-compact BCDs, which may undergo intermittent very violent 
SF episodes, but rather with average BCDs/dIs where star formation 
{\sl may} proceed at a lower rate over longer periods.

%
\subsection{Mkn 178 and VII Zw 403\label{S3a}}
\begin{figure*}[!ht]
\begin{picture}(16,13.1)
\put(0,5.6){{\psfig{figure=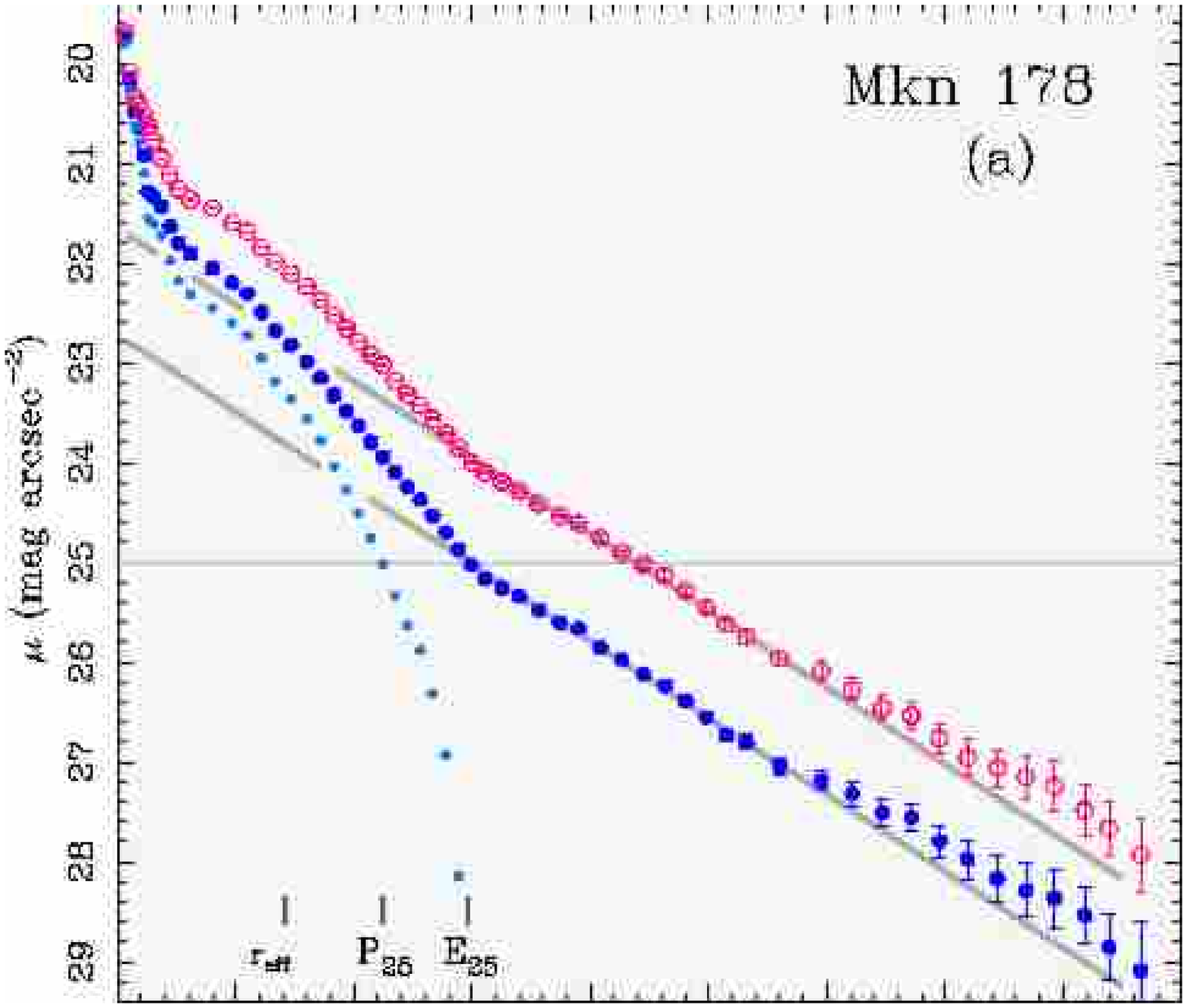,width=8.7cm,angle=0.,clip=}}}
\put(0,0.){{\psfig{figure=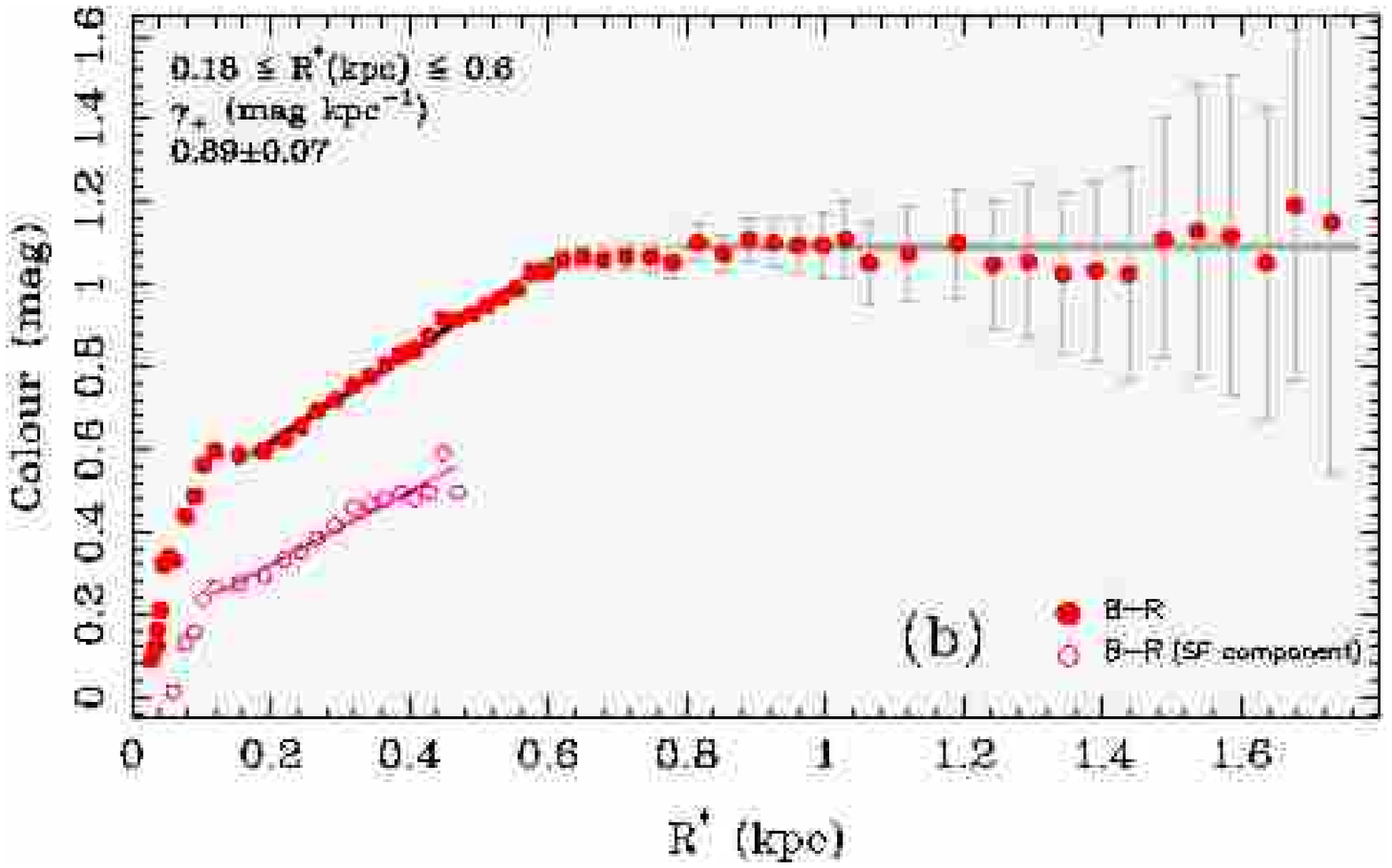,width=8.7cm,angle=0.,clip=}}}
\put(9,5.6){{\psfig{figure=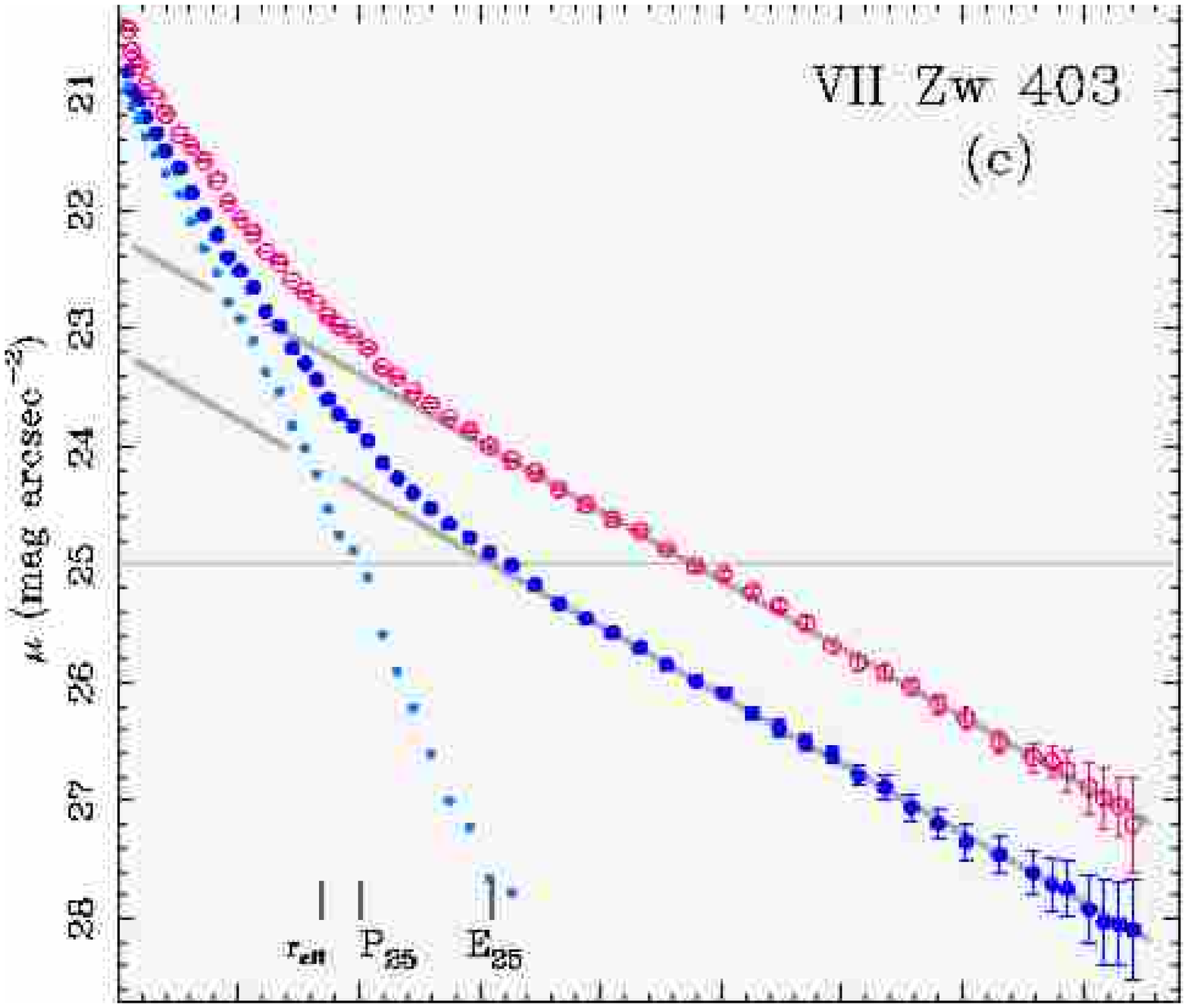,width=8.7cm,angle=0.,clip=}}}
\put(9,0.){{\psfig{figure=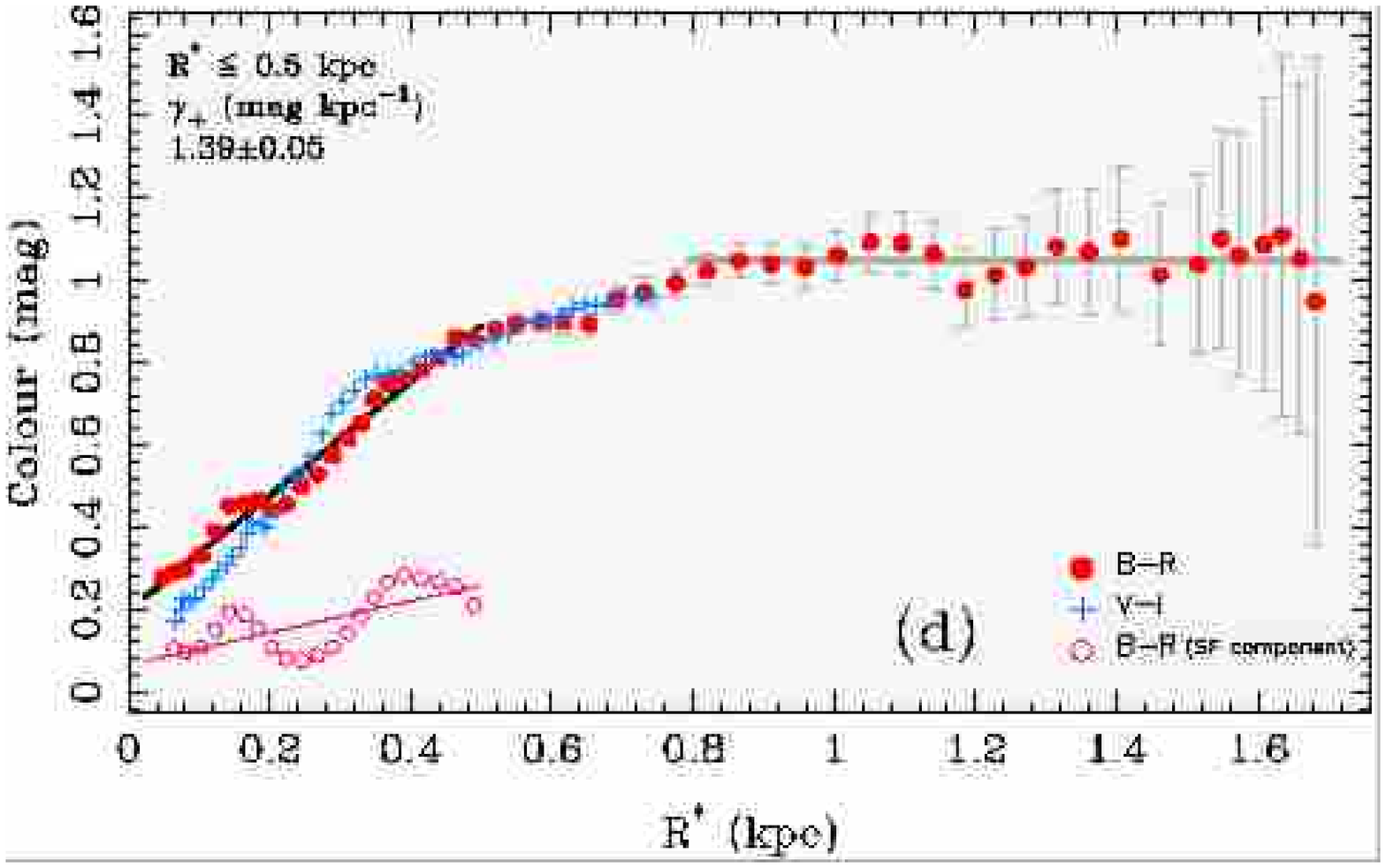,width=8.7cm,angle=0.,clip=}}}
\end{picture}
   \caption[]{{\bf (a)} Extinction-corrected $B$ (filled circles)
and $R$ (open circles) surface brightness profiles (SBPs) of Mkn 178 
vs. equivalent radius $R^*$. 
Thick-gray lines show exponential fits to
the LSB component in the range 0.6~$\leq R^*\,{\rm (kpc)}\leq$~1.2. 
The slight flattening with respect to the exponential fit for 
$R^*>$~1.2 kpc is caused by the southeastern extension of the BCD
(Fig. \ref{f2}a), which accounts for $\la$~2\% of the LSB light in $B$.
The intensity profile which is represented by small filled circles
is obtained by subtracting the exponential fit from the $B$ band SBP.
This excess emission is due to the young and intermediate-age 
stellar populations within the plateau component. It contributes 
$\sim$~60\% of the $B$ light of Mkn 178 inside its 25 $B$ \sbb\ isophote.
The isophotal radius of the plateau (\p25), that of the LSB 
component (\e25) and the effective radius $r_{\rm eff}$
in $B$ are indicated. The line-of-sight contribution of the 
plateau population to the $B$ intensity decreases from 
$<$~40\% at \p25\ to $\la$~4\% at \e25. 
{\bf (b)} $B-R$ profile of Mkn 178. For 0.2 kpc~$\la R^*\la$~\e25\
the $B-R$ increases linearly with a gradient $\gamma_+\approx$~0.9 mag
kpc$^{-1}$. For $R^*>$~\e25\ the colour becomes constant at
$B-R\approx$~1.1 mag and is due to the red stellar LSB population. 
The solid grey line results from the subtraction 
of the fits to the $B$ and $R$ SBPs of the LSB component.
Open circles show the colour distribution within \p25\ after 
correction for the luminosity contribution of the LSB host (see discussion in
the text). While this correction has no effect on $\gamma_+$, it shifts 
the $B-R$ profile by $\approx$~--0.4 mag.
{\bf (c)} and {\bf (d)} Surface brightness and colour profiles of VII Zw 403.
Symbols have the same meaning as in panels {\bf a} and {\bf b}.
Panel {\bf d} also shows the $V-I$ profile derived from archival 
\hst/WFPC2 data. Correction for the
emission of the LSB component results in a downwards shift 
of the $B-R$ colour profile and a decrease in $\gamma_+$ from 
$\sim$~1.4 mag kpc$^{-1}$ to $\sim$~0.4 mag kpc$^{-1}$.
}
  \label{f5}
  \end{figure*}

%
The SBP of Mkn 178 (Fig. \ref{f5}a) reveals three main
intensity regimes, as is commonly seen in iE BCDs. 
At small radii ($R^*\leq$150 pc), it exhibits a narrow excess 
due to the luminosity contribution of regions {\it a} and {\it b} 
(cf. inset A in Fig. \ref{f1}). Region {\it a} ($M_B$ = --12 mag), 
contributes $\sim$ 40\% of the $B$ light in excess to the LSB component. 
The large colour contrast between this bright assembly of SF sources 
and its surroundings is reflected on the inner part of the colour 
profile (Fig. \ref{f5}b), which rises steeply within $R^*\la$ 0.2 kpc 
from 0.1 mag to $\sim$ 0.6 mag.
The fainter ($M_B\approx$ --10.6 mag) source {\it b} ($B-R$ = 0.67 mag) 
is barely seen on the colour map (Fig. \ref{f2}b). This is in agreement with 
the results by Gonz\'alez-Riestra et al. (\cite{gr88}) who found also a 
colour difference between regions {\it a} and {\it b} 
(cf. their Fig. 3a -- 3c) and established spectroscopically 
that region {\it b} is part of Mkn 178. 
Interestingly, NIR imaging by Noeske et al. (\cite{Kai01b}) 
shows the optically faint source {\it b} to be coinciding with 
the surface brightness maximum in the $H$ band (inset B in 
Fig. \ref{f1}), whereas region {\it a} is nearly absent on NIR images.

The second intensity regime, with surface brightnesses between $\sim$ 22 and 
$\sim$ 25 $B$ \sbb, represents the total luminosity of compact and diffuse 
sources in the inner $\sim$ 0.65 kpc of Mkn 178 (Fig. \ref{f1}).
This pronounced bump, referred to as {\sl plateau} in 
\cite{P96a}\footnote{The same feature 
has been called {\sl platform} by Telles et al. (\cite{tmt97}) or {\sl core} 
by Marlowe et al. (\cite{Mar99}, see their Fig. 1). 
SBPs of BCDs that cannot be approximated by a simple fitting law 
(e.g. a de Vaucouleurs or exponential distribution) 
because of a pronounced {\sl plateau} feature at intermediate 
intensity levels, have been classified as {\sl dd (double)} 
by Telles et al. (\cite{tmt97}) or {\sl composite} by 
Doublier et al. (\cite{Doubl99}) and Cair\'os et al. (\cite{LM01a}).}, 
is a common feature in SBPs of iE BCDs. 
Its intensity distribution has been approximated 
in the decomposition scheme of \cite{P96a} 
by a S\'ersic profile (S\'ersic \cite{S68}) with exponent $\eta$ in 
the range 1 $\la\eta\la$ 5. The isophotal radius \p25\ of the plateau 
at the surface brightness level of 25 \sbb\ (cf. panels a and c of Fig. 
\ref{f5}) measures the radial extent of the SF component in BCDs (P96a). 
The plateau radius of $\sim$ 450 pc of Mkn 178 (Fig.~\ref{f5}a) 
together with the moderately red $B-R$ = 0.6 -- 0.8 mag at HSB levels (Fig. 
\ref{f2}b) suggests a mild SF activity on a spatial scale of $\sim$ 1 kpc. 
%
%
%
\begin{table*}
\caption{Structural properties of the iE BCDs Mkn 178 and VII Zw 403$^{\rm a}$}
\label{iEphotom}
\begin{tabular}{lccccccccccc}
\hline
\hline
Name & Band & $\mu_{E,0}$ & $\alpha $ & $m_{\rm LSB}^{\rm fit}$ &
\p25\  & $m_{P_{25}}$ & \e25\ & $m_{E_{25}}$  & $m_{\rm SBP}$ & $m_{\rm tot}$ & $r_{\rm eff}$,$r_{80}$  \\
     &     & \sbb\ &  pc   & mag    & pc       &  mag             &  pc
          &  mag             &    mag & mag & pc \\
    (1) &   (2)         &   (3)     &  (4)      &   (5)            &
 (6)    &  (7)             &  (8)     &  (9)  & (10) & (11) & (12) \\
\hline
Mkn 178 & $B$ & 22.72$\pm$0.04  & 283$\pm$4 & 15.01 & 447  & 15.00  & 593  & 15.53  & 
 14.25$\pm$0.02  & 14.25 & 283,569 \pano \\
  & $R$ & 21.66$\pm$0.03  & 284$\pm$4 & 13.94 & 479  & 14.78  & 873  & 14.17  & 
 13.52$\pm$0.02  & 13.51 & 350,704 \kato \\
\hline
VII Zw 403 & $B$ & 23.19$\pm$0.07  & 372$\pm$10 & 14.99 & 402  & 15.50  & 620  & 15.75  &
 14.47$\pm$0.02  & 14.46 & 336,776 \pano \\
           & $R$ & 22.23$\pm$0.20  & 374$\pm$18 & 14.01 & 422  & 15.34  & 955  & 14.36  &
 13.77$\pm$0.02  & 13.76 & 440,896 \kato \\
%
\hline
\end{tabular}

\parbox{16.5cm}{$^{\rm a}$
Distances of 4.2 Mpc and 4.4 Mpc have been adopted for Mkn 178 and VII Zw 403,
respectively (Schulte-Ladbeck et al. \cite{Reg99b}; Schulte-Ladbeck et al. \cite{Reg00}). 
The tabulated values are corrected for Galactic extinction ($A_B$=0.078 mag for Mkn 178
and 0.09 mag for VII Zw 403).}
\end{table*}
%
%
\begin{figure}
\begin{picture}(8.6,10.6)
\put(0.,0.){{\psfig{figure=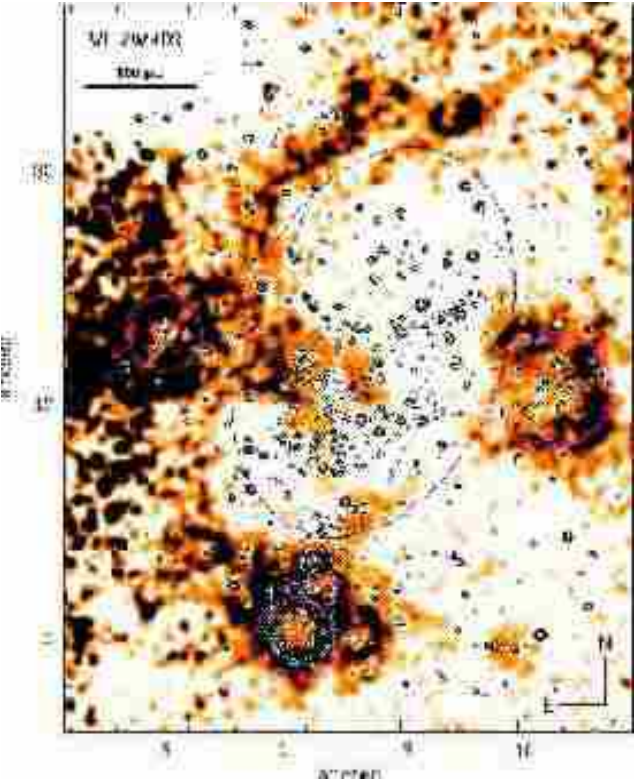,width=8.4cm,angle=0,clip=}}}
\end{picture}
\caption[]{\ha\ map of VII Zw 403 from archival \hst/WFPC2 images. 
Three of the bright stellar assemblies studied by Lynds et al. (\cite{Lynds98})
are indicated by circles. Contours show the spatial distribution of bright stellar 
sources in $V$ (cf. Fig. \ref{f3}c).
}
\label{f4}
\end{figure}

Signatures of a spatially extended, young stellar population are also present 
on the $B-R$ map of VII Zw 403 (Fig. \ref{f3}b), revealing blue colours 
inside its 24 $B$ \sbb\ isophote.
The case of VII Zw 403 illustrates the situation where, depending on the 
optical colour considered, the brightest region within the plateau may not 
coincide with the bluest one. The $B-R$ colour of the three 
bright SF regions indicated in Fig. \ref{f3}b (0.33 -- 0.55 mag) 
is on average redder than that of the diffuse stellar 
population surrounding them (0.12 -- 0.23 mag). 
The opposite trend is observed on the $V-I$ map (Fig. \ref{f3}c), 
where the surroundings of bright stellar clusters are bluer
(--0.18 -- --0.38 mag) than the smoothly distributed stellar population 
within the ellipse ($V-I$ = 0.1 -- 0.3 mag). Comparison of Figs. \ref{f3}c 
and \ref{f4} reveals that the regions with blue $V-I$ are spatially correlated
with the \ha\ emission and are more extended than the ionizing stellar 
sources. The combination of blue $V-I$ ($<$ 0 mag) with red 
$B-R$ ($\sim$ 0.4 -- 0.6 mag) colours is a sign of 
intense nebular emission on scales of a tens of 
parsecs around SF regions in VII Zw 403 (cf. e.g. 
Izotov et al. \cite{ILCFGK97c}; Papaderos et al. \cite{P98}; 
Noeske et al. \cite{Kai00}; Guseva et al. \cite{Nat01}). 

The plateau component of Mkn 178 (0.2 $\la$ $R^*\,{\rm (kpc)}$ $\la$ 0.65) 
shows a roughly linear colour increase with a gradient $\gamma_+$ of $\approx$ 0.9 
mag kpc$^{-1}$ (Fig. \ref{f5}b). 
This is also the case for VII~Zw~403 (Fig. \ref{f5}d) where $\gamma_+$
$\approx$ 1.4 mag kpc$^{-1}$, in the upper range 
of values for iE/nE BCDs (0.2 -- 1.6 mag kpc$^{-1}$) in the P96a sample. 
A large $V-I$ gradient (1.67 $\pm$ 0.1 mag kpc$^{-1}$) is also obtained 
from archival \hst/WFPC2 data in the radius range 
0.1 $\la$ $R^*\,{\rm (kpc)}$ $\la$ 0.5.

The gradual colour increase inside the plateau comes along with a 
population gradient in VII Zw 403.
CMD studies by Schulte-Ladbeck et al. (\cite{Reg99b}) reveal that 
the surface density of red stars in this BCD decreases slower 
with radius than other types of stars, implying that they dominate 
in the outer parts of VII Zw 403. Such a spatial segregation with 
young blue stars in the inner part ($R^*\!\!\la$\p25) of BCDs,
and old red stars in the outer regions has also been reported 
in CMD studies of Mkn 209 (Schulte-Ladbeck et al. \cite{Reg00}), 
NGC 1569 (Aloisi et al. \cite{Alo01}), UGCA 290 (Crone et al. \cite{Crone01}) 
and NGC 1705 (Tosi et al. \cite{Tosi01}).

We turn next to the exponential LSB component which dominates the light for
$\mu\ga$ 25 $B$ \sbb. The evolved state of the host galaxy of Mkn 178 
is evident from its red $B-R$ colour which levels off 
at 1.08 $\pm$ 0.04 mag for $R^*\ga$ \e25\ (cf. Fig. \ref{f5}b). 
The red optical colour, together with $J-H$ = 0.53 $\pm$ 0.13 mag 
and $H-K$ = 0.25 $\pm$ 0.12 mag (Noeske et al. \cite{Kai01b})
for the LSB component support the conclusion by Schulte-Ladbeck 
et al. (\cite{Reg98}) that Mkn 178 is not young.
This is also the case for the LSB host of VII Zw 403 for which 
we derive a mean $B-R$ colour of 1.05 $\pm$ 0.04 mag, in good agreement 
with the value of $\approx$ 1.12 mag of Schulte-Ladbeck \& Hopp 
(\cite{RU98}), after correction for extinction ($A_B=0.09$ mag).
The $B-R$ colour is also consistent with $V-I$ = $0.92\pm 0.05$ mag
obtained from \hst/WFPC2 data at $R^*$ = 0.75 kpc (cf. Fig. \ref{f5}d).

A substantial colour gradient, as typical property of BCDs 
inside a galactocentric distance slightly larger than \p25\ (P96a)
may reflect both, a true stellar age gradient and an outwards 
decreasing line-of-sight contribution of a young stellar population 
relative to the red underlying LSB host. 
By contrast, the majority of dIs exhibit no or minor colour 
gradients. van Zee (\cite{vZ01}) reports strong colour gradients 
(up to $\gamma_+(B-V)$ = 0.5 mag kpc$^{-1}$) only for those dIs with 
significant ongoing SF. Interestingly, these are compact systems, the scale 
length of their LSB host being comparable to those of BCDs ($\la$ 0.5 kpc).
However, the majority of dIs are quiescent and have small colour gradients, 
not exceeding $\gamma_+(B-V)$ = 0.2 mag kpc$^{-1}$ (Patterson \& Thuan
\cite{PT96}; van Zee \cite{vZ01}).

The effect of the light of the evolved stellar LSB host on the
luminosity-weighted BCD colour within \p25\ is illustrated in 
Figs. \ref{f5}b and \ref{f5}d. We show with open symbols 
the $B-R$ profile as computed from the light in excess to 
the exponential fit, i.e.  with the LSB emission subtracted out.
For Mkn 178 this correction has virtually no effect 
on $\gamma_+$, it shifts, however, the colour profile 
by about --0.4 mag.
The situation is different for VII Zw 403,
where subtraction of the contribution of the LSB component changes 
the colour gradient from $\gamma_+$~$\sim$~1.4 mag kpc$^{-1}$ to 
$\sim$~0.4 mag kpc$^{-1}$.
These corrections depend sensitively on the model adopted for 
the intensity distribution of the LSB host and, in particular on 
whether or not the exponential slope in the outer parts is 
extrapolated all the way to $R^*=0$ (see discussion in e.g. 
\cite{P96a} and Cair\'os et al. \cite{LM02}). 
They cannot be neglected, however, in the majority of BCDs, 
as the central surface brightness of their red LSB host 
is typically brighter than 22 $B$ \sbb\ (cf. Fig. \ref{f12}).

In Table\ \ref{iEphotom} we summarize the derived photometric 
properties of Mkn 178 and VII Zw 403. Cols.\ 3 and 4 list respectively 
the central surface brightness $\mu_{\rm E,0}$ and scale length $\alpha$ of 
the LSB host as obtained from exponential fits to the outer part of SBPs,
and weighted by the photometric uncertainty of each point. 
The corresponding total magnitude 
$m_{\rm LSB}^{\rm fit}\approx\mu_{\rm E,0}-5\log(\alpha\arcsec)-2$ of the LSB host for
a pure exponential distribution is given in col. 5.
Cols. 6 through 9 list quantities obtained from profile decomposition.
Cols.\ 6 and 8 give respectively the isophotal radii \p25\ and \e25\
of the SF and LSB components, both determined at 25\ \sbb.
The respective apparent magnitudes of each component within 
\p25\ and \e25\ are listed in cols.\ 7 and 9.
Total magnitudes of the BCD obtained from SBP integration out to 
the last point, and from flux integration within a polygonal aperture
are listed respectively in cols. 10 and 11. Note that the magnitudes 
in cols. 10 and 11 have been derived after removal of compact moderately 
bright sources in the LSB component.
Col.\ 12 gives the effective radius $r_{\rm eff}$ and the radius $r_{80}$ 
which encircles 80\% of the galaxy's total flux.
%
\subsection{The star-forming component in BCDs \label{S3b}}
%
In addition to the ongoing SF in several stellar clusters,
Mkn 178 and VII Zw 403 contain also in the inner part of their 
LSB host a more diffuse relatively blue stellar population 
contributing to the light of the plateau in optical SBPs.
Taking into account the effects of nebular 
line emission, patchy dust absorption and the red underlying LSB 
background on broad-band colours, both surface photometry and CMD studies
suggest that the stellar population responsible for the plateau in
Mkn 178 and VII Zw 403 cannot have formed in a single burst. 
Synchronization of SF over a spatial scale of $\sim$ 1 kpc
($\sim$ 2 \p25) requires a time scale of $\sim$ 50 Myr, assuming a 
sound speed velocity in the warm interstellar medium (ISM)
of the order of 10 km s$^{-1}$. 
CMD studies of Mkn 178 (Schulte-Ladbeck et al. \cite{Reg00}) and 
VII Zw 403 (Schulte-Ladbeck et al. \cite{Reg99b}) also suggest
that both systems experienced extended (several $10^8$ yr) periods 
of star formation, in agreement with theoretical predictions of the evolution 
of gas-rich dwarfs (e.g. Rieschick \& Hensler \cite{RiHe00}; 
Noguchi \cite{Noguchi01}). This is in contrast to the ``standard'' picture of 
SF in BCDs, of short ( $\sim$ 5 Myr ) bursts separated by long ($\sim$ 1 Gyr) 
quiescent phases (see e.g. Thuan \cite{Thuan91}). 

On the other hand, it is not yet clear whether the extended star formation 
scenario can be generalized to all BCDs. So far, no consensus has 
been reached concerning the intrinsic or environmental properties 
which control SF in these systems (see e.g. Papaderos et al. \cite{P96b}; 
Marlowe et al. \cite{Mar99}; Meurer et al. \cite{Meurer98}; 
V\'{\i}lchez \cite{Vi95}; Popescu et al. \cite{popescu99}; 
Pustilnik et al. \cite{pustilnik00}; Noeske et al. \cite{Kai01a}). 
Evolutionary synthesis models still lend support to the hypothesis of short
episodic bursts (Mas-Hesse \& Kunth \cite{MHK99}) with an amplitude being 
possibly inversely related to the mass of the BCD (Kr\"uger et al. \cite{Harald95}). 
Also, Papaderos \& Fricke (\cite{PF98}) have suggested that the burst duration
in BCDs is anticorrelated with the central mass density $\rho_*$ 
of their underlying LSB component. $\rho_*$ was found in the BCD sample 
of \cite{P96a} to increase with decreasing mass, being typically by a factor 
$\sim$10 larger than in dIs. It is thus conceivable that the 
large range covered by BCDs with respect to $\rho_*$ may result 
in a wide diversity in their SF histories. 

The structural properties of the LSB component may also 
influence the spatial extent of SF regions in BCDs. 
Papaderos et al. (\cite{P96b}) have remarked that the plateau radius \p25\ 
is typically $\sim$1/2 of the isophotal radius \e25\ of the LSB 
component. They also reported a trend for the fractional area 
of the SF component of BCDs to increase with decreasing mass of 
the stellar LSB host.
The structural properties of the stellar LSB component appear to influence 
global SF processes in dIs also.
Hunter et al. (\cite{Hunter98}; see also Youngblood \& Hunter \cite{YH99};
Heller et al. \cite{Heller00}) have found that the azimuthally averaged 
star formation rate (SFR) per unit area in dIs scales with the surface density 
of their stellar LSB component, with the \hh\ regions being typically confined within 
\e25\ (Roye \& Hunter \cite{RoyeHunter00}). 
Perhaps a key result reported recently for dIs is a trend between the average 
SFR per unit area and the central surface brightness of the stellar LSB
population (van Zee \cite{vZ01}).

These empirical lines of evidence suggest that the temporal and 
spatial occurrence of SF activity in gas-rich dwarfs is not entirely 
dictated by their Dark Matter halo (cf. e.g. Meurer et al. \cite{Meurer98})
but is significantly influenced by the gravitational potential of their 
evolved host galaxy as well.
In view of the age discussion for I Zw 18 (Sect. \ref{dis3}), it 
is important to bear in mind that SF does not occur uniformly over the 
whole optical extent of a typical BCD, but is generally confined to
$R^*$~$<$~\e25, as shown by both surface photometry and CMD studies.
The confinement of SF activity to the central part of BCDs 
and compact dIs manifests itself in an appreciable 
colour contrast between the SF component and the LSB periphery 
(Sect. \ref{S3b}).
%
\section{The photometric structure of I Zw 18 \label{S4}}
\subsection{Surface photometry of the main body \label{S4a}}
In this section we derive SBPs of the main body of I~Zw~18 
without correction for ionized gas emission. 
For iE/nE BCDs, this is generally well justified for 
$R^*\!\!>\!\!r_{\rm eff}$, as equivalent widths (\ew) of the nebular emission lines, 
when averaged over the SF component,
are typically smaller than the width of the broad-band filters 
(e.g., \ew(\ha)~$<$~500~\AA; Salzer et al. \cite{Salzer89}; Terlevich et 
al. \cite{Terl91}; Thuan et al. \cite{TIL95}; 
Izotov \& Thuan \cite{IT98b}; Guseva et al. \cite{Nat00}; Hopp et al. \cite{Hopp00}). 

Since we wish to study the unresolved LSB emission in I~Zw~18, we first 
replaced compact sources in its periphery ($R^*\ga$ 8\arcsec\ for ground-based
images) by the mean intensity level of adjacent regions (cf. Sect. \ref{S2b}). 
No attempt has been made to screen out filamentary features in the LSB envelope
of I Zw 18. Since the identification of diffuse \ha\ shells depends on the resolution 
and depth of a given data set, doing so could affect colours derived 
from intercomparison of ground-based and \hst\ SBPs.
\begin{figure*} 
\begin{picture}(16.4,17)
\put(0,8.7){{\psfig{figure=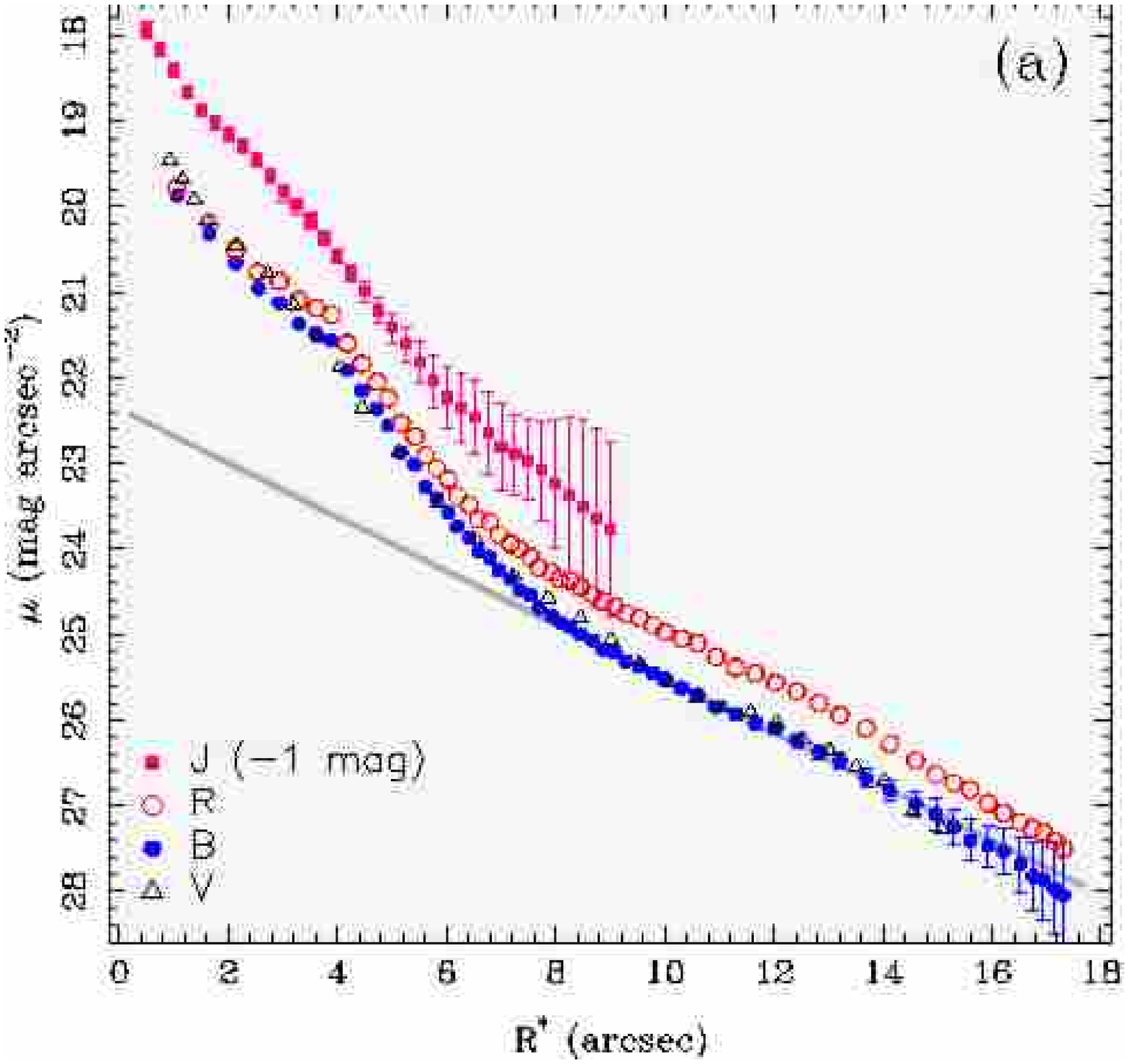,height=8.4cm,angle=0.,clip=}}}
\put(9.25,8.7){{\psfig{figure=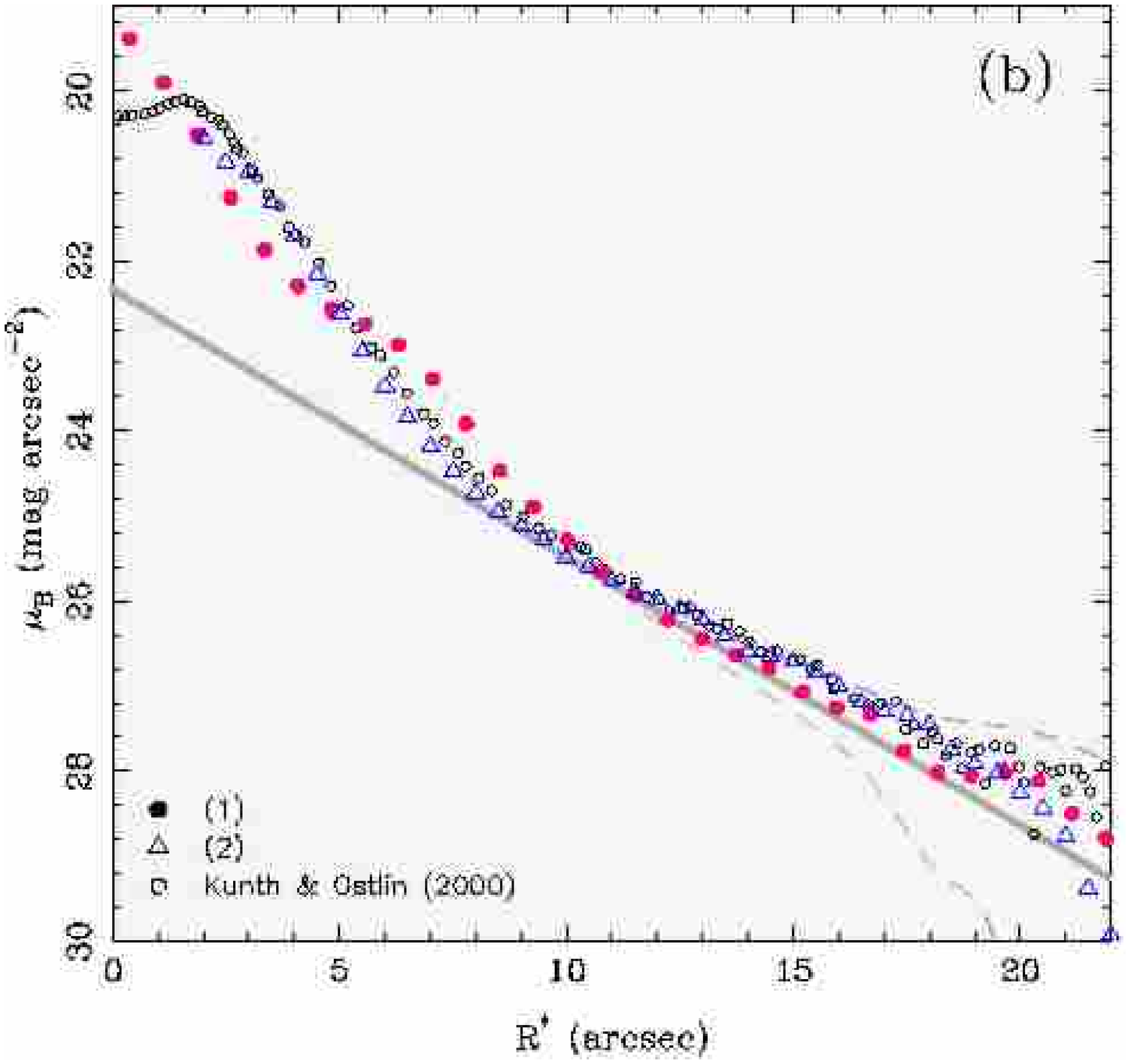,height=8.4cm,angle=0.,clip=}}}
\put(9.25,0){{\psfig{figure=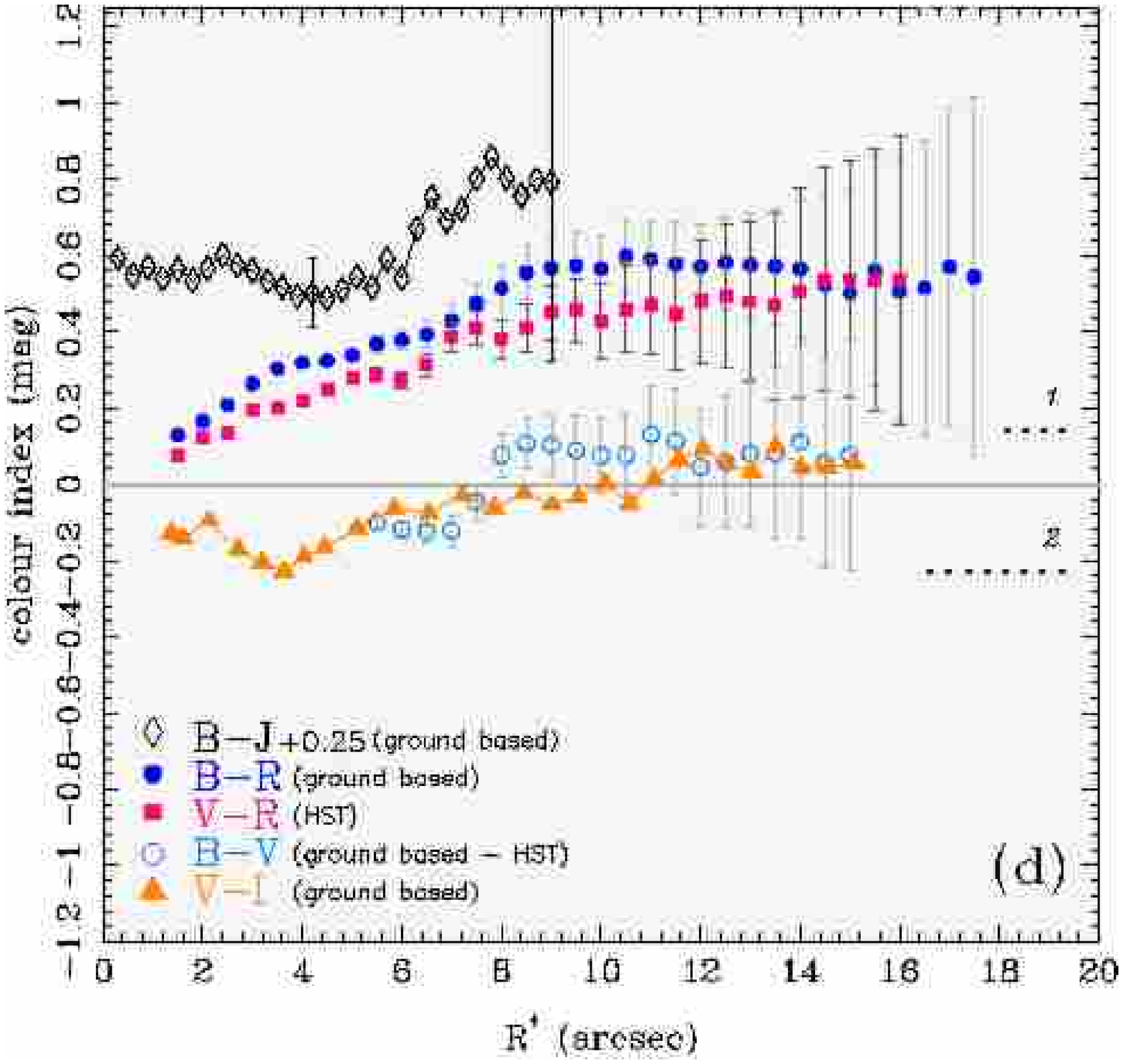,height=8.4cm,angle=0.,clip=}}}
\put(0,0){{\psfig{figure=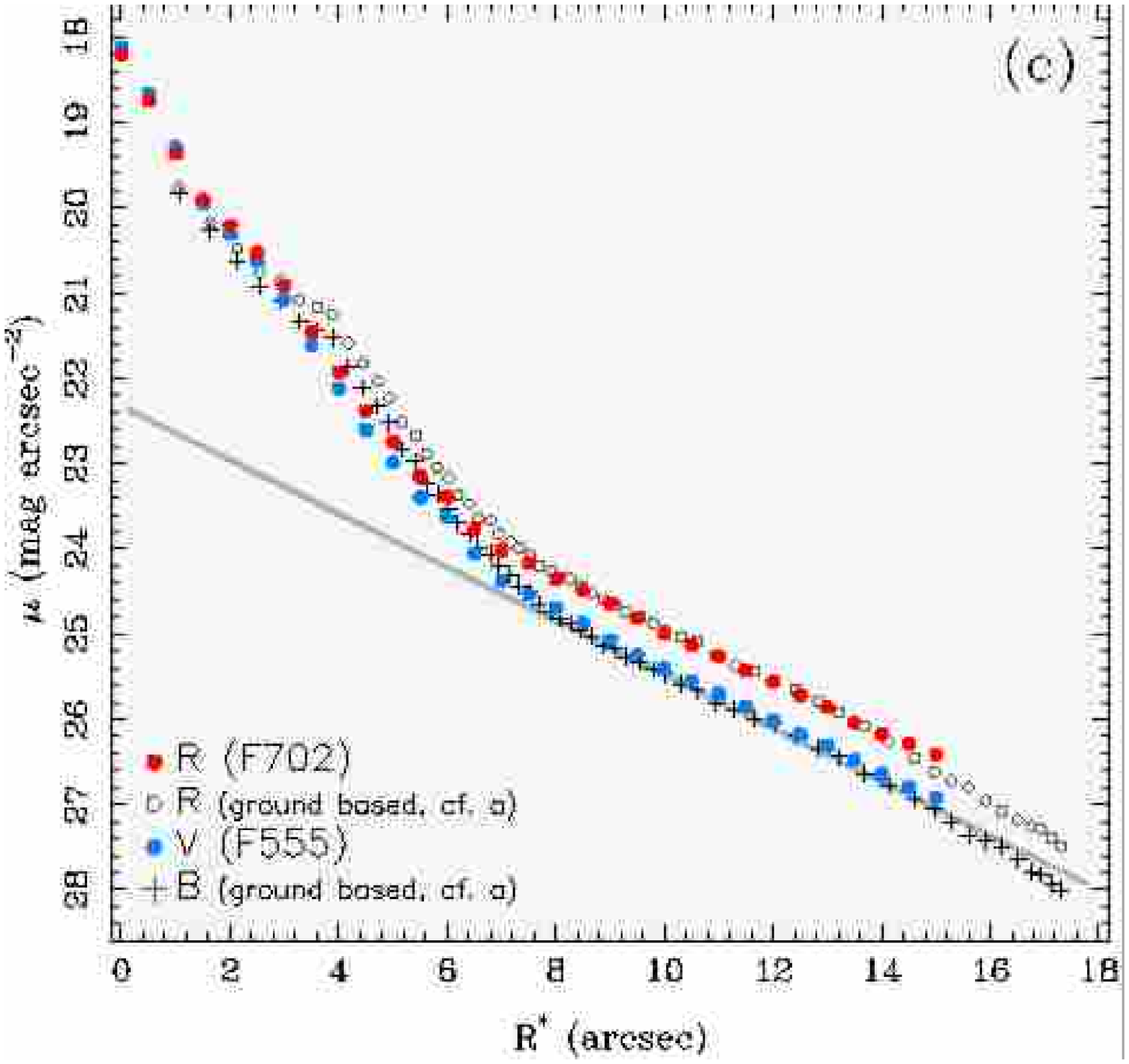,height=8.4cm,angle=0.,clip=}}}
\end{picture}
\caption[]{
{\bf (a)} $B$, $V$, $R$ and $J$ SBPs of 
I Zw 18 as derived from ground-based images. All SBPs have been
corrected for extinction ($A_V$~=~0.13 mag). The thick-grey line shows a linear 
fit to the $B$ SBP for $R^*\geq$~8\arcsec.
{\bf (b)} $B$ SBP of I Zw 18 as obtained from 
ground-based data using: (1) concentric circular apertures centered 
on the \nw\ region (filled points) and (2) ellipse fitting to the isophotes
(triangles), 
without removal of compact sources in the LSB envelope (see discussion 
in the text). 
Dashed curves show the effects on the SBPs obtained with method (1) when 
the sky background is over- or underestimated by $\pm$~1 count.
The linear fit to the LSB envelope in panel {\bf a} and the $B$ SBP of 
Kunth \& \"Ostlin (\cite{KO00}) (open circles) are included for comparison. 
{\bf (c)} Overlay of ground-based $B$ and $R$ SBPs on \hst/WFPC2 $V$ (F555W) and
$R$ (F702W) SBPs. The solid line has the
same meaning as in panel {\bf a}.
{\bf (d)} Colour profiles obtained by subtraction of ground-based (panel {\bf a})
and \hst/WFPC2 (panel {\bf c}) SBPs. 
Dotted lines indicate the average $B-R$ (denoted {\it 1}) and $B-V$ (denoted {\it 2})
colours obtained in the radius range 6~\arcsec$\leq R^*\leq$~12\arcsec, 
using \hst/WFPC2 images taken through the wide $B$ (F450W) filter.}
\label{f6}
\end{figure*}
%
We have evaluated the effect of this correction on the profile slope by masking
out extended filamentary features (such as the northwestern supershell; Fig.~\ref{f0}).
We have also compared SBPs derived prior 
to the subtraction of bright compact sources and after adding faint ($m_B\geq$ 22 mag) artificial stars 
at various locations in the LSB envelope. Depending on the method used 
(Sect. \ref{S2b}), a selective rejection of compact and diffuse features may 
result in a variation of up to 30\% in the SBP slope at low intensity levels.

Figure \ref{f6}a shows that the ground-based $B$, $V$, $R$ and $J$
SBPs of I Zw 18 can be well fitted by an exponential law in 
the radius range 8\arcsec~$\leq R^*\la$~16\arcsec\ (thick line). 
This gives for the LSB envelope a central surface brightness 
$\mu_{\rm E,0} = 22.3 \pm 0.1$ $B$ \sbb\ and an exponential scale 
length $\alpha\sim$ 250 pc (Table \ref{photom}). From extrapolation of the 
$B$ SBP slope to infinite radius, we obtain for the LSB envelope a total 
magnitude $m_B$ = 17.7 mag (equivalent to $\sim$ 24\% of the $B$ light). 
With a distance modulus of --30.88 mag, this corresponds to 
an absolute magnitude $M_{\rm LSB}$ = --13.2 mag. 
An exponential slope is also seen in the $J$ SBP
for $\mu\ga$ 23.5 $J$ \sbb, though with a significantly smaller 
scale length ($\alpha\sim$~165 pc). Since our data does not
allow for surface photometry beyond $\sim$ 25 $J$ \sbb, 
deeper imaging as that obtained by Hunt et al. (\cite{Hunt02})
is needed to assess the NIR properties of I Zw 18 
in its outermost regions.

The smooth exponential decrease in the LSB outer parts of 
I Zw 18 is surprising in view of its filamentary morphology. We have checked
that this is not an artifact introduced by methods {\sf iii} and {\sf iv}. 
Figure \ref{f6}b shows the overlay of the exponential slope 
in panel (a) on ground-based $B$ SBPs derived by: 
(1) flux averaging within concentric circular apertures centered on 
the \nw\ region (filled circles), and (2) ellipse fitting to isophotes 
prior to the removal of contaminating sources (triangles). 
The dashed curves show the shift that an offset of 
$\pm$1 count in the sky subtraction would introduce in the 
SBPs computed with method (1). We also show 
in Fig. \ref{f6}b the $B$ SBP of Kunth \& \"Ostlin (\cite{KO00}). 
All SBPs are in good agreement. Within the uncertainties, our derived
$B$ central surface brightness and scale length are consistent
with $\mu_{\rm E,0}=$~22.9 \sbb\ and $\alpha=$~310 pc 
read off Fig. 8 in Kunth \& \"Ostlin (\cite{KO00}). 
\hst\ $V$ (F555W) and $R$ (F702W) SBPs are also nearly indistinguishable from 
ground-based profiles in $B$ and $R$ in the radius range 
8\arcsec~$\leq R^*\leq$~14\arcsec\ (Fig. \ref{f6}c).  
Thus, although $\alpha$ may vary by $\sim$~20\% depending on the profile 
extraction method, the LSB envelope of I Zw 18 can be well fitted by an 
exponential law, despite its patchiness. 
\begin{figure}
\begin{picture}(8.6,13.)
\put(0,5.5){{\psfig{figure=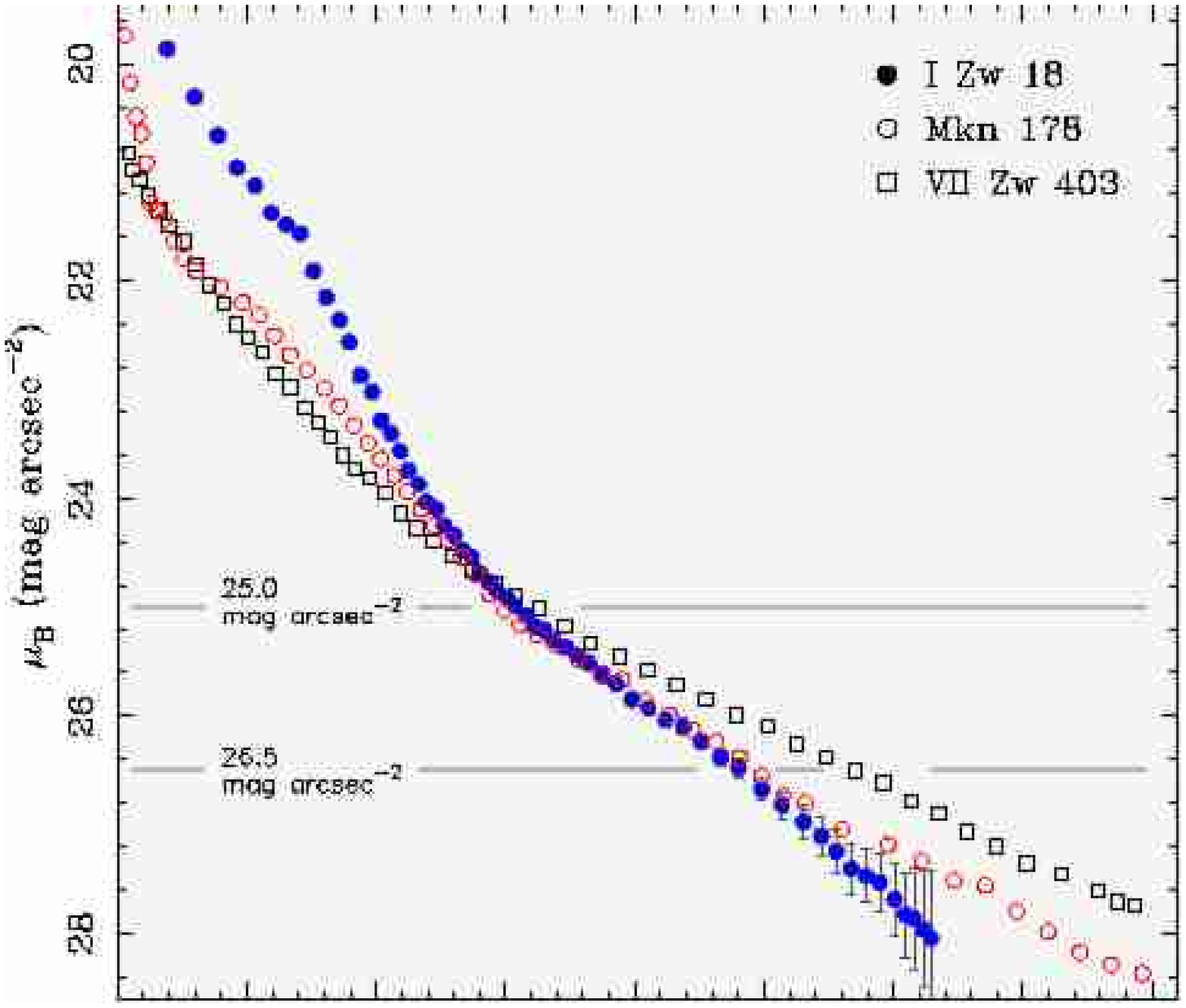,width=8.6cm,angle=0.,clip=}}}
\put(0,0.){{\psfig{figure=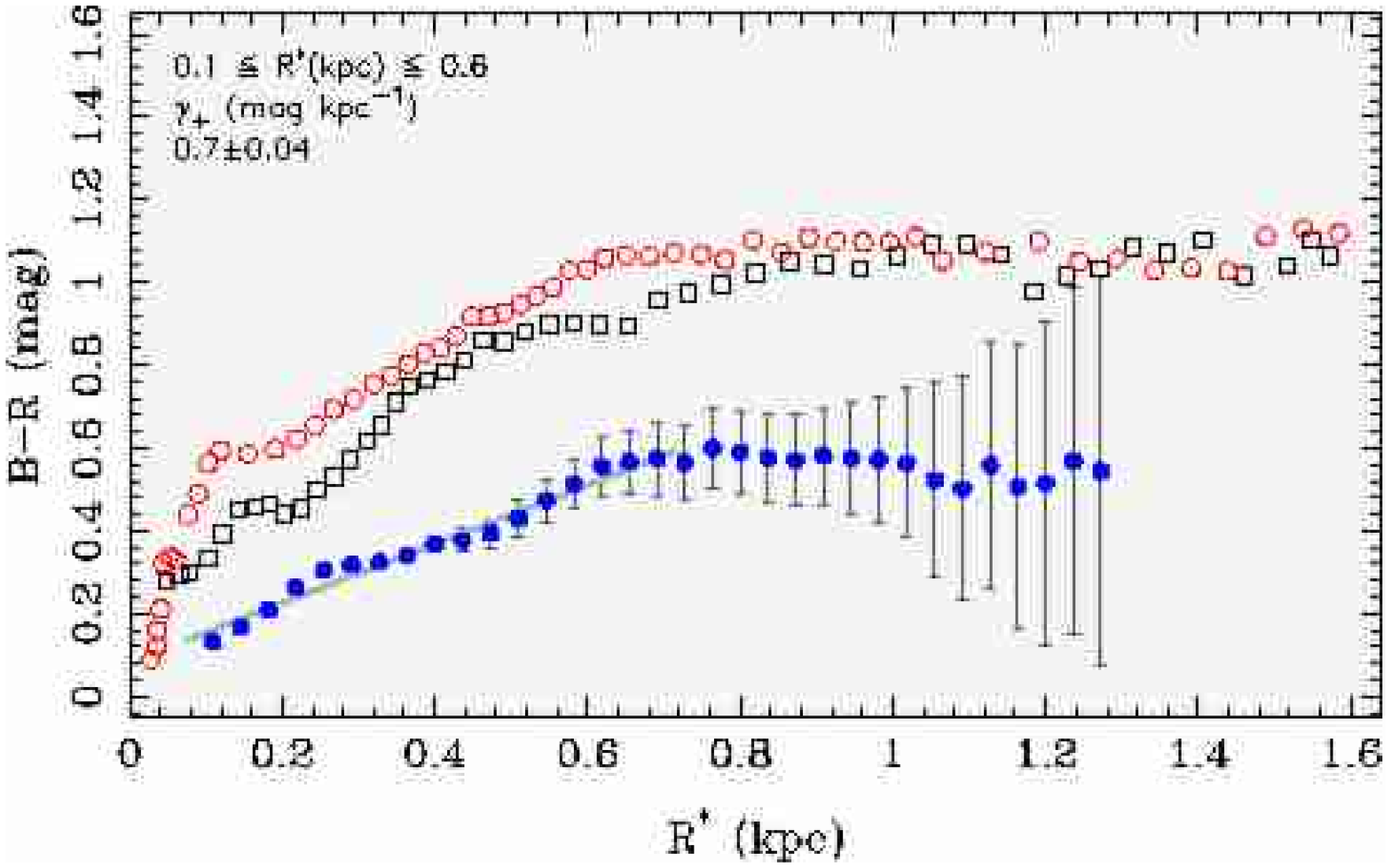,width=8.6cm,angle=0.,clip=}}}
\end{picture}
\caption[]{{\bf (top)} Comparison of the ground-based $B$ SBP 
of I Zw 18 with those of Mkn 178 and VII Zw 403.
All three BCDs have similar \e25\ and Holmberg 
radius, and show for $R^*>0.6$ kpc an exponential 
intensity decrease. 
The $B-R$ profile of I Zw 18 {\bf (bottom)} shows for 
$R^*<0.6$ kpc a linear colour increase with an average 
gradient $\gamma_+\approx 0.7$ mag kpc$^{-1}$, similar to that 
derived for Mkn 178 in Sect. \ref{S3a}. 
I Zw 18 is bluer throughout in comparison to the standard BCDs, 
the difference in the $B-R$ colour exceeding 0.5 mag in the
LSB regime, i.e. when $\mu >$25 $B$ \sbb.}
\label{f7}
\end{figure}

Figure \ref{f7} shows that the $B$ SBP of I Zw 18 
resembles closely that of an evolved iE BCD, 
such as Mkn 178 or VII Zw 403. All three systems have
similar isophotal sizes \e25\ and exponential 
scale lengths ($\sim$~0.25 -- 0.37 kpc). Furthermore, the absolute magnitude 
of the LSB envelope of I Zw 18 ($M_{\rm LSB}\approx$~--13.2 mag) is comparable
to that of Mkn 178 ($M_{\rm LSB}\approx$~--13.1 mag) and 
VII Zw 403 ($M_{\rm LSB}\approx$~--13.2 mag). 
The only notable difference between standard BCDs and I~Zw~18 
becoming apparent from Fig. \ref{f7} (bottom) is that the latter 
is considerably bluer throughout, the difference 
getting more pronounced in the outer regions.

The $B-R$ and $V-R$ colours of I Zw 18 (Fig. \ref{f6}d) increase smoothly, 
with gradients 0.74 mag kpc$^{-1}$ and 0.65 mag kpc$^{-1}$ for radii 
$R^*\leq$~9\arcsec\ ($\sim$~0.65 kpc). 
In the LSB regime ($\mu\geq$~25.2 $B$ \sbb) both profiles level off at 
$B-R\approx$~0.55 mag and $V-R\approx$~0.47 mag. 
The $B-R$ colour derived here is in good agreement with the 
value of $\sim$~0.6 mag of Kunth \& \"Ostlin (\cite{KO00}).

Assuming that the LSB emission in I Zw 18 is predominantly stellar, 
the observed $B-R$ colour would translate into an age $\tau$ of at 
least a few hundred Myr. From the PEGASE.2 evolutionary synthesis 
models (Fioc \& Rocca-Volmerange \cite{F97}) and adopting a Salpeter 
IMF with a lower and upper stellar mass limit of 0.1 and 120 \msun\ 
and a stellar metallicity of \zsun/50, one infers for an 
instantaneously formed stellar population a $\tau\sim$0.8 Gyr.
For an exponentially decreasing SF rate with an e-folding time of 1 Gyr, 
$\tau$ increases to $\sim$~2 Gyr, comparable to previous age estimates by 
Aloisi et al. (\cite{Alo99}; 0.5 -- 1.0 Gyr), \"Ostlin (\cite{Gor00}, 
$\geq$~1 Gyr) and Kunth \& \"Ostlin (\cite{KO00}; $\sim$~1 -- 5 Gyr).
%
\begin{figure*}[!ht]
\begin{picture}(16.4,6.25)
\put(12.1,0){{\psfig{figure=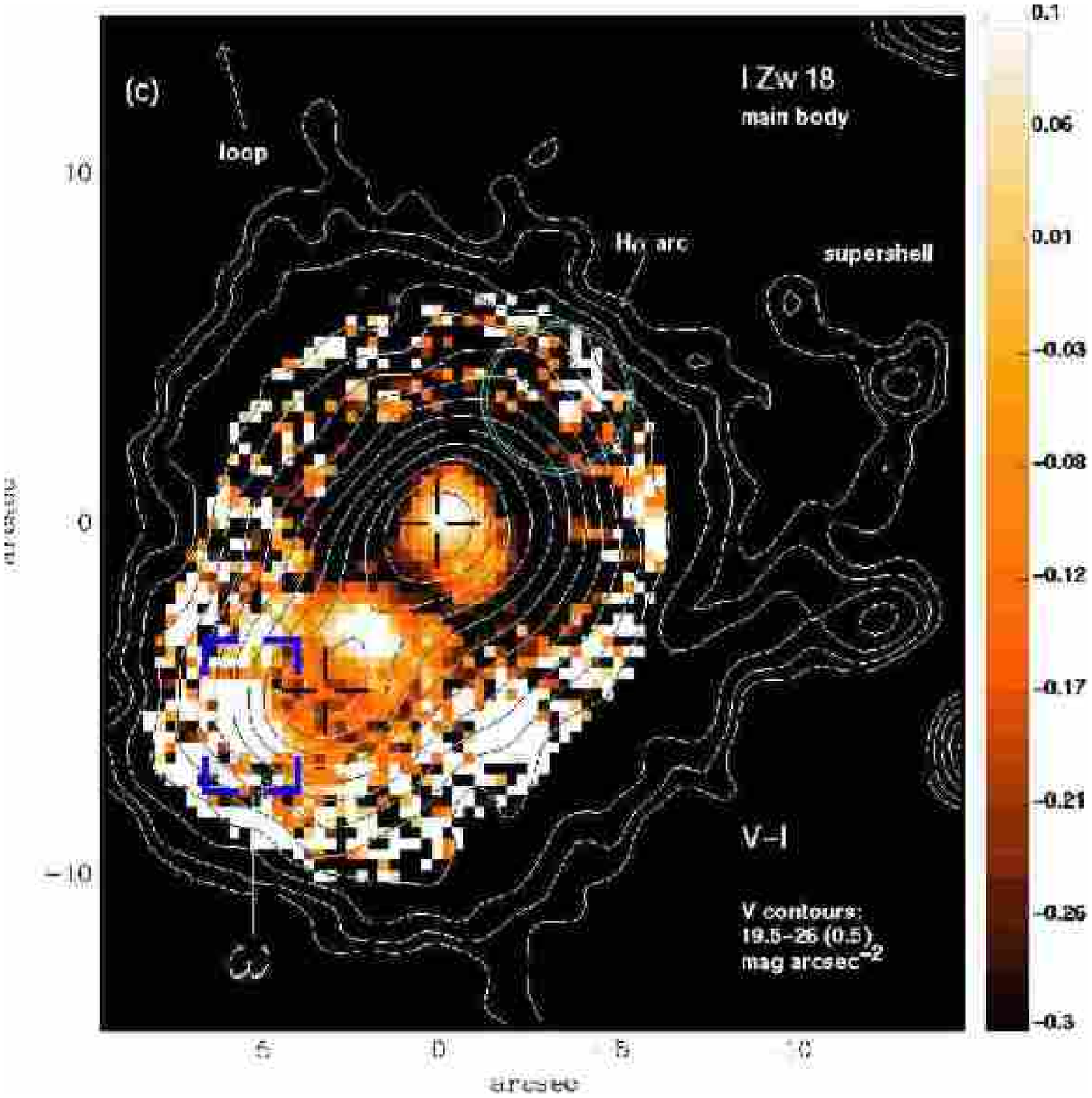,height=6.25cm,angle=0.,clip=}}}
\put(6.,0){{\psfig{figure=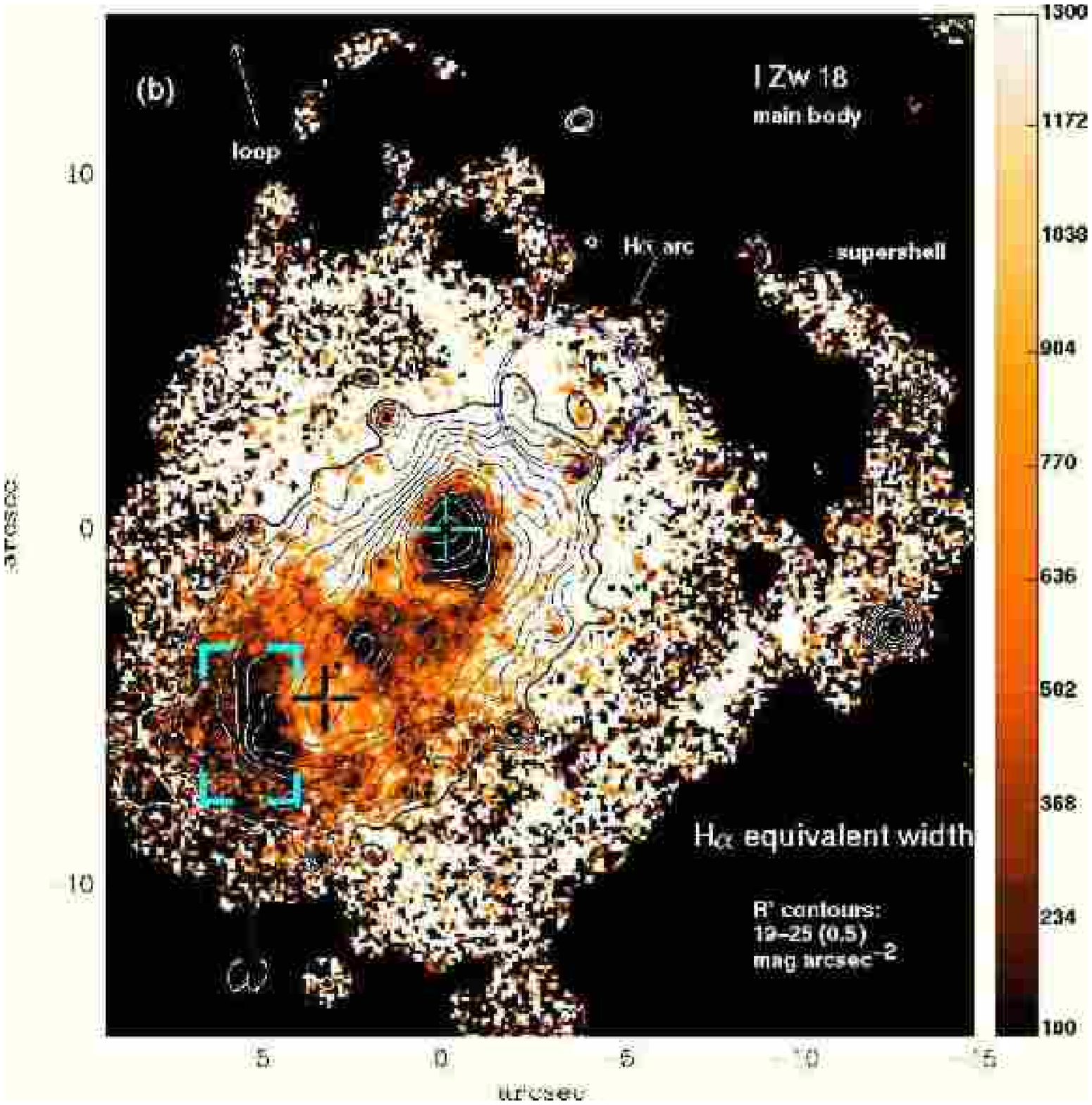,height=6.25cm,angle=0.,clip=}}}
\put(-0.2,0){{\psfig{figure=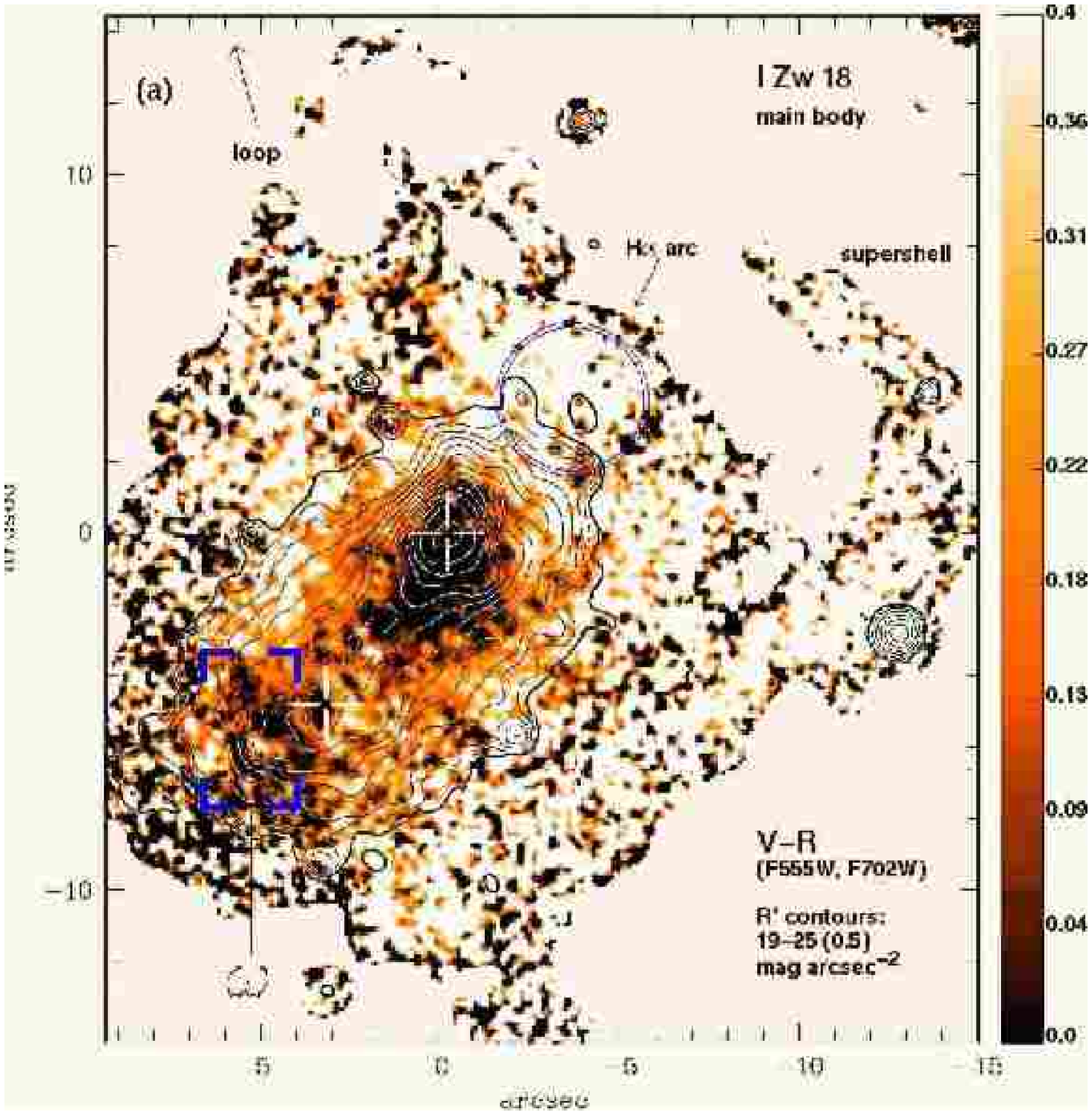,height=6.25cm,angle=0.,clip=}}}
\end{picture}
\caption[]{{\bf (a)} $V-R$ map of I Zw 18, as derived from \hst/WFPC2 F555W
and F702W images, assuming a uniform extinction of $A_V=0.13$ mag. 
Crosses mark the positions of the \nw\ and \se\ star-forming regions. 
The region labelled ``{\sl \ha\ arc}'' 
and the direction of the northern tip of the filamentary envelope 
(``{\sl loop}''), discussed by Izotov et al. (\cite{Yu01a}), are indicated.
Contours show the morphology of the stellar and ionized gas continuum
emission (referred to as $R$\arcmin), as 
obtained after subtraction of \ha\ emission from \hst/WFPC2 F702W images. 
Contours go from 19 to 25 $R$\arcmin\ \sbb\ in steps of 0.5 mag. 
A relatively smooth stellar 
host (denoted \lsb, cf. Sect. \ref{S4c}) underlying the \nw\ and \se\ 
regions, is seen. 
Note the featureless red ($V-R\ga$~0.4 mag) envelope surrounding the 
stellar body beyond the 25 \sbb\ $R$\arcmin\ isophote. 
The southeastern tip of the \lsb\ component where ionized 
gas emission is negligible is labelled $\omega$.
{\bf (b)} \ha\ equivalent width (\ew(\ha)) map of I Zw 18 
in the range 100 -- 1300 \AA\ (cf. Izotov et 
al. \cite{Yu01a}). Contours have the same meaning as in panel {\bf a}.
Note that the \ew(\ha) increases sharply in the periphery of region \nw, outside
the 22 \sbb\ $R$\arcmin\ isophote, exceeding 1300 \AA\ throughout the northwestern 
part of I Zw 18.
{\bf (c)} $V-I$ map of I Zw 18, as derived from ground-based data. 
The colour map is displayed in the range --0.3 -- 0.1 mag and overlaid with
19.5 to 26 \sbb\ $V$ contours. 
Note that the morphology of the blue ($V-I\sim$~--0.5 mag) region surrounding 
the \nw\ component correlates well with the $V-R$ and 
\ew(\ha) morphologies (panels {\bf a} and {\bf b}).
The reddest $V-I$ colours are observed 3\farcs6 northwest and southeast 
of the \se\ region, both coinciding with local minima in the \ew(\ha) distribution 
(cf. panel {\bf b}). The intermediate region, located between the \se\ and \nw\ 
regions, shows $V-I\approx$~0.1 mag while $V-I$ in 
region $\omega$ ranges between 0.17 and 0.32 mag.}
\label{f8}
\end{figure*}

However, such an evolutionary scenario for I Zw 18 cannot account for
all colours derived for the LSB envelope (Fig.~\ref{f6}d). 
The observed $V-R$ colour of 0.47 mag would require an age of $\sim$~20 Gyr
for a decreasing SF rate with an e-folding time of 1~Gyr.
Such a $V-R$ colour has been observed for the LSB host of 
old more metal-rich BCDs, such as Mkn 5, Mkn 33, Mkn 35, Mkn 370 
(Cair\'os et al. \cite{LM01a}) and Mkn 86 (Gil de Paz et al. \cite{Gil00a}).
However, these systems also show red $B-V$, $B-R$ and $V-I$ colours equal to
$\sim$~0.5 mag, $\geq$~1.0 mag and $\sim$~0.9 mag, respectively, in contrast  
to the $B-V$ and $V-I$ colours of the LSB envelope of I Zw 18 which are 
both very blue, equal to $\sim$~0.1 mag and $\sim$~0 mag, respectively. 
For an exponentially decreasing SF, such colours would respectively translate 
into stellar ages of $\sim$0.8 Gyr and $<$100 Myr, much lower than those obtained before
(an instantaneous burst model would result in even lower ages).
As for the $B-J$ colour, its value of $\sim$~0.6 mag at $R^*=9$\arcsec\ 
suggests a stellar age $\la$~250 Myr.
Beyond that radius corresponding to $\mu\sim$~25 $J$ \sbb, noise and source 
confusion prevent a reliable determination of optical--NIR colours.

Since neither a relatively young stellar population formed 
instantaneously nor an older one formed through an exponentially 
declining SF process can simultaneously account for all observed 
LSB colours, it is to be expected that also a purely stellar 
emission connected with other schematic SF histories 
(e.g., constant SF, periodic bursts) will not allow for a 
consistent solution. Therefore we have to abandon in the case of I Zw 18 
the usual assumption that stellar emission dominates the light
of the LSB envelope of BCDs.

As noted in Sect. \ref{S3a}, the combination of blue $V-I$ ($\la$~0 mag) 
with moderately red ($\sim$~0.3 -- 0.6 mag) $B-R$ colours 
suggests a significant nebular line contamination. 
There are a few examples of BCDs with colours severely affected by 
ionized gas emission on a spatial scale comparable to the LSB envelope of I Zw 18,
the magnitude of the effect depending on the telescope-filter 
transmittance and the spectroscopic properties and redshift of the BCD.
%
\begin{figure*} 
\begin{picture}(16.4,9.4)
\put(-0.1,0){{\psfig{figure=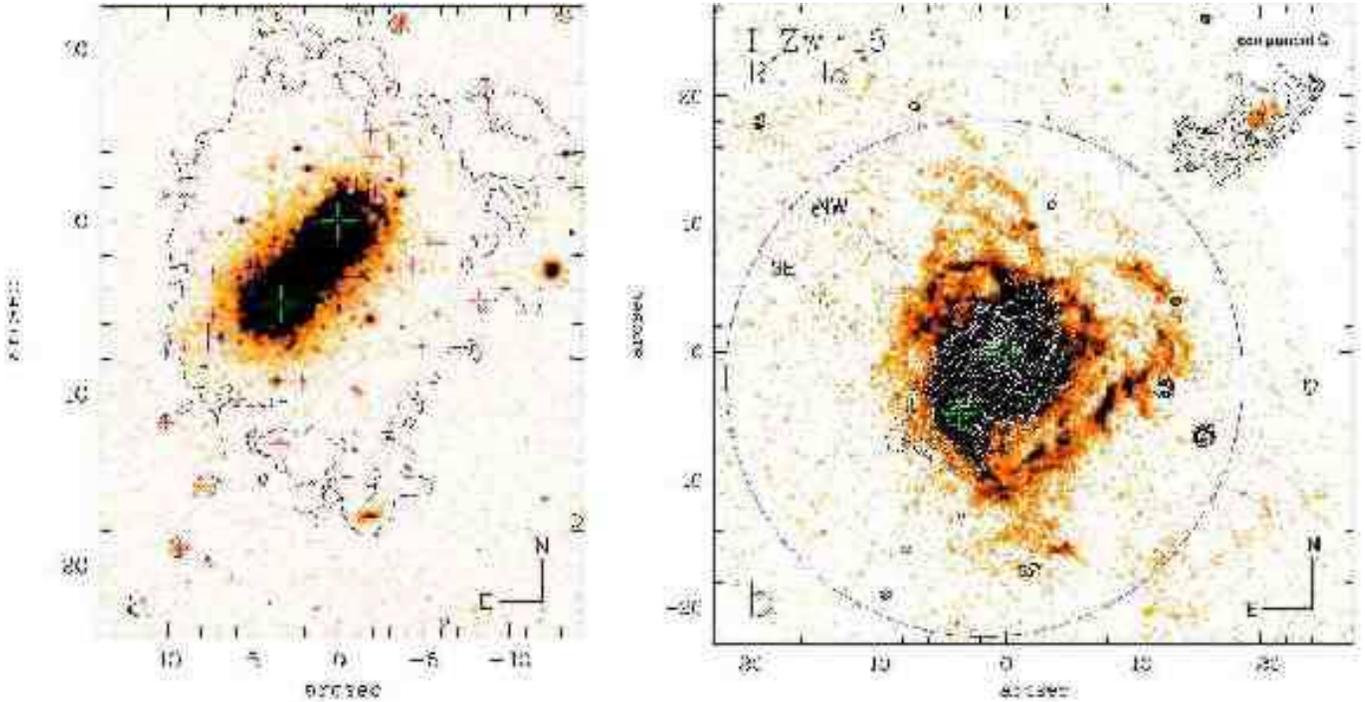,height=9.4cm,angle=0.,clip=}}}
\end{picture}
\caption[]{
{\bf (a)} $R$\arcmin\ map of I Zw 18.
The contour shows the morphology of the BCD at 
26 $R$ \sbb, prior to subtraction of \ha\ emission. 
Large crosses indicate the positions of the \se\ and \nw\ 
star-forming regions. Small crosses mark the location of compact sources in the 
outer parts of I Zw 18 that have been removed before the
derivation of surface brightness profiles (Figs. \ref{f5}a, \ref{f5}c and \ref{f10}a).
The mean $B$\arcmin~--~$V$\arcmin\ colour decreases from $\sim$~0.17 mag 
in region 1 to 0.05~--~0.1 mag in region 4 and inside the southeastern 
region $\omega$ (cf. Fig. \ref{f8}). 
The $V$\arcmin~--~$R$\arcmin\ colour is still significantly affected by 
ionized gas continuum emission in region 1
($\sim$~0.3 mag), being $\sim$~0.12~--~0.18 mag in the southeastern tip
of I Zw 18.
{\bf (b)} $R$\arcmin\ contours overlaid on an \ha\ map of I Zw 18, in the range
18.5 to 25.5 \sbb\ in steps of 0.5 mag. The circle, 40\arcsec\ in diameter, 
is centered between the \nw\ and \se\ regions.} 
\label{f9}
\end{figure*}
%
One example is the starburst region of the BCD Tol 1214--277 with 
\ew(H$\beta$)~$\approx$~324~\AA, $U-B$ $\sim$~--0.9 mag 
and $B-R$ $\sim$0.4 mag (Fricke et al. \cite{Fricke01}). 
Another is the filamentary LSB envelope of the BCD SBS 0335--052\,E where 
ionized gas accounts for $\sim$ 1/3 of the optical emission 
(Izotov et al. \cite{ILCFGK97c,Yu01b}). 
Just as in I Zw 18, the outer parts of SBS 0335--052 have a 
$B-R$ index of $\sim$~0.7 mag together with very blue 
$U-B$ $\sim$ --0.8 and $V-I$ $\sim$ --0.06 mag 
(Thuan et al. \cite{TIL97}; Papaderos et al. \cite{P98}),
a combination of colours which points to a substantial
contribution of ionized gas emission.

The ionized gas emission hypothesis is also supported by the fact that 
the apparent magnitude of I Zw 18 as determined from \hst\ images with 
the F450W $B$ filter, $m_B=$~15.83 mag, is $\sim$~0.3 mag brighter than 
$m_B$~=~16.12 mag derived from ground-based Johnson $B$ images. 
This is not the case for the \hst\ F555W and ground-based Johnson $V$ 
magnitudes which agree to within 0.02 mag.
Contrary to Johnson $B$, the transmittance of the F450W filter 
at the H$\beta$ and [\ion{O}{iii}]$\lambda\lambda$4959,5007 wavelengths is 
comparable to that at 4300 \AA, suggesting that the brighter 
F450W $B$ magnitude is caused by prominent emission lines.
Consequently, the $B-V$ = --0.23~$\pm$~0.06 mag and $B-R$ = 0.14~$\pm$~0.04 mag 
derived in the LSB envelope from the \hst\ images (dotted lines {\it 1} and {\it 2}
in Fig. \ref{f6}d), are significantly bluer than those derived from ground-based images.

Recently, deep Keck spectra by Izotov et al. (\cite{Yu01a}) have shown that
nebular line emission is present as far as 30\arcsec\ 
($\sim$~2.2~kpc) from the main body of I Zw 18, and that the equivalent width
of the H$\alpha$ line exceeds 1300 \AA\ 
over the whole northwestern half of the BCD and along the western supershell.
In the regions labelled {\sl loop} and {\sl H$\alpha$ arc} in 
Fig. \ref{f0} the \ew(\ha) attains values of $\approx$~1400 \AA\ 
and $\approx$~1700 \AA, respectively (Izotov et al. \cite{Yu01a}).
These results are in agreement with the high ($\sim$~1500 \AA) \ew(\ha) 
measured over a large part of the optical extent of I Zw 18 by 
\"Ostlin et al. (\cite{Gor96}) from high resolution ground-based images.

It is instructive to compare the $V-R$ and $V-I$ maps
with the \ew(\ha) distribution (Fig.\ \ref{f8}).
The reddest $V-R$ colours (0.2 -- 0.5 mag) are observed 
along the supershell and the extended \ha\ rim intersecting
the {\sl H$\alpha$ arc} region (Fig. \ref{f8}a). 
On the other hand, the $V-I$ colours in the corresponding 
regions (panel c) show an opposite trend, being the bluest 
($\sim$ --0.5 mag). 
Inspection of Fig. \ref{f8}b reveals that extended regions 
showing both red $V-R$ and blue $V-I$ colours are spatially 
correlated with \ew(\ha), but are anticorrelated with the 
surface density of the stellar background (contours in panels a and b). 
This situation resembles that in the periphery of bright SF 
regions in VII Zw 403 (Sect. \ref{S3a}) or the one described by 
Fricke et al. (\cite{Fricke01}) in the BCD Tol 1214--277.
In the absence of a significant underlying stellar component, 
the \ew(\ha) in the extended ionized gas emission 
raises in the latter to $\sim$~2000~\AA\ as far out as 2.7 kpc away 
from the starburst region. 

The conspicuous spatial anticorrelation between stellar and nebular
emission together with the tight spatial coupling of \ew(\ha) with colours 
renders a meaningful study of any stellar LSB background component
of I Zw 18 from broad-band data impossible.
Therefore it is necessary to subtract the ionized gas emission prior 
to the broad-band photometry.

\subsection{Subtraction of nebular line emission \label{S4b} }
%
One way to subtract out the contribution of the line and continuum emission
of the ionized gas is to use long-slit spectra (Papaderos et al. \cite{P98}; 
Thuan et al. \cite{TIF99}; Noeske et al. \cite{Kai00}; Izotov et
al. \cite{Yu01a}; Guseva et al. \cite{Nat01}). 
However, this allows correction only along the slit position,
i.e. it can hardly be used to study the 2-dimensional properties 
of the stellar LSB component in I Zw 18. 

Therefore we decided to use continuum-subtracted narrow-band \ha\ and 
[\ion{O}{iii}]~$\lambda$5007 maps to correct broad-band \hst\ images for 
the strongest nebular emission lines. 
We first correct the $R$ (F702W) image where \ha\ is the major contributor of 
ionized gas emission by subtracting from the 
$R$ frame the continuum-subtracted \ha\ (F658N) image, scaled 
by a factor $c_R\approx 0.7$. The resulting image is called $R$\arcmin\ (Fig. \ref{f9}a).
The scaling factor has been 
determined by taking into account the telescope/detector throughput 
with the F658N and F702W filters at the redshifted wavelength 
of the \ha\ line, the respective exposure times and the width of 
the F702W filter. The latter has been calculated by integrating the 
system/filter transmittance curve over the wavelength range 
5700\AA\ --\ 9000\AA. 

\begin{figure*}[!ht]
\begin{picture}(16.4,13.25)
\put(0,5.55){{\psfig{figure=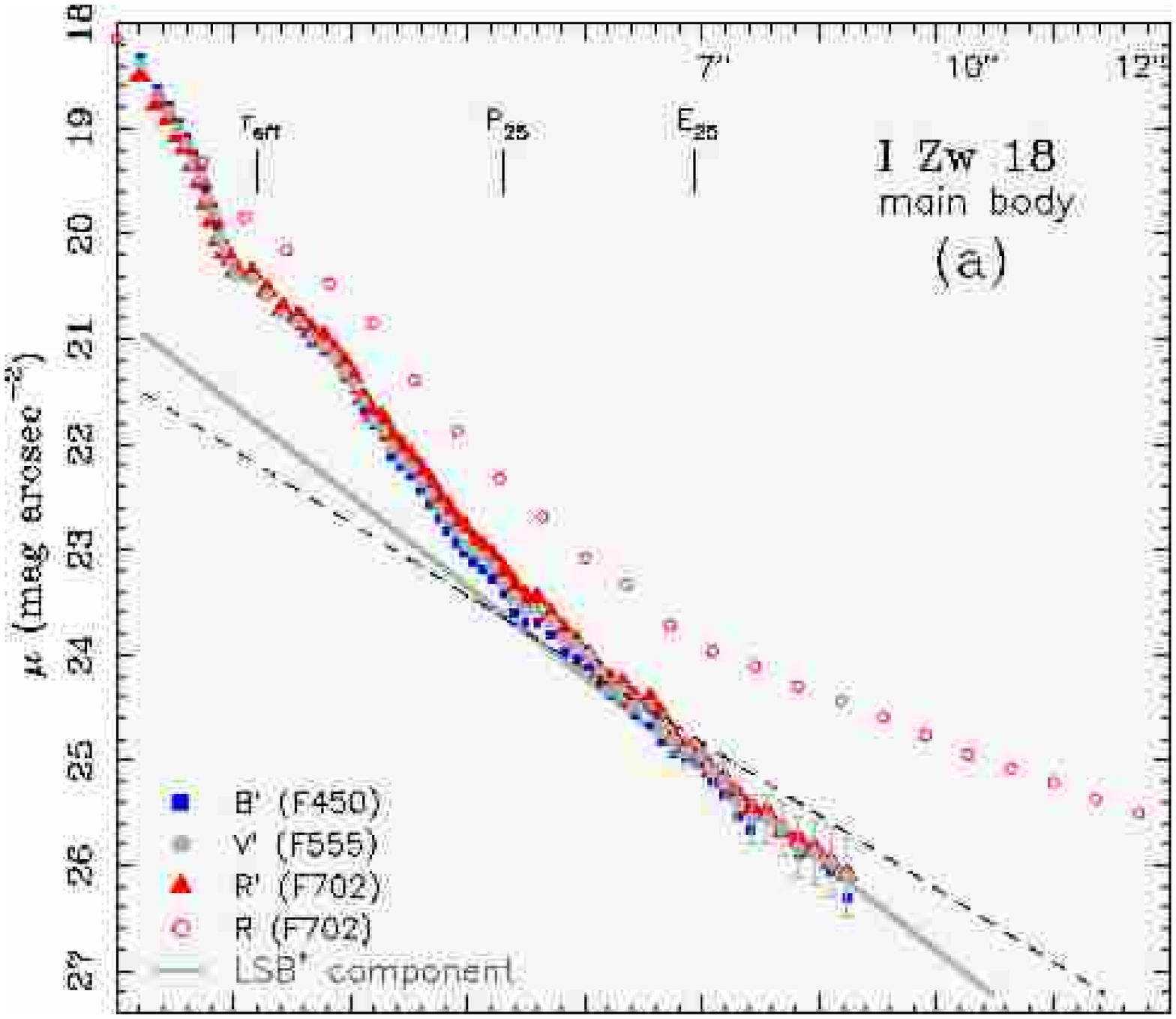,width=8.7cm,angle=0.,clip=}}}
\put(0.05,0.){{\psfig{figure=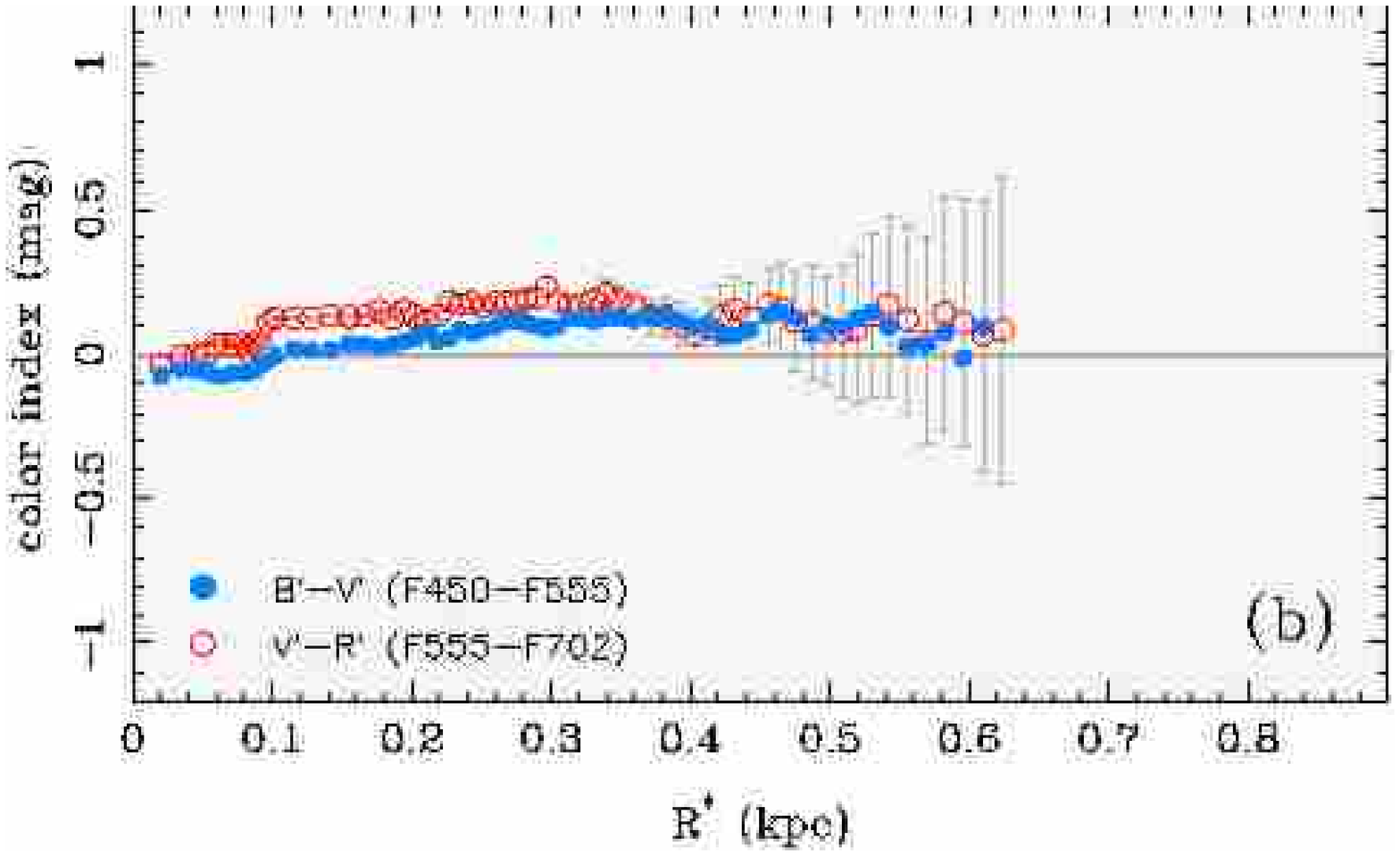,width=8.7cm,angle=0.,clip=}}}
\put(9,5.55){{\psfig{figure=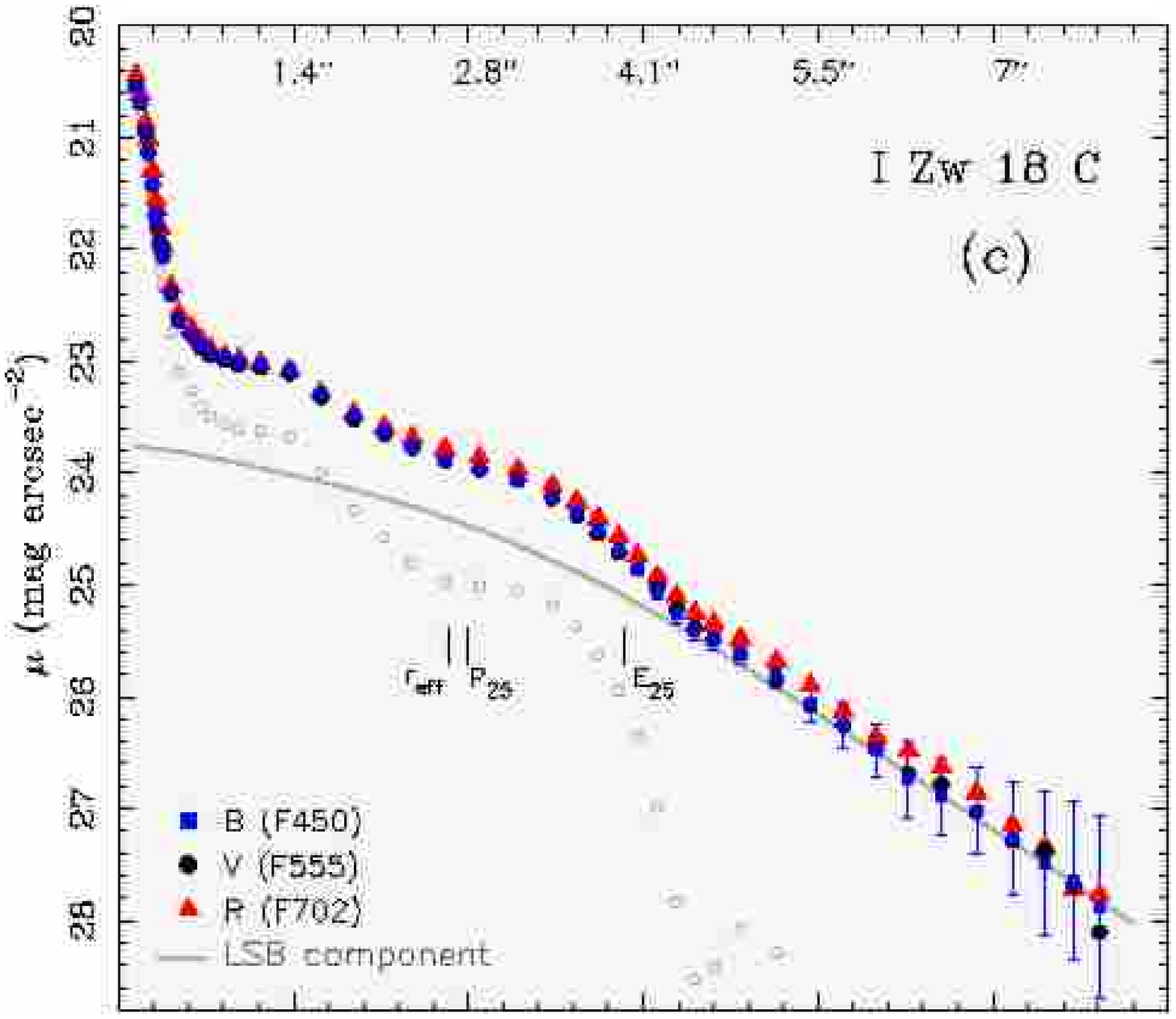,width=8.7cm,angle=0.,clip=}}}
\put(9.05,0.){{\psfig{figure=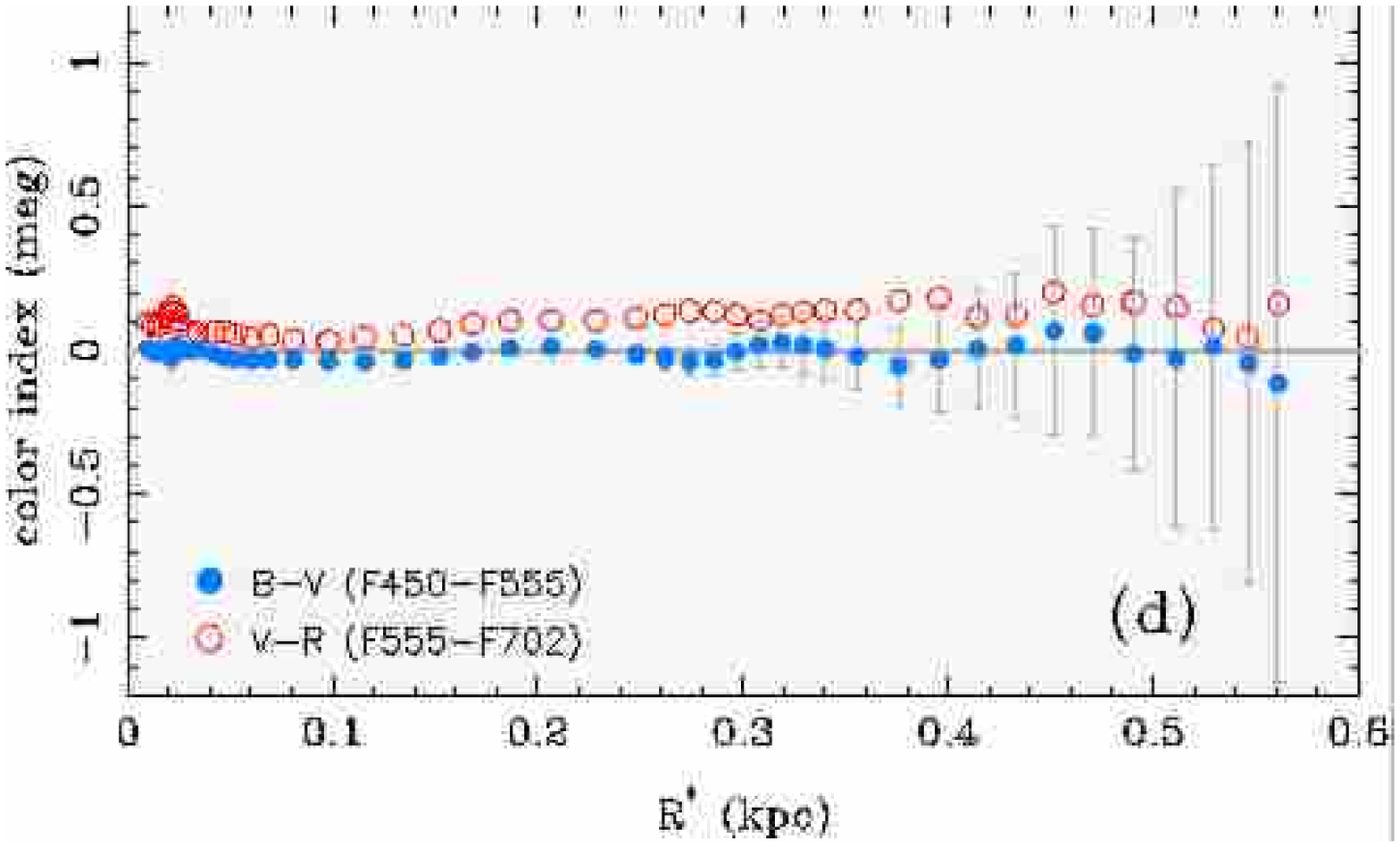,width=8.8cm,angle=0.,clip=}}}
\end{picture}
\caption[]{{\bf (a)} Surface brightness profiles of the main body of I\ Zw\ 18
derived from archival \hst/WFPC2 $B$ (F450W), $V$ (F555W) and $R$ (F702W) images, 
after subtraction of nebular line emission. These SBPs, labelled $B$\arcmin, 
$V$\arcmin and $R$\arcmin, show the radial intensity distribution
of the stellar and ionized gas continuum emission.
For comparison, we show with open circles the $R$ SBP, 
as derived prior to subtraction of nebular line emission (Fig \ref{f6}c). 
A linear fit to the outermost part ($R^*\geq$~0.5 kpc) 
of the $V$\arcmin\ SBP (labelled ``\lsb\ component'') is 
shown by the thick-grey line. The slope of the dashed line is an upper 
limit to the exponential scale length of the \lsb\ component. 
It has been derived from fitting ellipses to its outer parts, without 
removing isolated spots of residual ionized continuum emission along the 
northwestern supershell. The effective radius $r_{\rm eff}$ and
the isophotal radii \p25\ and \e25\ of the star-forming and LSB 
components at 25 $V$ \sbb\ are indicated. 
{\bf (b)} 
$B$\arcmin~--~$V$\arcmin\ and $V$\arcmin~--~$R$\arcmin\ colour
profiles of the main body of I Zw 18, obtained by subtraction of the 
SBPs in panel {\bf a}. 
{\bf (c)} SBPs of I Zw 18\,C derived from archival \hst/WFPC2 images 
assuming $A_V=0$ mag. 
The thick grey curve (labelled ``LSB component'') shows a model of the $B$
intensity distribution 
of the LSB component, according to Eq. (\ref{eq:p96a}) with ($b$,$q$)~=~(2.4,0.8). 
The surface brightness distribution of the luminosity in excess of the 
fit is shown by small open circles. The radii $r_{\rm eff}$, \p25\ and \e25\ 
refer to the $B$ SBP.
{\bf (d)} $B$~--~$V$ and $V$~--~$R$ colour
profiles of the C component.}
\label{f10}
\end{figure*}

The subtraction of the \ha\ emission results in a nearly complete removal 
of the patchy LSB envelope of I Zw 18. While this correction 
has practically no effect on the C component (its flux is reduced by 
less than 2\%), it removes about 1/3 of the $R$ light in the main body. 
Some residual ionized continuum emission is still visible on the $R$\arcmin\ 
image, especially along the northwestern supershell. Since these faint 
(25.2 \sbb\ $\la \mu_{R\arcmin} \la$~26.7 \sbb) residuals are well separated 
from the main body and can be easily masked out before computing SBPs 
(Fig. \ref{f10}a), no attempt has been made to remove them by adjusting 
$c_R$. A nearly complete subtraction of both the \ha\ and continuum emission 
was possible however, by decreasing $c_R$ in small steps to about 0.5. Each 
resulting image was visually inspected after adaptive smoothing and checked 
for non-negative background. From the image with $c_R$ = 0.5,
we estimate the total ionized continuum and \ha\ contribution 
to be $\sim$~50\% of the $R$ light of I Zw 18.
The subtraction of the \ha\ emission gives the LSB component underlying the NW
and SE SF regions of I Zw 18 a more regular and compact appearance
(Fig. \ref{f9}b, $R$\arcmin\ contours).
The compactness of the stellar host galaxy can also be seen in Fig. 1c
by Dufour et al. (\cite{Duf96b}) who were the first to
present \hst/WFPC2 $V$ images with the ionized gas emission removed.

Correction for nebular line emission is more uncertain for
the $B$ (F450W) and $V$ (F555W) images, since both are
contaminated by several prominent Balmer and oxygen 
emission lines. While e.g. the [\ion{O}{ii}]$\lambda$3727, H$\delta$, H$\gamma$ 
emission lines are included in F450W only, both filters have nearly the same 
transmittance near the H$\beta$ and [\ion{O}{iii}]$\lambda\lambda$4959,5007.
Because of the lack of complete 2-D information, we have taken 
the line ratios tabulated for the region ``{\sl \ha\ arc}'' by 
Izotov et al. (\cite{Yu01a}) to be representative for the whole LSB envelope. 

First we determined for each filter the ratio of the flux $f_{\rm i}$ 
of the [\ion{O}{iii}]~$\lambda$5007 line to the sum of the fluxes of all 
other prominent emission lines. In doing so, we have taken into 
account the intensity of each emission line and the 
system's transmittance at the corresponding redshifted wavelength. 
From each $f_{\rm i}$ and the estimated flux contribution
of the [\ion{O}{iii}]~$\lambda$5007 line to the filters F450W and 
F555W, we computed the factors $c_B$ and $c_V$ which we used to scale 
the F502N image prior to its subtraction from broad-band images. 
Alternatively, $B$ and $V$ images were 
corrected using both F502N and F658N exposures to subtract 
separately oxygen and Balmer emission lines. 
Visual inspection of the corrected images after adaptive filtering 
shows that both approaches give comparable results, with uncertainties in 
the SBP slope comparable to 1$\sigma$ Poisson errors. 

\subsection{Surface photometry of the stellar and ionized gas continuum
in the main body of I Zw 18 \label{S4c} }

In the following we use broad-band \hst/WFPC2 images, corrected for 
nebular line emission (henceforth denoted $B$\arcmin, $V$\arcmin\ and
$R$\arcmin), to compute the surface brightness and colour distribution
of the {\sl unresolved} LSB background surrounding the \nw\ and \se\ regions. 
As in Sect. \ref{S3a}, we have first replaced compact 
sources in the outer parts of I Zw 18 (indicated by crosses in Fig. 
\ref{f9}a) by the mean value of adjacent regions.
We refer to the residual stellar and ionized continuum LSB emission 
of I~Zw~18, obtained after subtraction of the nebular line contribution, 
as \lsb, in order to distinguish this from the filamentary LSB envelope 
studied from raw $B$, $V$ and $R$ images in Sect. \ref{S4a}.

In Fig. \ref{f10}a we compare the $B$\arcmin, $V$\arcmin\ and $R$\arcmin\ SBPs 
of I Zw 18 with its $R$ SBP (Fig. \ref{f6}c). 
The $B$\arcmin, $V$\arcmin\ and $R$\arcmin\ profiles 
are nearly indistinguishable from one another, suggesting negligible 
colour gradients over the optical extent of the BCD. They display an 
exponential fall-off with a scale length $\alpha\sim$100 pc for 
$R^*\ga$~3\arcsec\ ($\sim$~0.2 kpc). 
To ensure that residual gaseous continuum emission in the vicinity of the 
SF regions does not affect the derived structural properties of the 
\lsb\ component, we fit only its outermost regions ($R^*\geq~$0.5 kpc) 
where the surface brightness drops in all bands to less than 25 \sbb,
and where in standard BCDs, the emission is dominated by the luminosity
of the evolved stellar LSB host (cf. Sect. \ref{S3b}).
Linear fits give in all bands $\alpha\sim$~120 pc and a $B$\arcmin\ central surface 
brightness $\mu_{\rm E,0}\approx$~20.7 \sbb\ (cf. Table \ref{photom}). 
The dashed line in Fig. \ref{f10}a represents the SBP of the \lsb\ component,
estimated by fitting ellipses to the $B$\arcmin\ image, without masking out 
compact sources or spots of residual continuum emission 
along the northwestern supershell. Its scale length of 160 pc represents a
formal upper limit to the true scale length of the LSB$^*$ component.

\begin{figure}
\begin{picture}(16.4,7)
\put(0,0){{\psfig{figure=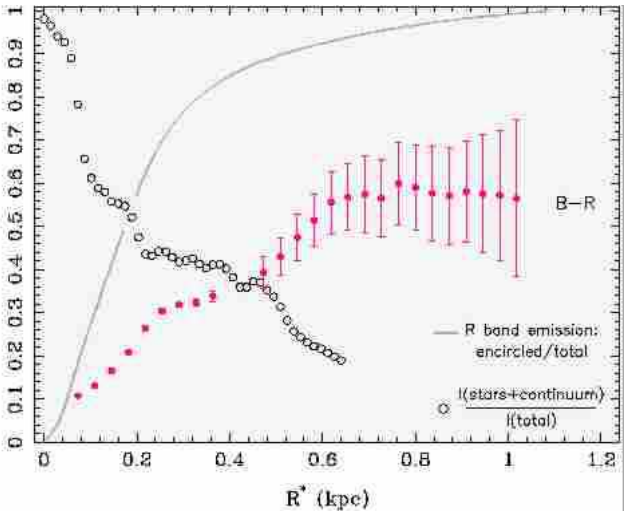,width=8.4cm,angle=0.,clip=}}}
\end{picture}
\caption[]{Fractional contribution of the stellar and 
ionized gas continuum ($R$\arcmin) to the $R$ emission of 
the main body of I Zw 18 as a function of the galactocentric distance
$R^*$ (open circles). The run of the ratio of the luminosity within a
circle of radius $R^*$ to the total luminosity
is shown by the light curve. The contribution of the underlying stellar 
component drops to $<$~50\% beyond the effective radius ($\sim$~0.2 kpc, 
cf. Table \ref{photom}). 
For $R^*>$~0.65 kpc ($\approx$~9\arcsec), ionized gas emission accounts 
for more than 80\% of the $R$ line-of-sight intensity and is responsible 
for the relatively red $B-R$ colour of
the filamentary LSB envelope of I Zw 18 (cf. Fig. \ref{f6}d).}
\label{f11}
\end{figure}

From the comparison of the $R$ and $R$\arcmin\ SBPs (Figs. \ref{f6}c and 
\ref{f10}a), we can derive a one-dimensional representation 
of the relative contribution of the stellar and 
ionized gas continuum to the $R$ intensity as a 
function of the galactocentric radius $R^*$. 
Figure \ref{f11} shows that the stellar and ionized gas continuum emission
dominates only within the inner 0.2 kpc ($\sim r_{\rm eff}$; 
cf. Table \ref{photom}) of I Zw 18. 
For larger radii, its contribution gradually decreases, 
becoming less than 20\% at $R^*\!\!\sim$~0.65 kpc (9\arcsec). 
Beyond that radius, where the surface brightness is $\ga$~26 $R$\arcmin\ \sbb,
the data become too noisy to allow to pin down the 
line-of-sight contribution of the stellar component to the total luminosity.
However, since nebular line emission accounts for at least 80\% of the $R$ 
flux at large radii, it determines the colour in the outer parts of I~Zw~18
(cf. Fig.~\ref{f6}d).

The \lsb\ underlying component of I Zw 18 shows an important difference as 
compared to the main class of evolved BCDs. From both surface 
photometry and CMD studies, it is known that these systems show a substantial 
reddening from the SF region to the LSB periphery (Sect. \ref{S3a}).
The $B'-V'$ and $V'-R'$ profiles of I Zw 18 (Fig. \ref{f10}b) do show 
relatively strong gradients of respectively 0.72~$\pm$~0.03 mag kpc$^{-1}$ and 
0.77~$\pm$~0.05 mag kpc$^{-1}$ within the inner $\sim$~0.2 kpc, only.
However, at larger radii both colours remain blue
($B'-V'$~=~0.09~$\pm$~0.04 mag and $V'-R'$~=~0.12~$\pm$~0.04 mag), 
implying a minor colour contrast between the SF and the \lsb\ component.
Because of this, subtraction of the \lsb\ emission (cf. Sect. \ref{S3a}) 
has practically no effect on the colour profile inside \p25\ ($\sim$~0.3~kpc).
This correction shifts the $B$\arcmin~--~$V$\arcmin\ 
and $V$\arcmin~--~$R$\arcmin\ colours at $R^*\approx$~60~pc to --0.07 mag and 
--0.03 mag, respectively, not significantly bluer than those obtained 
from direct SBP subtraction. 
This is to be compared with the situation for the old BCDs 
Mkn 178 and VII Zw 403 (Sect. \ref{S3a}, Fig. \ref{f5}b and \ref{f5}d), where 
correction for the red LSB background shifts optical colours 
by --0.4 - --0.6 mag over a galactocentric distance of $\sim$~0.5 kpc.

Admittedly, surface photometry allows to derive only a 
luminosity-weighted average colour. By this fact, a nearly 
constant blue colour (Fig. \ref{f10}b) over a surface brightness 
span of some 8 mag does not rule out {\sl per se} the presence 
of a faint old stellar background. However, inspection of colour 
maps does not show any evidence for such a population. 
Even after subtraction of nebular line emission, the colours of 
I Zw 18 in its northern half, especially along the {\sl \ha\ arc} 
($B$\arcmin~--~$V$\arcmin\ $\approx$ 0.16 mag, 
$V$\arcmin~--~$R$\arcmin$\approx$ 0.3 mag) appear to 
be significantly affected by ionized continuum emission 
for which $B-V = 0.34$ mag and $V-R =0.64$ mag (Kr\"uger \cite{Harald92}).
%
\begin{table*}
\caption{Structural properties of I Zw 18 (main body) and I Zw 18\,C$^{\rm a}$}
\label{photom}
\begin{tabular}{lccccccccclc}
\hline
\hline
Component & Band & $\mu_{E,0}$ & $\alpha $ & $m_{\rm LSB}^{\rm fit}$ & \p25\  & 
$m_{P_{25}}$ & \e25\ & $m_{E_{25}}$ & $m_{\rm SBP}$ & $m_{\rm tot}$ & $r_{\rm eff}$,$r_{80}$\pano  \\
   &       & \sbb\ &  pc   & mag    & pc       &  mag             &  pc
          &  mag             &    mag & mag & pc \\
    (1) &   (2)         &   (3)     &  (4)      &   (5)            &
 (6)    &  (7)             &  (8)     &  (9)  & (10) & (11) & (12)\kato \\
\hline
I Zw 18 & $B$ & 22.33$\pm$0.14  & 249$\pm$10 & 17.67 & 468  & 16.44  & 611  & 18.05  &
 16.11$\pm$0.01  & 16.12$^{\rm b}$ & 200,344\pano \\
{\sf main body} & $V$ & 22.23$\pm$0.17  & 246$\pm$16 & 17.59 & 469  & 16.29  & 627  & 17.91  & 
 16.01$\pm$0.01  & 16.01 & 181,319\\
{\sf ground-based} & $R$ & 21.71$\pm$0.17  & 244$\pm$12 & 17.08 & 462  & 16.33  & 741  & 17.31  & 
 15.90$\pm$0.02  & 15.89 & 220,377\\
 & $I$ & 22.28$\pm$0.24  & 247$\pm$27 & 17.63 & 471  & 16.33  & 619  & 18.00  & 
 16.04$\pm$0.03  & 16.00 & 174,323\\
 & $J$ & 20.41$\pm$0.60  & 165$\pm$39 & 16.64 & 409  & 16.45  & 697  & 16.73  & 
 15.83$\pm$0.13  & 15.80 & 180,311\\
\hline
I Zw 18 & $B$\arcmin\ & 20.72$\pm$0.23  & 123$\pm$6 & 17.58 & 323  & 16.88  & 485  & 17.69  & 
 16.43$\pm$0.02  & 16.42 & 114,221\pano\\
{\sf main body} & $V$\arcmin\ & 20.73$\pm$0.13  & 126$\pm$4 & 17.55 & 331  & 16.90  & 494  & 17.66  & 
16.42$\pm$0.03  & 16.40 & 121,230\\
{\sf HST}  & $R$\arcmin\ & 20.50$\pm$0.16  & 123$\pm$9 & 17.37& 343  & 16.82  & 509  & 17.46  & 
 16.31$\pm$0.02  & 16.31 & 129,240\\
\hline
I Zw 18$^{\rm c}$  & $B$ & 22.01$\pm$0.22  & 105$\pm$4 & 19.22 & 199  & 20.52  & 289  & 20.19  & 
 19.18$\pm$0.05  & 19.19 & 188,298 \pano \\
{\sf C component} & $V$ & 21.99$\pm$0.27  & 105$\pm$6 & 19.19 & 187  & 20.64  & 292  & 20.15  & 
 19.19$\pm$0.06  & 19.20 & 189,300\\
{\sf HST} & $R$ & 21.73$\pm$0.31  & 103$\pm$8 & 18.98 & 159  & 20.88  & 310  & 19.85  & 
 19.1~\,$\pm$0.05  & 19.08 & 193,305 \\
\hline
\end{tabular}

\parbox{17cm}{$^{\rm a}$ A distance of 15 Mpc has been adopted for I~Zw~18 (Izotov et al. \cite{Yu01a}).
The tabulated values assume an extinction $A_V=0.13$ mag for I~Zw~18 and no
extinction for the C component. \\
$^{\rm b}$ The $B$ magnitude, as obtained from \hst/WFPC2 
F450W images is 15.83 mag, nearly 0.3 mag brighter than the magnitude 
derived from ground-based Johnson $B$ images. \\
$^{\rm c}$ The intensity distribution of the LSB component of I Zw 18~C has been 
modelled adopting an exponential 
distribution which flattens inward (Eq.~\ref{eq:p96a}) with ($b$,$q$)~=~(2.4,0.8). 
Note that this 
model implies for the LSB component a total magnitude 0.5 mag 
fainter than the integrated magnitude of a purely exponential distribution 
(Col. 5).}
\end{table*}
%
%
Towards the southeastern tip of I Zw 18, however, where, 
from the \ew(\ha) map (Fig. \ref{f8}b), ionized gas emission 
is weak, both colours become progressively bluer.
In regions 3 and 4 (Fig. \ref{f9}a) and throughout the 
southeastern tip of the \lsb\ body (region $\omega$ 
in Fig. \ref{f8}), the $B$\arcmin~--~$V$\arcmin\ and 
$V$\arcmin~--~$R$\arcmin\ colours decrease to $\sim$~0.05 -- 0.1 mag 
and 0.12 -- 0.18 mag, respectively. 

At the same time, region $\omega$ is identifiable with the reddest 
part of the \lsb\ host with respect to the $V-I$ colour 
(cf. Fig. \ref{f8}c), determined to be 0.17\dots 0.32 mag.
Thus, the \lsb\ component of I Zw 18 in its southeastern tip is by $\sim$0.3 mag 
redder than the ionized envelope regarding its $V-I$ colour (Fig. \ref{f6}d).
It is, however, $\sim$~0.6 mag bluer than the LSB component of normal BCDs, such 
as VII~Zw~403 (Fig. \ref{f5}d) or that of the nE BCD Mkn 996 (Thuan et al. \cite{TIL95}).

In summary, the {\sl unresolved} \lsb\ host of I Zw 18 which is well 
approximated by the colour profiles in Fig. \ref{f10}b has significantly 
bluer colours than in standard evolved BCDs.
%
\subsection{The C component \label{S4d} }
%
Before we proceed to the discussion of the implications of our photometric
analysis for the evolutionary state of I Zw 18 we briefly discuss the
photometric structure of the faint northwestern component I Zw 18\,C.  
Surface brightness profiles for this system (Fig. \ref{f10}c) were derived
from \hst/WFPC2 $B$, $V$ and $R$ exposures after removal of compact sources 
in its periphery (Sect. \ref{S4c}). 
Figure \ref{f10}c shows that the faint ($>$ 26 $B$ \sbb) outer parts of its 
LSB component are well fitted by an exponential law.
However, profile extrapolation to the center results in a higher intensity
than is observed in the radius range 0.05~$\la R^*\,{\rm (kpc)}\la$~0.2, 
implying a significant flattening of the exponential profile inside $\sim$~0.3 kpc.
This type of convex profile has frequently been reported for early- and
late-type dwarfs (Vennik et al. \cite{vennik00}; Guseva et al. \cite{Nat01} 
and references therein). Such a distribution can be approximated by 
a S\'ersic profile with an exponent $\eta>$1 (e.g. Caon et al. \cite{caon93}; 
Cellone et al. \cite{cellone94}) and an exponential distribution flattening 
inwards, such as the one proposed in \cite{P96a}:
\begin{equation}
I(R^*) = I_0\,\exp\left( -\frac{R^*}{\alpha}\right)
\big[1-q\,\exp(-P_3(R^*))\big],
\label{eq:p96a} 
\end{equation}
where $P_3(R^*)$ is
\begin{equation}
P_3(R^*) = \left(\frac{R^*}{b\,\alpha}\right)^3+\left(\frac{R^*}{\alpha}\,\frac{1-q}{q}\right).
\label{eq:p96b} 
\end{equation}
Near the center, the intensity distribution (Eq. \ref{eq:p96a}) depends
on the relative central intensity depression 
$q=\Delta I/I_0$, where $I_0$ is the central intensity obtained 
by extrapolation to $R^*=0$ of the outer exponential slope with scale length 
$\alpha$, and $b \alpha$ is the cut-off radius inside which 
the flattening occurs.

Following the procedure described in Guseva et al. (\cite{Nat01}), we 
fitted Eq. (\ref{eq:p96a}) to the SBPs of I Zw 18 C for $R^*\geq 0.33$ kpc 
adopting $b=2.4$ and $q=0.8$. This fit (grey curve in Fig. \ref{f10}c)
gives for the LSB population an absolute $B$ magnitude of --11.15 mag and
$\alpha\sim$~100 pc, close to the scale-length obtained in 
Sect. \ref{S4c} for the main body. 

Figure \ref{f10}d shows that, for $R^*\geq$~0.2 kpc, the mean $B-V$ and $V-R$ 
colours are respectively 0~$\pm$~0.04 mag and 0.1~$\pm$~0.04 mag. 
Because of its compactness, 
the radial colour distribution of component C cannot be derived from the 
available ground-based data. We measured a global $V-I$ colour of 0.12 mag 
for the C component, in good agreement with the value of 0.1 mag of 
van Zee et al. (\cite{vZ98}). 
As noted in Izotov et al. (\cite{Yu01a}), I~Zw~18~C shows a conspicuous 
colour gradient along its major axis, becoming markedly bluer from its
northwestern tip ($B-V$ = 0.05 mag, $V-I$ = 0.2 mag) to its southeastern tip 
($B-V$ = --0.07 mag and $V-I$ = --0.2 mag). 
%
\section{Discussion \label{S5} }
\subsection{The structural properties of the stellar LSB component in I Zw 18
\label{dis1}}
\begin{figure*} 
\begin{picture}(16.4,9)
\put(-0.3,0){{\psfig{figure=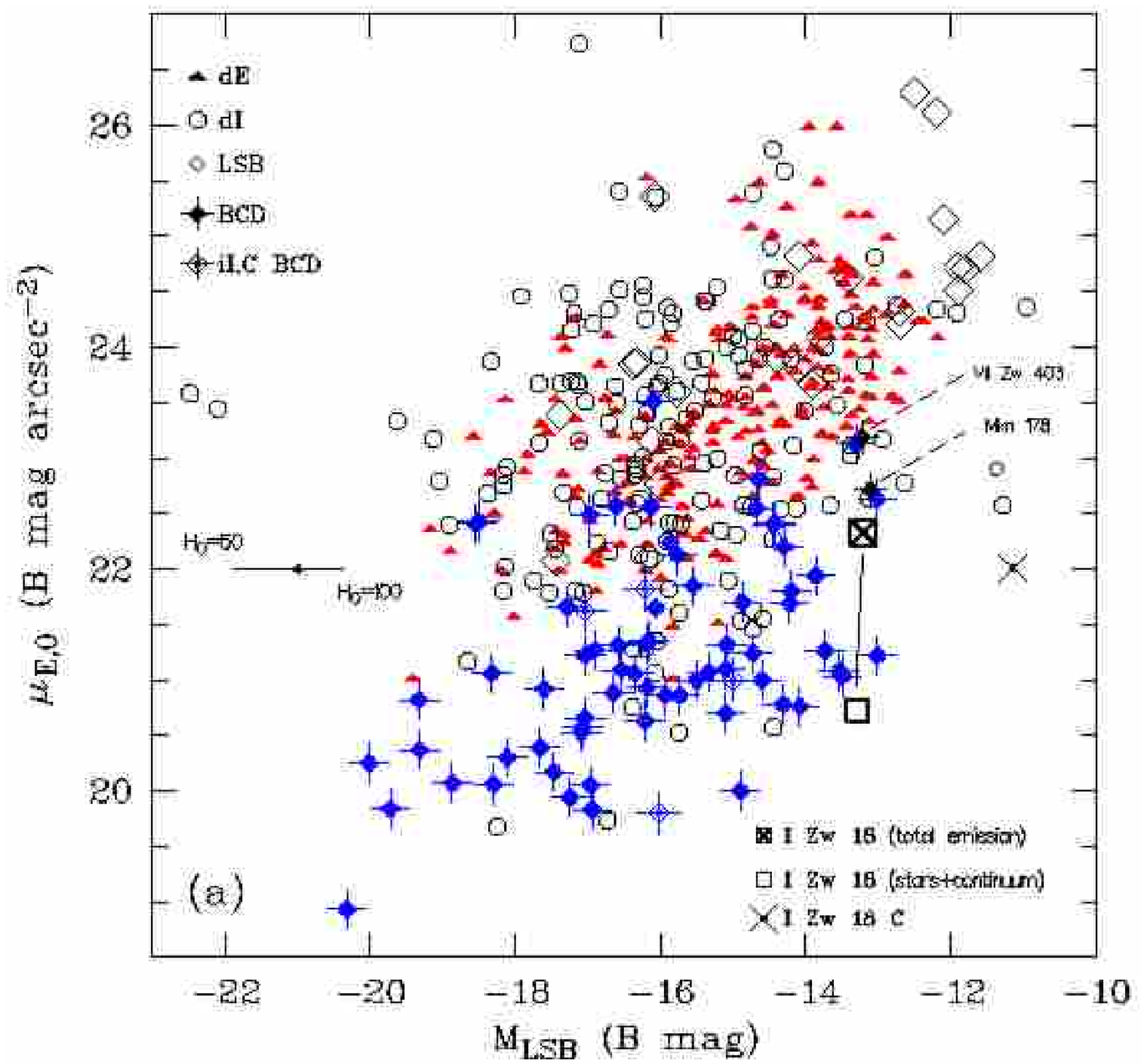,height=8.8cm,angle=0.,clip=}}}
\put(9,0){{\psfig{figure=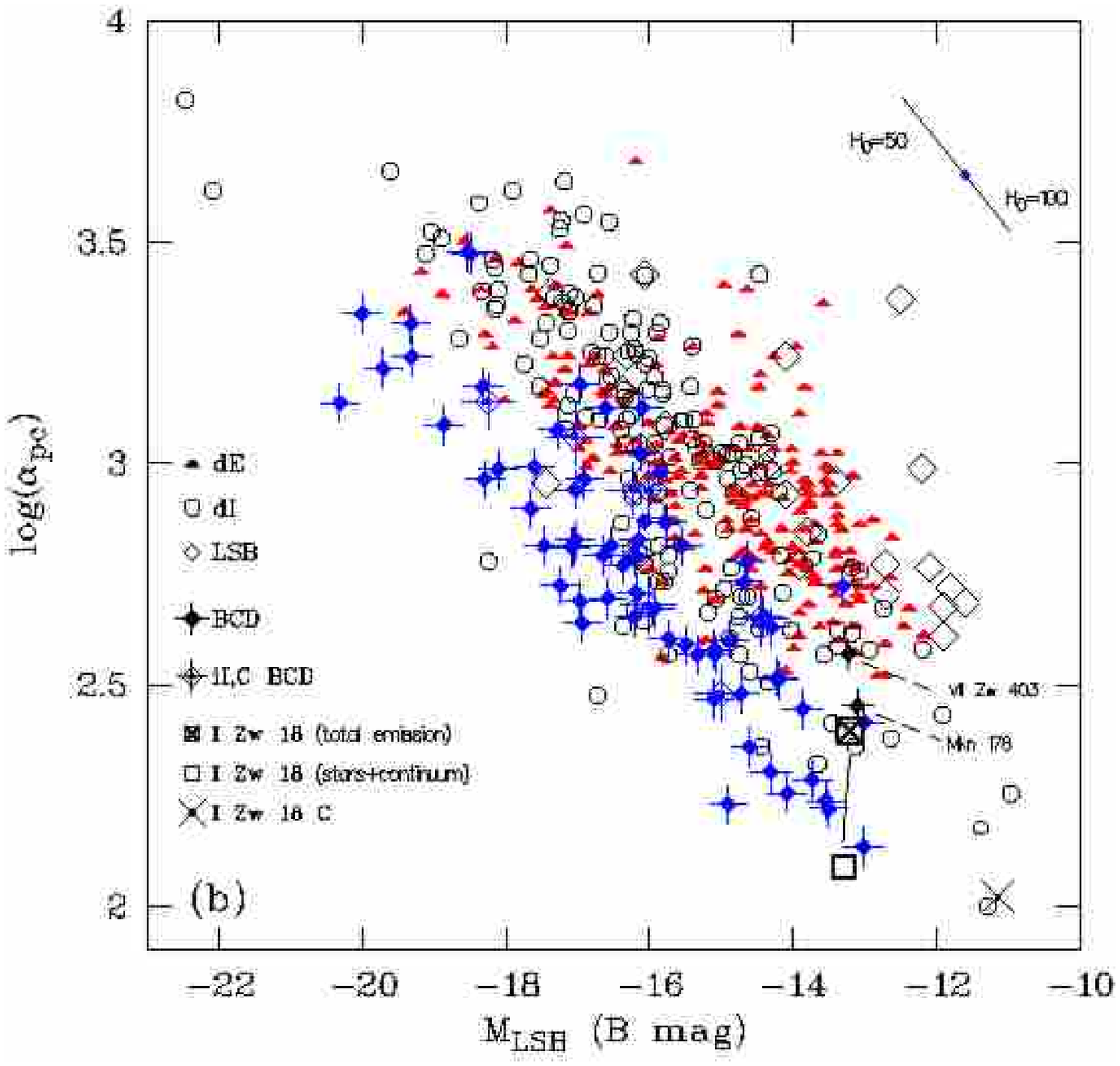,height=8.8cm,angle=0.,clip=}}}
\end{picture}
\caption{Comparison of the structural properties of the stellar LSB component 
of BCDs with those of other types of dwarf galaxies. 
Data for iE/nE BCDs are compiled from 
Cair\'os et al. (\cite{LM01a}), Drinkwater \& Hardy (\cite{drinkwater91}),
Marlowe et al. (\cite{Mar97}), Noeske (\cite{noeske99}), 
Noeske et al. (\cite{Kai01b}) and \cite{P96a}, and for
{\sl cometary} iI,C BCDs from Noeske et al. (\cite{Kai00}). 
Data for other types of dwarf galaxies (dE, dI and LSB) are 
taken from Binggeli \& Cameron 
(\cite{binggeli91}, \cite{binggeli93}), 
Bothun et al. (\cite{bothun91}), Caldwell \& Bothun (\cite{caldwell87}), 
Carignan \& Beaulieu (\cite{carignan89}), Hopp \& Schulte--Ladbeck (\cite{Hopp91}), 
Patterson \& Thuan (\cite{PT96}), 
Vigroux et al. (\cite{vigroux86}) and van Zee (\cite{vZ00}).
The horizontal line shows
the shift of the data points caused by a change of the Hubble constant 
from 75 to 50 and 100 km s$^{-1}$ Mpc$^{-1}$.
{\bf (a)} Central surface brightness $\mu _{\rm E,0}$ vs. 
absolute $B$ magnitude $M_{\rm LSB}$ of the LSB component.
{\bf (b)} Logarithm of the exponential scale length $\alpha $ in pc 
vs. $M_{\rm LSB}$. Without correction for nebular line emission, I~Zw~18 is 
located close to Mkn 178 and 
VII Zw 403 (Sect. \ref{S3a}), between the regions populated by 
iE/nE BCDs and other types of dwarf galaxies. Subtraction of nebular line
emission shifts I Zw 18 by --1.6 mag and --0.31 dex in panels a and b, making
it one of the most compact BCD known. 
The absolute $B$ magnitude of the LSB component of I Zw 18\,C (--11.15 mag) 
has been determined by extrapolating the fitting formula Eq. (\ref{eq:p96a}) 
to infinity. Note that the actual central surface brightness of 
I Zw 18 C, as given by Eq. (\ref{eq:p96a}) is $\approx$~1.75 mag fainter than 
the extrapolated value $\mu_{\rm E,0}\approx$~22 $B$ \sbb.}
\label{f12}
\end{figure*}

%
In Fig. \ref{f12} we compare the absolute $B$ magnitude $M_{\rm LSB}$, 
the central surface brightness $\mu_{\rm E,0}$ (panel a) and the
exponential scale length $\alpha$ (panel b) of the LSB component of 
different types of dwarf galaxies. 
I Zw 18 is nearly indistinguishable from Mkn 178 and VII Zw 403 
in the $\mu_{\rm E,0}$~--~$\alpha$ plane prior to subtraction of 
ionized gas emission. All three systems lie in the intermediate 
zone between the regions populated by dIs and iE/nE BCDs. 
However, subtraction of nebular line emission shifts I~Zw~18 
by --1.6 mag and --0.31 dex in panels a and b, respectively. 
Note that the absolute $B$ magnitude of the stellar \lsb\ host (--13.3 mag) is close to the value
derived for the patchy envelope (--13.2 mag) prior to subtraction 
of nebular line emission (Sect. \ref{S4a}). This is because, although the \lsb\ 
component is smaller by a factor $\sim$~2 in $\alpha$, it is brighter by
$\approx$~1.6 mag in $\mu_{\rm E,0}$.
As for I Zw 18 C, despite a difference $\sim$~2 mag in $M_{\rm LSB}$, 
it is comparably compact ($\alpha\approx$~100 pc) as the main body.
Thus, for $R^*\ga3\alpha$ the main body and the C component of I~Zw~18 are very 
similar with respect to both the surface brightness distribution and the mean 
colour of their stellar LSB host galaxy (Fig. \ref{f10}b and \ref{f10}d).
%

As evident from Fig. \ref{f12}, the \lsb\ component of I Zw 18 does 
not show exceptional structural properties as compared to other 
intrinsically faint dwarf galaxies. 
It does not have an unusually bright $\mu_{\rm E,0}$ or an extraordinarily small $\alpha$.
This conclusion still holds even if the commonly assumed distance of 10 Mpc 
to I~Zw~18 is adopted (Hunter \& Thronson \cite{HT95}; Dufour et
al. \cite{Duf96b}; Aloisi et al. \cite{Alo99}). Assuming such a distance would
merely result in a shift of 0.88 mag in $M_{\rm LSB}$ and of --0.18 dex in 
$\log\alpha$. 
Note that there are a few dIs in Fig. \ref{f12} that are as compact as I~Zw~18
and I~Zw~18~C regarding their exponential scale length. Examples are UGC 8091 
(Patterson \& Thuan \cite{PT96}), DDO 210 (van Zee \cite{vZ00}) 
and UGC 5272\,B (Hopp \& Schulte-Ladbeck \cite{Hopp91}). The BCD 
HS 0822+3542 ($\alpha\approx$~85 pc, Kniazev et al. \cite{Alex00}), the  
Ursa Minor and Leo I dwarfs in the Local Group ($\alpha\ga$~100 pc, 
Caldwell et al. \cite{Caldw92}) and a number of compact ($\alpha\sim$~100 pc) dwarfs 
in the Fornax Cluster (Hilker et al. \cite{Hilker99}; Phillips et al. \cite{Phill01};
Drinkwater et al. \cite{Drink01}) have also comparable scale length. 

The archival \hst\ images, after subtraction of nebular line emission 
(Sect. \ref{S4b}), do not go deep enough to allow to directly test
the hypothesis of a faint ($\overline{\mu}_V\sim$~28 \sbb) and extended stellar 
disc, forming continuously over a Hubble time (Legrand \cite{Leg00}). 
From its predicted magnitude ($m_V\sim$~20 mag, or $\sim$~10\% of the 
\lsb\ light), and assuming that at least half of its mass 
is within 22\arcsec\ from the main body of I Zw 18 (the angular distance 
between the \nw\ region and the southeastern tip of I~Zw~18~C), we deduce an 
upper limit $\alpha_{\rm max}\la$~13\arcsec\ for its exponential scale length. 
Such a putative disc would escape detection at the outermost point of our 
SBPs ($\mu$~=~26.2~$\pm$~0.4 $V$ \sbb\ at $R^*$~=~8\farcs6) at 
the 1$\sigma$ level if its central surface brightness $\mu_{\rm E,0}>27.1$ 
\sbb\ and its scale length $\alpha>$~10\farcs 5. 
%
\begin{figure*} 
\begin{picture}(16,11.0)
\put(0,0){{\psfig{figure=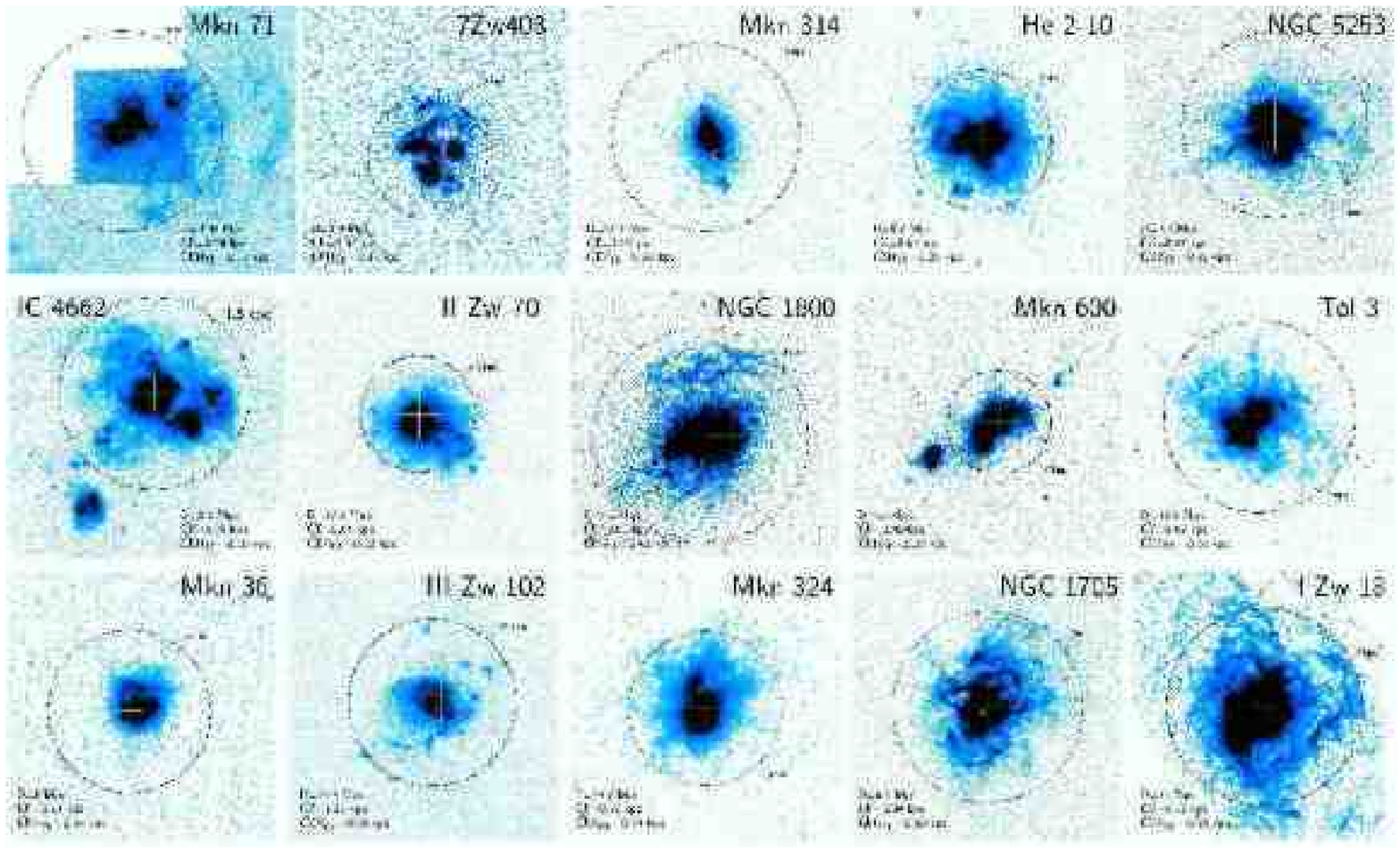,height=11.0cm,angle=0,clip=}}}
\end{picture}
   \caption[]{Continuum-subtracted \ha\ images of a sample of nearby BCDs, 
the \ha\ surface brightness profiles of which are shown in Fig. \ref{f14}.
Each sample galaxy is labelled with the adopted distance and the $B$ exponential 
scale length $\alpha$ of its stellar LSB host. $\alpha_{\rm H\alpha}$ is the 
scale length of the outer exponential part of the corresponding \ha\ profile. 
Each circle is labelled with its diameter in kpc.}
\label{f13}
  \end{figure*}

While the postulated disc can neither be confirmed nor ruled out 
with the present data, its stability over a Hubble time, in the presence of 
I Zw 18 C, has yet to be demonstrated by numerical simulations. 
Moreover, the disc hypothesis does not appear to be in accord with recent
CMD studies. The few red ($V-I\geq$~1 mag) point sources detected 
in I Zw 18 (Aloisi et al. \cite{Alo99}) are all located within the 
\lsb\ host (Izotov \& Thuan \cite{IT01}). This is in contrast with CMD 
studies of standard BCDs which, in agreement with surface photometry, 
reveal a sizeable population of old stars lying well beyond
the plateau radius \p25\ (Sect. \ref{S3a}).
As the detectability of a point source above the photon noise 
decreases with the square root of the local intensity level, evolved 
stars would be traceable, if present, in the periphery ($\mu>$27 \sbb) 
of the hypothetical disc rather than in front of the by two orders 
of magnitude brighter ($\mu<$22 \sbb) background in the central part 
of the \lsb\ host.
The observed concentration of red point sources toward the center of 
I Zw 18 suggests that their red colours are not due to a large age. 
Hunt et al. (\cite{Hunt02}) have argued that these sources 
may be red supergiants or H {\sc ii} regions reddened by dust.
%
\subsection{The ionized gas component of I Zw 18\label{dis2}}
%
We have seen that ionized gas emission dominates 
the light of I~Zw~18 for $R^*>$~0.2 kpc (Fig. \ref{f11}), and that its
intensity falls off exponentially in the radius range 0.6 $\la$ $R^*$
(kpc) $\la$ 1.3 (Fig. \ref{f6}a and \ref{f6}c).
We wish to explore here whether this exponential fall-off is particular 
to I Zw 18 or it is a common property of the ionized halo of BCDs on scales 
of several kpc away the SF region. 
%
\begin{figure}[!t]
\begin{picture}(8.4,17.)
\put(0,0){{\psfig{figure=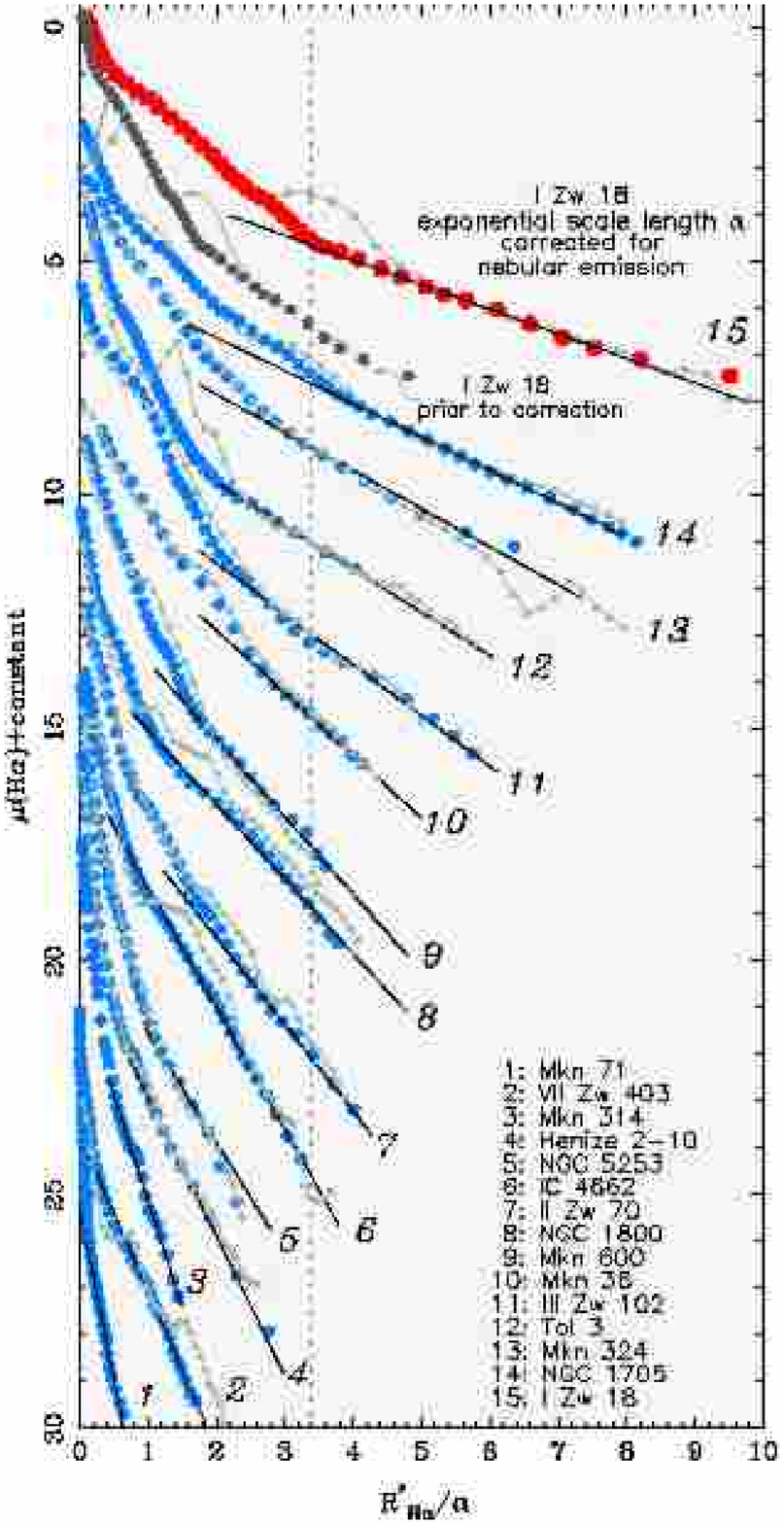,width=8.6cm,angle=0,clip=}}}
\end{picture}
   \caption[]{\ha\ surface brightness profiles of the BCDs shown in
Fig.~\ref{f13}, derived with method {\sf iv} (large filled symbols) 
and with concentric circular apertures (small interconnected symbols). 
The profiles have been shifted vertically by an arbitrary amount for clarity. 
The \ha\ surface brightness $\mu{\rm (H\alpha)}$ is plotted against
$R^*_{\rm H\alpha}/\alpha$ where $R^*_{\rm H\alpha}$ is
the equivalent radius of the \ha\ light and $\alpha$ is the $B$
exponential scale length of the stellar LSB component. For I Zw 18,
the \ha\ profile with $R^*_{\rm H\alpha}/\alpha$ 
obtained from the scale length of the profile of the LSB envelope, prior 
to correction for nebular line emission is also included for comparison.
Linear fits to the outer exponential part of each SBP are
shown by black lines.
The vertical dotted line indicates the mean \e25/$\alpha$ ratio
of 3.4~$\pm$~0.14 derived by Papaderos et al. (\cite{P96b})
for a sample of evolved BCDs.}
  \label{f14}
  \end{figure}

We have examined the \ha\ surface brightness profiles of a 
small sample of BCDs. The profiles were derived from 
continuum-subtracted \ha\ images (Fig. \ref{f13}),
based partly on data taken with the Danish 1.54m telescope at
La Silla (Papaderos \& Fricke \cite{PF98}; Noeske et al. \cite{Kai01b}). 
The La Silla sample includes the BCDs He 2-10, NGC 5253 and the BCD/dI 
IC 4662. 
We have also used archival \ha\ images of the BCDs Tol 3 
(ESO/NTT, PI: Vacca, Program 53.1-0086), NGC 1705 and NGC 1800 
(ESO/VLT, PI: T\"ullmann, Program 64.N-0399A; cf. Bomans \cite{Dom01}), 
as well as \hst/WFPC2 narrow-band images of the 
\ion{H}{ii} region Mkn 71 in the nearby galaxy 
NGC 2366 (Noeske et al. \cite{Kai00} and references therein).
In addition, we included H$\alpha$ images for VII Zw 403 (Sect. \ref{S3a})
as well as for a subset of objects in Cair\'os et al. (\cite{LM01b}),
consisting of Mkn~36, Mkn 314, Mkn 324, Mkn 600, III~Zw~102 and II~Zw~70. 
The H$\alpha$ images are shown in Fig. \ref{f13}, each labelled with the 
adopted distance to the BCD and the $B$ exponential scale length $\alpha$ 
of its stellar LSB component, taken from Cair\'os et al. (\cite{LM01b}) for Mkn 36, Mkn 314, 
Mkn 324, Mkn 600, III Zw 102, II Zw 70, from Marlowe et al. (\cite{Mar97}) 
for NGC 1705, NGC 1800, from Noeske et al. (\cite{Kai00}) for Mkn 71, 
from Papaderos \& Fricke (\cite{PF98}) for He 2-10, from Noeske
et al. (\cite{Kai01b}) for NGC 5253, Tol 3, IC 4662 and 
from Table \ref{iEphotom} for VII Zw 403.

In Fig. \ref{f14} we plot the \ha\ surface brightness 
distribution of each BCD, shifted vertically by an arbitrary value,
versus the equivalent radius $R^*_{\rm H\alpha}$ of the \ha\ emission, 
normalized to the scale length $\alpha$.
SBPs were derived using method {\sf iv} (large filled circles) 
and concentric circular apertures centered on the intensity maximum 
in each \ha\ image (small interconnected circles). 
We apply also the latter relatively crude technique to make sure that
the derived SBP slope does not depend sensitively on the 
profile extraction method. The SBPs were derived from 
the total \ha\ emission, without distinguishing between emission from 
HSB \hh\ regions and from the diffuse ionized medium. 
Note that the \ha\ profiles of Mkn 600 do not include the 
detached \ha\ sources indicated by rectangles in Fig. \ref{f13}. Furthermore, 
in computing the \ha\ profiles of IC 4662, we have excluded region {\sl d} 
(Heydari-Malayeri et al. \cite{HMM91}; see also 
Hidalgo-Gamez et al. \cite{HMO01}) $\sim$~1\arcmin\ southwest of the galaxy
(rectangle in Fig. \ref{f13}). 

From Fig. \ref{f14} it is evident that the \ha\ surface brightness profiles
show for all sample galaxies a nearly exponential regime in the outer parts, 
typically at surface brightness levels fainter by $\ga$~5 mag than the \ha\ 
maximum. 
The galactocentric radius where the exponential behaviour appears 
shows no dependence on the structural properties of the 
LSB galaxy, ranging from $<$~1$\alpha$ in VII Zw 403 to $\sim$~4$\alpha$ 
in NGC 1705. Given an average \e25/$\alpha$ ratio of $\sim$~3.4 for BCDs 
(Papaderos et al. \cite{P96b}) where \e25\ is the isophotal radius at a
surface brightness level of 25 \sbb, our sample (Fig.~\ref{f14})
includes quite a few systems where the exponential \ha\ slope 
continues well beyond the optical size of the galaxy.
The most extreme case is NGC 1705 where 
deep VLT data allow us to trace the exponential intensity fall-off of its
ionized gas halo out to $\sim$~8$\alpha$ (3.1 kpc), or more than twice its 
isophotal radius \e25, in agreement with the H$\alpha$ extent of $\sim$~2 kpc 
found by Meurer et al. (\cite{Meurer92}) for this galaxy.

\begin{figure}[h]
\begin{picture}(8.4,6.7)
\put(0,0){{\psfig{figure=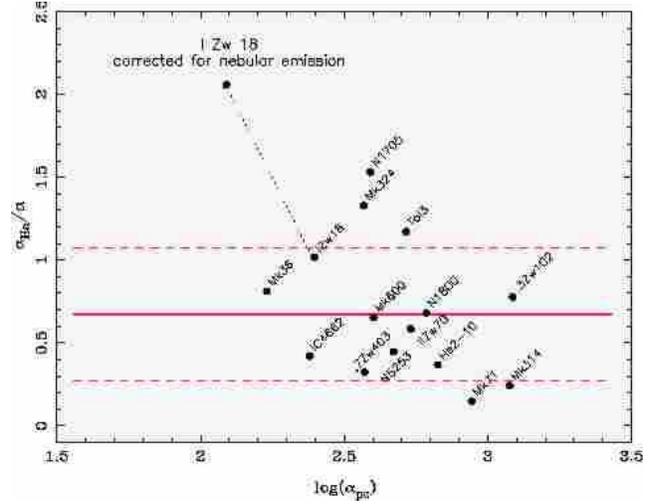,width=8.4cm,angle=0,clip=}}}
\end{picture}
   \caption[]{Ratio of the exponential scale length $\alpha_{\rm H\alpha}$
of the outer parts of the \ha\ profiles (Fig. \ref{f14}) to 
the $B$ exponential scale length $\alpha$ of the stellar LSB component 
for the BCDs shown in Fig. \ref{f13}. The thick line
indicates the mean $\alpha_{\rm H\alpha}/\alpha$ ratio of 0.67~$\pm$~0.4 
derived for all sample BCDs, excluding I Zw 18. Dashed lines show the
1$\sigma$ lower and upper bounds. Correction for nebular line emission
pushes I Zw 18 upwards to a $\alpha_{\rm H\alpha}/\alpha$ ratio 
of 2.06.}
  \label{f15}
  \end{figure}

Figure \ref{f15} reveals that the scale length of the ionized 
halo $\alpha_{\rm H\alpha}$ (0.1 -- 1 kpc) does not depend 
on the scale length of the stellar LSB host. 
Excluding I Zw 18, we obtain a mean $\alpha_{\rm H\alpha}/\alpha$ ratio of 0.67~$\pm$~0.4 
(thick gray line in Fig. \ref{f15}), implying that the \ha\ intensity decreases 
in the outer regions of BCDs more steeply than the stellar light. 
This fact suggests that ionized gas contamination is generally of no concern in surface 
photometry studies of the LSB component of standard BCDs. 
%

This {\sl may} not be the case, however, in BCDs with 
$\alpha_{\rm H\alpha}\ga\alpha$, i.e. for those systems located 
close to the line representing the 1$\sigma$ upper bound in Fig.~\ref{f15}. 
I Zw 18 falls into that category. We obtain for it,
prior to subtraction of nebular line emission, 
$\alpha_{\rm H\alpha}=253\pm10$ pc, i.e. $\alpha_{\rm H\alpha}/\alpha$
$\approx$~1 (cf. Table \ref{photom}). Adopting the actual scale length 
of the stellar \lsb\ host ($\alpha=$~123 pc) doubles the 
$\alpha_{\rm H\alpha}$/$\alpha$ ratio, shifting I Zw 18 to 
the upper left corner of Fig. \ref{f15}. 

The cursory study here suggests that an exponential intensity 
distribution may evolve in the warm ISM in the presence of SF activity
on both, small and large spatial scales.
Kennicutt (\cite{Kennicutt84}) has remarked that the emission measure of 
individual \hh\ regions follows a scalable exponential profile. 
This was found to also approximate well the surface brightness profiles 
of \hh\ regions in nearby face-on galaxies (Rozas et al. \cite{Rozas98})
and of the \ha\ halo of several Markarian starburst galaxies 
(Chitre \& Joshi \cite{Chitre01}).

From the evidence of Sect. \ref{S3}, it appears that extended 
ionized emission in BCDs does not imply significant
star formation beyond the \e25\ radius, although detached SF regions
may exist at the periphery of these systems (Mkn 600, 
Cair\'os et al. \cite{LM01b}; IC 4662, Hidalgo-Gamez et al. \cite{HMO01};
He 2-10, Beck \& Kovo \cite{BK99}). Instead, the exponential \ha\ distribution
traceable to several kpc away from the SF region, is probably the result
of the radiative and mechanical output of the starburst into
the ambient ISM. As such, it deserves a closer study, since the \ha\ extent 
and its scale length $\alpha_{\rm H\alpha}$ may give clues to 
the amplitude and temporal progression of SF activity in BCDs.

As for I Zw 18, its exponential \ha\ envelope is
not exceptional for BCDs, either by its extent or scale length. 
The central point is that I Zw 18 does not differ from standard BCDs 
in the properties of its ionized gas halo, but in its 
abnormally large $\alpha_{\rm H\alpha}/\alpha$ ratio.
In the absence of an appreciable stellar underlying component, 
ionized gas emission dominates entirely in the periphery of I Zw 18,
and its exponential profile mimicks that of a moderately red
($B-R\sim$~0.6 mag) stellar disc in optical SBPs. 
This has led some previous investigators 
to conclude erroneously that an evolved stellar disc exists in I Zw 18.
Complementing optical photometry with $V-I$ or $B-I$ colours (see the
discussion in Sect.~\ref{S3}; Izotov et al. \cite{Yu01a}) is essential for assessing 
the importance of ionized gas emission in the periphery of I Zw 18.

The case of I Zw 18 suggests caution in the search of 
more distant young galaxy candidates. Intense SF activity 
in the early phase of dwarf galaxy formation may result in an 
extended ionized gas halo which can be mistaken for an evolved 
stellar disc by studying only its exponential SBP.
%
\subsection{Clues to the evolutionary state of I Zw 18 \label{dis3}}
%
We have seen in Sect. \ref{S4b} that subtraction of nebular line emission 
reveals a stellar host galaxy extending slightly beyond the SF region. 
The structural properties and formation history of this underlying population 
are central to the age debate of I~Zw~18. 
Its blue colours are consistent with either the superposition of a 
young stellar population on an underlying old stellar LSB host, or
a moderately young stellar population formed instantaneously or over an
extended time period.

In the first scenario, the light of the old host galaxy is swamped 
by the light of an evenly distributed young $\sim$ 100 Myr stellar population.
By definition, photometry cannot disprove the presence of a massive though
faint old stellar substrate down to an arbitrarily low mass limit, since the 
latter would cause a barely detectable colour shift.
For instance, Papaderos et al. (\cite{P98}) have argued that a 10 Gyr old 
population making up to 90\% of the stellar mass may be present 
within the blue ($V-I\approx$ 0 mag) LSB envelope of the metal-poor 
BCD SBS~0335--052\,E. While such an old population cannot be entirely ruled out 
on photometric grounds (see also Vanzi et al. \cite{Vanzi00}), there is as yet 
no compelling reason to be postulated either.

In some BCDs a range of star-formation histories 
(SFHs) with very different mass fractions of young-to-old stars can account 
equally well for the observed colours, equivalent widths of Balmer 
absorption lines and spectral energy distributions.
For instance, population synthesis models by Guseva et al. (\cite{Nat01}) 
show that the above properties of the main body of the {\sl cometary} iI BCD 
SBS 0940+544 can be well reproduced either by a young stellar 
population forming continuously since less than 100 Myr ago, or by an old
stellar population which started to form 10 Gyr ago. 
In the latter case, however, the SFR over the last 100 Myr must have 
increased by a factor $\beta$ of at least five.
Likewise, in the case of I Zw 18, it is conceivable that a ratio of 
recent-to-past star formation rate $\beta > 1$ 
(cf. Guseva et al. \cite{Nat01}) may adequately explain its blue 
\lsb\ colours (Fig. \ref{f10}b). 
Enhanced SF activity on a radial extent of $\sim 0.65$ kpc 
does not contradict empirical evidence for BCDs (Sect. \ref{S3}); 
the \lsb\ host of I Zw 18 is comparable in size to the plateau component of
standard BCDs, where photometric signatures of a diffuse and relatively 
young stellar population are present (cf. Sect. \ref{S3a}).

If this was the case, however, the ratio $\beta$ must have been remarkably 
constant throughout the \lsb\ component, so as to eliminate any colour gradient. 
A constant $\beta$ would imply that the amplitude of recent 
star formation is directly proportional to the surface density of the 
underlying old stellar background. Although there is empirical evidence 
that the structural properties of the stellar LSB host influence 
the global SF process in BCDs and dIs (cf. Sect. \ref{S3b}), 
such a tight spatial coupling between the old
and young stellar populations does not seem plausible.
From these considerations and lacking solid photometric 
evidence for a red stellar LSB envelope usually seen
in standard BCDs for $\mu\ga$~25 $B$ \sbb, we do not consider
models with a dominant old host galaxy for I Zw 18.

Alternatively, if the \lsb\ component has formed in a single, 
instantaneous or extended SF episode, then its blue mean colours 
(Fig. \ref{f10}b) are consistent with an age not greater than a few 100 Myr.
The colours of the stellar \lsb\ population can be most reliably deduced 
from region $\omega$ (Fig. \ref{f8}b), where the contribution of ionized gas 
emission is negligible. The colours $B$\arcmin~--~$V$\arcmin~= 0.05 -- 0.08 mag, 
$V$\arcmin~--~$R$\arcmin~= 0.12 -- 0.18 mag, and $V-I$~= 0.17 -- 0.32 mag 
of this region translate for an instantaneously formed stellar population 
with the model assumptions described in Sect. \ref{S4a} to an age of $\sim$100 Myr.
The age increases to $\la$~0.5 Gyr for a constant or 
exponentially decreasing SF model with an e-folding time 
$\tau=$~1 Gyr, still considerably lower than 
the ages derived for standard iE/nE BCDs.

Similar considerations give for I Zw 18 C an age lower than or 
approximately equal to the age of the main body. The optical colours of the 
northwestern tip of I Zw 18 C ($B-V$~=~0.05 mag, $V-R$~=~0.12 mag, 
$V-I$~=~0.2 mag) are comparable to those of region $\omega$.
The limiting cases of an instantaneous burst and of a continuous SF give
stellar ages between $\la$~100 and 350 Myr. Such estimates are 
consistent with the upper age limit of 100 Myr derived recently 
for I Zw 18 C by Izotov et al. (\cite{Yu01a}), based on its broad-band 
colours, the equivalent width of Balmer emission and absorption lines and 
the spectral energy distribution of its stellar continuum.
To summarize, simple SFHs do not indicate a substantial stellar population 
with cosmological age in either the main body or its C component. 
They suggest instead that I Zw 18 formed coevally or successively 
within the last $\la$0.5 Gyr.

From the above considerations, we suggest that at least 
the two following conditions have to be fulfilled for 
a BCD to be a young galaxy candidate. It must have:
a) a single-burst stellar age of the order of the sound 
crossing time in its warm gas; and b) lack the photometric 
signatures of a red and uniformly distributed stellar population 
down to a very faint surface brightness level (e.g. $\sim$~29 $B$ \sbb).
The first criterion is consistent with the hypothesis
that the formation of the stellar population dominating 
the light is due to a single triggering signal propagating 
through the gaseous halo. 
The diameter of the \lsb\ component of I~Zw~18 being $\sim$~1.3 kpc,
corresponds to a timescale of $\sim$~130 Myr, assuming a sound velocity 
in its warm ISM of $\sim$~10 km s$^{-1}$ (approximately the H{\small I} 
velocity dispersion in I Zw 18 (van Zee et al. \cite{vZ98})
and in other dwarf galaxies (cf. e.g. van Zee et al. \cite{vZ98a},
Walter \& Brinks \cite{WB99}, Ott et al. \cite{Jurgen01})).
This timescale is close to the stellar age estimated
above for region $\omega$, assuming an instantaneous burst.
The second criterion ensures that an old extended stellar component
is either absent or dynamically insignificant. This condition is 
consistent with the notion that young stars had not yet time to 
diffuse far enough from their initial locus to gradually form 
an LSB envelope.

Stellar diffusion, together with propagating star formation 
(see Noeske et al. \cite{Kai00} and references therein), may play 
a key role in the build-up of the LSB component of BCDs. 
This is suggested by the recent CMD study 
of the relatively unevolved ($\leq$~2 Gyr) {\it cometary} BCD UGC 4483 
(Izotov \& Thuan \cite{IT01}). The main sequence stars in this system
are spatially confined to the SF region, while those with 
progressively larger ages (red supergiants and red giant branch stars)
are distributed more evenly and occupy successively larger 
volumes of the galaxy. This suggests stellar diffusion 
of young stars to the periphery (Izotov \& Thuan \cite{IT01}). 
As noted by Izotov \& Thuan (\cite{IT01}), the fact that 
red point sources in I Zw 18 still share the same volume 
with the young stellar component suggests that these stars have not had time to
diffuse out, and that I Zw 18 is considerably younger than UGC 4483.

A study of the role of stellar diffusion on the build-up and structural 
evolution of the LSB component of dwarf galaxies is beyond the scope of 
this paper. However, if this process is primarily driving the formation of the 
stellar LSB host in young BCDs, then the stars observed in the outermost 
regions of these systems cannot be younger than a diffusion 
timescale $\tau_{\rm diff}$.
Since at constant diffusion velocity the maximal galactocentric radius that a star 
can reach scales with the --2.5 power of its mass, stellar diffusion 
acts as a ``mass filter'', preventing stars with a lifetime shorter than
$\tau_{\rm diff}$ from contributing to the light of the LSB outer regions.
In that case, interpreting their colours assuming continuous SF 
lasting until the present would result in a systematic age overestimate. 
The maximum stellar age, as inferred from continuous SF models
assuming no SF since a time span $\tau_{\rm diff}$ or instantaneous 
burst models at various positions along the outermost parts of the LSB 
host may be a more reliable indicator of the evolutionary state of 
young BCD candidates.
Thus, the estimated single-burst stellar age of $\sim$ 100 Myr 
for region $\omega$ may be representative for all the \lsb\ population 
of I Zw 18, making the BCD an excellent candidate for being a truly 
young galaxy.
%
\section{Summary and conclusions \label{Conclusions}}
%
The central question to this study is (i) the evolutionary state of I Zw 18 
and (ii) whether this galaxy is qualitatively different as compared to 
standard blue compact dwarf (BCD) galaxies concerning its photometric structure.

We first discuss the photometric structure of two nearby {\sl bona fide} old 
BCDs, Mkn 178 and VII Zw 403, to provide a comparison basis for I Zw 18.
Both galaxies exemplify that star formation (SF) in BCDs does not extend out 
to a fortuitous galactocentric distance but, as consistently implied by
surface photometry and CMD studies, occurs within the 25 $B$ \sbb\ 
isophotal radius \e25\ of the underlying low-surface-brightness (LSB) host.
The confinement of star-forming activity to the inner part of a BCD results in
an appreciable colour contrast ($\sim$0.6--1 $B-R$ mag) 
between the star-forming and LSB component. 

Using ground-based $B$, $V$, $R$, $I$ and $J$ and archival \hst/WFPC2 
images we investigate next the photometric structure of I Zw 18.
Our surface brightness profiles (SBPs) confirm the exponential intensity 
fall-off and the moderately red $B-R$ colour ($\sim$~0.6 mag) found in 
the outer parts of I Zw 18 by Kunth \& \"Ostlin (\cite{KO00}) which led 
them to conclude that I Zw 18 possesses an evolved ($\sim$~5 Gyr) and 
spatially extended stellar disc underlying its star-forming regions. 

While the $B$ and $R$ SBPs alone suggest {\sl prima facie} 
that I Zw 18 resembles most other BCDs in having an underlying 
old stellar population, inspection of more colours severely 
challenges this interpretation.
The moderately red $B-R$ colour of the LSB envelope 
cannot be reconciled with the red $V-R$ ($>$~0.4 mag) 
and is in conflict with the blue $V-I$~$\sim$~0 mag, $B-V$~$\sim$~0.1 mag and 
$B-J$~$\sim$~0.6 mag. No evolved population model is consistent with these 
colours, irrespective of star formation history, age or metallicity.
This inconsistency can be removed by taking into account strong contamination of 
the optical light over galactocentric distances as large as $\sim$~1.3 kpc
by ionized gas emission.
This procedure is further supported by the large ($>$~1300 \AA) \ha\ equivalent 
widths observed in the periphery of the BCD, as well as by the good spatial 
correlation of the large H$\alpha$ equivalent widths with $V-R$ and $V-I$ colours.

We have used archival \hst\ narrow-band [\ion{O}{iii}]~$\lambda$5007 and
\ha\ images to correct broad-band $B$, $V$ and $R$ \hst/WFPC2 images for 
nebular line emission. 
While complete removal of ionized gas continuum and line emission is not possible
with the available narrow-band images, subtraction of the 
most prominent nebular emission lines is sufficient to allow us to
explore the hypothesis of a stellar host galaxy in I Zw 18.
We find that ionized gas contributes $\ga$~30\% to 50\% of the $R$ emission 
of I Zw 18. 
The stellar and ionized gas continuum emission 
dominates the $R$ light only within the effective radius ($R^*\approx$~0.2 kpc), 
whereas its line-of-sight contribution decreases to $<$~20\% at a galactocentric radius 
$R^*>$~0.65 kpc. Consequently, the exponential intensity decrease derived in the
LSB outer parts (0.6$\la R^*\,{\rm (kpc)}\la$1.3) of I Zw 18 from uncorrected
broad-band images is mainly due to ionized gas emission.

Subtraction of nebular line emission reveals a relatively smooth and very blue
stellar LSB envelope (denoted \lsb), extending not much beyond 
the star-forming regions of I~Zw~18. Although our surface photometry does 
not go below $\sim$~26 $B$ \sbb, it does allow for a quantitative study 
of the structural properties and colours of this underlying host galaxy.
We find that it is not exceptional when compared to intrinsically faint 
ultra-compact dwarfs, regarding its central surface brightness 
($\mu_{\rm E,0}=20.7$ $B$ \sbb) and exponential scale length 
($\alpha\approx$~120 pc). This is also the case for I~Zw~18~C, which shows in
its outer parts an exponential intensity decrease with $\alpha\approx$~100 pc 
and a marked flattening with respect to the exponential fit 
for $R^*\la$~0.3 kpc. 

Although it does not stand out in the $\mu_{\rm E,0}$~--~$\alpha$ 
parameter space, the \lsb\ host of I Zw 18, being blue on a radius 
range of $\sim$5 exponential scale lengths and showing little colour 
contrast to the star-forming regions, differs strikingly from the 
underlying component in standard BCDs.
The colours, as determined at its southeastern tip where ionized gas 
emission is negligible, are consistent with an instantaneous burst or 
continuous star formation starting less than 0.5 Gyr ago. 
Since I Zw 18 C shows comparably blue colours at its reddest northwestern 
tip, we conclude that most of the stellar mass in I Zw 18 has formed within 
the last 0.5 Gyr.

Finally, we show that the exponential intensity decrease observed in 
the filamentary envelope of I Zw 18 is typical of ionized gaseous 
halos in star-forming dwarf galaxies. I Zw 18 is not exceptional 
among BCDs, neither by the extent nor the exponential scale length 
of its ionized envelope. However, in the absence of a significant
underlying stellar population, extended ionized gas emission 
dominates the light in the periphery 
of I Zw 18, mimicking the SBP of a relatively red ($B-R\sim$0.6 mag) 
old stellar disc.

\begin{acknowledgements}
We thank L.-M. Cair\'os for making available 
to us \ha\ images of her BCD sample. Leslie Hunt graciously communicated 
to us her NIR results on I Zw 18 in advance of publication.
We are indebted to the referee, Dr. K. O'Neil, for her comments 
which helped clarifying the paper.
Research by P.P. and K.J.F. has been supported by the
Deutsches Zentrum f\"{u}r Luft-- und Raumfahrt e.V. (DLR) under
grant 50\ OR\  9907\ 7. Y.I.I. and N.G.G. have been partially supported 
by INTAS 97-0033 and Swiss SCOPE 7UKPJ62178 grants. K.G.N. acknowledges the 
support of DFG grant FR325/50-1.
Y.I.I. acknowledges the G\"ottingen Academy of Sciences
for a Gauss professorship and N.G.G. has been supported by the Deutsche 
Forschungsgemeinschaft (DFG) grant 436 UKR 17/14/01.  Those authors 
acknowledge support by the Volkswagen Foundation under grant No. I/72919. 
Y.I.I. and T.X.T have been partially supported by NSF grant AST-9616863. 
T.X.T. also acknowledges the support of {\sl HST} grant GO-08769.01-A.
This research has made use of data acquired with the 
European Southern Observatory telescopes, and obtained from 
the ESO/ST-ECF Science Archive Facility and of the NASA/IPAC 
Extragalactic Database (NED) which is operated by the 
Jet Propulsion Laboratory, CALTECH, under 
contract with the National Aeronautic and Space Administration.
\end{acknowledgements}


\begin{thebibliography}{}
\bibitem[1999]{Alo99} Aloisi, A., Tosi, M., \& Greggio, L. 1999, AJ, 118, 302
\bibitem[2001]{Alo01} Aloisi, A., Clampin, M., Diolaiti, E., et al. 2001, AJ, 
121, 1425
\bibitem[1999]{BK99}Beck, S. C., \& Kovo, O. 1999, AJ, 117, 190
\bibitem[1991]{binggeli91} Binggeli, B., \& Cameron, L. M. 1991, A\&A, 252, 27
\bibitem[1993]{binggeli93} Binggeli, B., \& Cameron, L. M. 1993, A\&AS, 98, 
297
\bibitem[2001]{Dom01}Bomans, D. J. 2002, in Dwarf Galaxies and their 
Environment, ed. K. S. de Boer, R.-J. Dettmar, \& U. Klein (Shaker Verlag), 
145
\bibitem[1991]{bothun91} Bothun, G. D., Impey, C. D., \& Malin, D. F. 1991, 
ApJ, 376, 404
\bibitem[2001a]{LM01a}Cair\'os, L. M., V\'{\i}lchez, J. M., 
G\'onzalez-P\'erez, J. N., Iglesias-P\'aramo, J., \& Caon, N. 2001a, 
ApJS, 133, 321
\bibitem[2001b]{LM01b}Cair\'os, L. M., Caon, N., V\'{\i}lchez, J. M., 
G\'onzalez-P\'erez, J. N., \& Mu\~noz-Tu\~n\'on, C. 2001b, ApJS, 136, 393 
\bibitem[2002]{LM02}Cair\'os, L. M., Caon, N., Garc\'\i a Lorenzo, B., 
V\'{\i}lchez, J. M., Mu\~noz-Tu\~n\'on, C. 2002, ApJ, in press (astro-ph/0204343)
%
\bibitem[1987]{caldwell87} Caldwell, N., \& Bothun, G. D. 1987, AJ, 94, 1126
\bibitem[1992]{Caldw92}Caldwell, N., Armandroff, T. E., Seitzer, P., \&
Da Costa, G. S. 1992, AJ, 103, 840 
\bibitem[2002]{Cannon01}Cannon, J. M., Skillman, E. D., Garnett, D. R., 
\& Dufour, R. J. 2002, ApJ, 565, 931
\bibitem[1993]{caon93}Caon, N., Capaccioli, M., \& D'Onofrio, M. 1993, 
MNRAS, 265, 1013
\bibitem[1989]{carignan89} Carignan, C., \& Beaulieu, S. 1989, ApJ, 347, 760
\bibitem[1994]{cellone94}Cellone, S. A., Forte, J. C., \& Geisler, D. 1994, 
ApJS, 93, 397
\bibitem[1999]{Chitre99} Chitre, A., \& Joshi, U. C. 1999, A\&AS, 139, 105
\bibitem[2001]{Chitre01} Chitre, A., \& Joshi, U. C. 2001, JApA, 22, 155 
\bibitem[1985]{Christian85} Christian, C. A., Adams, M., Barnes, J. V., et al.
 1985, PASP, 97, 363
\bibitem[2002]{Crone01} Crone, M. M., Schulte-Ladbeck, R. E., Greggio, L.,
\& Hopp, U. 2002, ApJ, 567, 258
\bibitem[1989]{Dav89} Davidson, K., Kinman, T. D., \& Friedman, S. D. 1989, 
AJ, 97, 1591
\bibitem[1997]{Doubl97} Doublier, V., Comte, G., Petrosian, A., Surace, C., 
\& Turatto, M. 1997, A\&AS, 124, 405
\bibitem[1999]{Doubl99} Doublier, V., Caulet, A., \& Comte, G.
  1999, A\&AS, 138, 213
\bibitem[1991]{drinkwater91} Drinkwater, M., \& Hardy, E. 1991, AJ, 101, 94
\bibitem[2001]{Drink01}Drinkwater, M. J., Gregg, M. D., Holman, B. A., \&
Brown, M. J. I. 2001, MNRAS, 326, 1076
\bibitem[1990]{Duf90} Dufour, R. J., \& Hester, J. J. 1990, ApJ, 350, 149
\bibitem[1996a]{Duf96a}Dufour, R. J., Esteban, C., \& Casta\~neda, H. O. 
1996a, ApJ, 471, L87
\bibitem[1996b]{Duf96b}Dufour, R. J., Garnett, D. R., Skillman, E. D., 
\& Shields, G. A. 1996b, in ASP Conf. Ser. 98, From Stars To Galaxies, ed. 
C. Leitherer, U. Fritze-v. Alvensleben, \& J. Huchra, 358
\bibitem[1997]{F97} Fioc, M., \& Rocca-Volmerange, B. 1997, A\&A, 326, 950
\bibitem[1980]{Fre80} French, H. B. 1980, ApJ, 240, 41
\bibitem[2001]{Fricke01} Fricke, K. J., Izotov, Y. I., Papaderos, P., 
Guseva, N. G., \& Thuan, T. X. 2001, AJ, 121, 169
\bibitem[2000]{Gil00a}Gil de Paz, A., Zamorano, J., Gallego, J., \& de
  B. Dom\'{\i}nguez, F. 2000, A\&AS, 145, 377
\bibitem[1988]{gr88}Gonz\'alez-Riestra, R., Rego, M., \& Zamorano, J. 1988,
  A\&A, 202, 27
\bibitem[2001]{Grebel00} Grebel, E. K. 2001, in Dwarf Galaxies and their 
Environment, ed. K. S. de Boer, R.-J. Dettmar, \& U. Klein (Shaker Verlag), 45
\bibitem[2000]{Nat00}Guseva, N. G., Izotov, Y. I., \& Thuan, T. X. 2000, ApJ,
531, 776
\bibitem[2001]{Nat01}Guseva, N. G., Izotov, Y. I., Papaderos, P., 
et al. 2001, A\&A, 378, 756
\bibitem[2000]{Heller00}Heller, A. B., Brosch, N., Almoznino, E., van Zee, L.,
\& Salzer, J. J. 2000, MNRAS, 316, 569
\bibitem[1990]{HMM91} Heydari-Malayeri, M., Melnick, J., \& Martin, J.-M. 
1990, A\&A, 234, 99
\bibitem[2001]{HMO01} Hidalgo-Gamez, A. M., Masegosa, J., \& Olofsson, K. 
2001, A\&A, 369, 797
\bibitem[1999]{Hilker99}Hilker, M., Infante, L., Vieira, G., 
Kissler-Patig, M., \& Richtler, T. 1999, A\&AS, 134, 75
\bibitem[1995]{Holtz95}Holtzman, J. A., Burrows, C. J., Casertano, S., et al. 
1995, PASP, 107, 1065 
\bibitem[1991]{Hopp91} Hopp, U., \& Schulte--Ladbeck, R. E. 1991, A\&A, 
248, 1
\bibitem[2000]{Hopp00} Hopp, U., Engels, D., Green, R. F., et al. 
2000, A\&AS, 142, 417
\bibitem[1998]{Hunt98} Hunt, L. K., Mannucci, F., Testi, L., et al. 1998, 
AJ, 115, 2594
\bibitem[2002]{Hunt02} Hunt, L. K., Thuan, T. X., \& Izotov, Y. I. 2002, 
ApJ, submitted
\bibitem[1995]{HT95} Hunter, D. A., \& Thronson, H. A., Jr. 1995, ApJ, 452, 
238
\bibitem[1998]{Hunter98}Hunter, D. A., Elmegreen, B. G., \& Baker, A. L. 1998,
ApJ, 493, 595
\bibitem[1998a]{IT98a} Izotov, Y. I., \& Thuan, T. X. 1998a, ApJ, 497, 227
\bibitem[1998b]{IT98b} Izotov, Y. I., \& Thuan, T. X. 1998b, ApJ, 500, 188
\bibitem[2002]{IT01}Izotov, Y. I., \& Thuan, T. X. 2002, ApJ, 567, 875
\bibitem[1997]{ILCFGK97c} Izotov, Y. I., Lipovetsky, V. A., Chaffee, F. H., 
et al. 1997, ApJ, 476, 698
\bibitem[1999]{Yu99} Izotov, Y. I., Chaffee, F. H., Foltz, C. B., et al.
1999, ApJ, 527, 757
\bibitem[2001a]{Yu01a} Izotov, Y. I., Chaffee, F. H., Foltz, C. B., 
et al. 2001a, ApJ, 560, 222 
\bibitem[2001b]{Yu01b}Izotov, Y. I., Chaffee, F. H., \& Schaerer, D. 2001b, 
A\&A, 378, L45
\bibitem[1984]{Kennicutt84} Kennicutt, R. C. 1984, ApJ, 287, 116
\bibitem[1981]{KD81}Kinman, T. D., \& Davidson, K. 1981, ApJ, 243, 127
\bibitem[2000]{Alex00}Kniazev, A. Y., Pustilnik, S. A., Masegosa, J., et al.
2000, A\&A, 357, 101
\bibitem[1992]{Harald92}Kr\"uger, H. 1992, PhD, Univ. G\"ottingen
\bibitem[1995]{Harald95}Kr\"uger, H., Fritze-v. Alvensleben, U., \& Loose,
  H.-H. 1995, A\&A, 303, 41
\bibitem[2000]{KO00} Kunth, D., \& \"Ostlin, G. 2000, A\&A Reviews, 10, 1 
\bibitem[1988]{KMV88} Kunth, D., Maurogordato, S., \& Vigroux, L. 1988, A\&A, 
204, 10
\bibitem[1992]{Landolt92} Landolt, A. U. 1992, AJ, 104, 340
\bibitem[2000]{Leg00} Legrand, F. 2000, A\&A, 354, 504
\bibitem[2000]{Leg00a} Legrand, F., Kunth, D., Roy, J.-R., Mas-Hesse, J. M., 
\& Walsh, J. R. 2000, A\&A, 355, 891
\bibitem[1979]{Leq79} Lequeux, J., Rayo, J. F., Serrano, A., Peimbert, M.,
\& Torres-Peimbert, S. 1979, A\&A, 80, 155
\bibitem[1986]{LT86} Loose, H.-H., \& Thuan, T. X. 1986, in Star Forming
  Dwarf Galaxies and Related Objects, ed. D. Kunth, T. X. Thuan \& 
J. T. T.Van (Editions Frontieres), 73 (LT86)
\bibitem[1998]{Lynds98} Lynds, R., Tolstoy, E., O'Neil, E. J., Jr., \& Hunter,
  D. A. 1998, AJ, 116, 146
\bibitem[1999]{Makarova99} Makarova, L. 1999, A\&AS, 139, 491
\bibitem[1997]{Mar97}Marlowe, A. T., Meurer, G. R., Heckman, T. M., 
\& Schommer, R. 1997, ApJS, 112, 285
\bibitem[1999]{Mar99}Marlowe, A. T., Meurer, G. R., \& Heckman, T. M. 1999, 
ApJ, 522, 183
\bibitem[1996]{Martin96} Martin, C. L. 1996, ApJ, 465, 680
\bibitem[1999]{MHK99}Mas-Hesse, J. M., \& Kunth, D. 1999, A\&A, 349, 765
\bibitem[1992]{Meurer92}Meurer, G. R., Freeman, K. C., Dopita, M. A., 
\& Cacciari, C. 1992, AJ, 103, 60
\bibitem[1998]{Meurer98}Meurer, G. R., Staveley-Smith, L., \& 
Killeen, N. E. B. 1998, MNRAS, 300, 705
\bibitem[1999]{noeske99} Noeske, K. G. 1999, Diploma Thesis, 
Universit\"at G\"ottingen
\bibitem[2000]{Kai00} Noeske, K. G., Guseva, N. G., Fricke, K. J., et al.
2000, A\&A, 361, 33
\bibitem[2001]{Kai01a}Noeske, K. G., Iglesias-P\'aramo, J., 
V\'{\i}lchez, J. M., Papaderos, P., \& Fricke, K. J. 2001, A\&A, 371, 806
\bibitem[2002]{Kai01b} Noeske, K.G., et al. 2002, in prep.
\bibitem[2001]{Noguchi01}Noguchi, M. 2001, ApJ, 555, 289
\bibitem[2000]{Gor00}\"Ostlin, G. 2000, ApJ, 535, 99
\bibitem[1996]{Gor96} \"Ostlin, G., Bergvall, N., \& R\"onnback, J. 1996, 
in The Interplay Between Massive Star Formation, the ISM and Galaxy Formation,
ed. D. Kunth, B. Guiderdoni, M. Heydari-Malayeri, \& T. X. Thuan 
(Gif-sur-Yvette: Edition Fronti\`eres), 605
\bibitem[2001]{Jurgen01}Ott, J., Walter, F., Brinks, E., van Dyk, S.D., Dirsch,
B. \& Klein, U. 2001, AJ, 122, 3070
\bibitem[1992]{Pagel92} Pagel, B. E. J., Simonson, E. A., Terlevich, R. J., 
\& Edmunds, M. G. 1992, MNRAS, 255, 325
\bibitem[1999]{PF98}Papaderos, P. \& Fricke, K. J. 1999, in Highlights 
in X-ray Astronomy, eds. B. Aschenbach \& M.J. Freyberg, MPE Report 272, p. 193 (astro-ph/9810101)
\bibitem[P96a]{P96a} Papaderos, P., Loose, H.-H., Thuan, T. X., \& Fricke,
  K. J. 1996a, A\&AS, 120, 207 (P96a)
\bibitem[1996b]{P96b} Papaderos, P., Loose, H.-H., Fricke, K. J., \& 
Thuan, T. X. 1996b, A\&A, 314, 59
\bibitem[1998]{P98} Papaderos, P., Izotov, Y. I., Fricke, K. J., 
Thuan, T. X., \& Guseva, N. G. 1998, A\&A, 338, 43
\bibitem[2001]{P01} Papaderos, P., Izotov, Y. I., Noeske, K. G., Thuan, T. X.,
\& Fricke, K. J. 2001, in Dwarf Galaxies and their Environment,
ed. K. S. de Boer, R.-J. Dettmar, \& U. Klein (Shaker Verlag), 111
\bibitem[1996]{PT96} Patterson, R. J., \& Thuan, T. X. 1996, ApJS, 107, 103
\bibitem[1997]{Petr97} Petrosian, A. R., Boulesteix, J., Comte, G., Kunth, D.,
\& LeCoarer, E. 1997, A\&A, 318, 390
\bibitem[2001]{Phill01}Phillipps, S., Drinkwater, M. J., Gregg, M. D., 
\& Jones, J. B. 2001, ApJ, 560, 201
\bibitem[1999]{popescu99} Popescu, C. C., Hopp, U., \& Rosa, R. 1999, A\&A, 
350, 414
\bibitem[2001]{pustilnik00} Pustilnik, S. A., Kniazev, A. Y., 
Lipovetsky, V. A., \& Ugryumov, A.V. 2001, A\&A, 373, 24
\bibitem[1999]{RiHe00}Rieschick, A. \& Hensler, G. 2000, in ASP Conference 
Proceedings, 215 3rd Guillermo Haro Astrophysics Conference, Cosmic Evolution 
and Galaxy Formation: Structure, Interactions, and Feedback, ed. J. Franco,
L. Terlevich, O. L\'opez-Cruz, \& I. Aretxaga, 130
\bibitem[2000]{RoyeHunter00} Roye, E. W., \& Hunter, D. A. 2000, ApJ, 119, 1145
\bibitem[1988]{Rozas98} Rozas, M., Casta\~neda, H. O., \& Beckman, J. E. 
1998, A\&A, 330, 873
\bibitem[1989]{Salzer89}Salzer, J. J., MacAlpine, G. M., \& Boroson, T. A. 1989, ApJS, 70, 447
\bibitem[1970]{SS70} Sargent, W. L. W., \& Searle, L. 1970, ApJ, 162, L155
\bibitem[1998]{RU98}Schulte-Ladbeck, R. E., \& Hopp, U. 1998, AJ, 116, 2886
\bibitem[1998]{Reg98} Schulte-Ladbeck, R. E., Crone, M. M., \& Hopp, U. 1998,
  ApJ, 493, L23
\bibitem[1999a]{Reg99a} Schulte-Ladbeck, R. E., 
Hopp, U., Crone, M. M., \& Greggio, L. 1999a, ApJ, 525, 709
\bibitem[1999b]{Reg99b} Schulte-Ladbeck, R. E., Hopp, U., Greggio, L., 
\& Crone, M. M. 1999b, AJ, 118, 2705
\bibitem[2000]{Reg00} Schulte-Ladbeck, R. E., Hopp, U., Greggio, L., 
\& Crone, M. M. 2000, AJ, 120, 1713
\bibitem[1972]{SS72} Searle, L., \& Sargent, W. L. W. 1972, ApJ, 173, 25
\bibitem[1968]{S68}S\'ersic, J. L. 1968, Atlas de Galaxias Australes 
(C\'{o}rdoba: Obs. Astron.)
\bibitem[1993]{SK93}Skillman, E. D., \& Kennicutt, R. C., Jr. 1993, ApJ, 411, 655
\bibitem[1999]{Smoker99} Smoker, J. V., Axon, D. J., \& Davies, R. D. 1999, 
A\&A, 341, 725
\bibitem[1997]{tmt97}Telles, E., Melnick, J., \& Terlevich, R. 1997, 
MNRAS, 288, 78
\bibitem[1991]{Terl91} Terlevich, R., Melnick, J., Masegosa, J., Moles, M.,
\& Copetti, M. V. F. 1991, A\&AS, 91, 285
\bibitem[1991]{Thuan91}Thuan, T. X. 1991, in Massive Stars in Starbursts,
ed. C. Leitherer, N. R. Walborn, T. M. Heckman, \& C. A. Norman (Cambridge
  University Press), 183 
\bibitem[1995]{TIL95}Thuan, T. X., Izotov, Y. I., \& Lipovetsky, V. A. 1995, 
ApJ, 445, 108
\bibitem[1997]{TIL97} Thuan, T. X., Izotov, Y. I., \& Lipovetsky, V. A. 
1997, ApJ, 477, 661
\bibitem[1999]{TIF99} Thuan, T. X., Izotov, Y. I., \& Foltz, C. B. 1999,
  ApJ, 525, 105
\bibitem[2001]{Tosi01}Tosi, M., Sabbi, E., Bellazzini, M., et al. 2001, AJ, 
122, 1271
\bibitem[2000]{vZ00} van Zee, L. 2000, AJ, 119, 2757
\bibitem[2001]{vZ01} van Zee, L. 2001, AJ, 121, 2003
\bibitem[1998a]{vZ98a}van Zee, L., Skillman, E.D. \& Salzer, J. J. 1998a, AJ 116, 1186
\bibitem[1998b]{vZ98} van Zee, L., Westphahl, D., Haynes, M., \& Salzer, J. J.
  1998b, AJ, 115, 1000
\bibitem[2000]{Vanzi00}Vanzi, L., Hunt, L. K., Thuan, T. X., \& Izotov, Y. I. 
2000, A\&A, 363, 493 
\bibitem[1996]{vennik96}Vennik, J., Hopp, U., Kovachev, B., Kuhn, B.,
\& Els\"asser, H. 1996, A\&AS, 117, 261
\bibitem[2000]{vennik00}Vennik, J., Hopp, U., \& Popescu, C. C. 2000,
A\&AS, 142, 399
\bibitem[1987]{Via87} Viallefond, F., Lequeux, J., \& Comte, C. 1987, in
Starbursts and Galaxy Evolution, ed. T. X. Thuan, T. Montmerle, \& 
J. Tran Thanh (Editions Fronti\`eres), 139
\bibitem[1986]{vigroux86} Vigroux, L., Thuan, T. X., Vader, J. P., 
\& Lachi\`eze--Rey, M. 1986, AJ, 91, 70
\bibitem[1995]{Vi95}V{\'\i}lchez, J. M. 1995, AJ, 110, 1090
\bibitem[1998]{ViIp98}V{\'\i}lchez, J. M., \& Iglesias-P\'aramo J. 1998, ApJ, 508, 248
\bibitem[1999]{WB99}Walter, F., \& Brinks, E. 1999, AJ, 118, 273 
\bibitem[1999]{YH99}Youngblood, A. J., \& Hunter, D. A. 1999, ApJ, 519, 55
\bibitem[1966]{Fritz66} Zwicky, F. 1966, ApJ, 143, 192
%
\end{thebibliography}
\end{document}